\documentclass[12pt]{article}
\usepackage{epsfig}

\def\simgt{\rlap{\lower 3.5 pt \hbox{$\mathchar \sim$}} \raise 1pt \hbox {$>$}}
\def\simlt{\rlap{\lower 3.5 pt \hbox{$\mathchar \sim$}} \raise 1pt \hbox {$<$}}
\catcode`\@=11
\def\@citex[#1]#2{\if@filesw\immediate\write\@auxout{\string\citation{#2}}\fi
  \def\@citea{}\@cite{\@for\@citeb:=#2\do
    {\@citea\def\@citea{,\penalty\@m}\@ifundefined
       {b@\@citeb}{{\bf ?}\@warning
       {Citation `\@citeb' on page \thepage \space undefined}}%
\hbox{\csname b@\@citeb\endcsname}}}{#1}}
\def\citer{\@ifnextchar [{\@tempswatrue\@citexr}{\@tempswafalse\@citexr[]}}
\def\@citexr[#1]#2{\if@filesw\immediate\write\@auxout{\string\citation{#2}}\fi
  \def\@citea{}\@cite{\@for\@citeb:=#2\do
    {\@citea\def\@citea{--\penalty\@m}\@ifundefined
       {b@\@citeb}{{\bf ?}\@warning
       {Citation `\@citeb' on page \thepage \space undefined}}%
\hbox{\csname b@\@citeb\endcsname}}}{#1}}

\begin{document}

\begin{flushright}
Edinburgh 2001/06\\
hep-ph/0106120
\end{flushright}

\vspace*{2cm}

\begin{center}

{\large\bf Quarkonium Production at High-Energy 
Colliders\footnote{To be published in 
Progress in Particle and Nuclear Physics, Vol.\ 47, issue 1.}}\\[1.3cm]

{\sc Michael~Kr\"amer}

Department of Physics and Astronomy\\ 
The University of Edinburgh\\
Edinburgh EH9 3JZ, Scotland

\end{center}

\vspace*{0.3cm}

\begin{abstract} 
\noindent
The theoretical description of heavy quarkonium production at
high-energy $p\bar{p}$ and $ep$ colliders is reviewed. Predictions
based on non-relativistic QCD factorisation are confronted with recent
charmonium and bottomonium data from the Tevatron and HERA. Potential
shortcomings of the present theoretical analyses are discussed, and
the prospects for quarkonium physics at the upgraded Tevatron and HERA
colliders and at the LHC are summarised.
\end{abstract}

\vfill

\eject
\tableofcontents
\eject
\listoffigures
\eject
\listoftables
\eject

\section{Introduction}
The production of heavy quarkonium states at high-energy colliders has
been the subject of considerable interest during the past few years. A
wealth of new experimental results has become available, some of which
revealed dramatic shortcomings of earlier quarkonium models. In
theory, progress has been made on the factorisation between the
short-distance physics of heavy-quark creation and the long-distance
physics of bound state formation. The colour-singlet
model~\cite{Berger:1981ni,Baier:1981uk} has been superseded by a
consistent and rigorous framework, based on the use of
non-relativistic QCD (NRQCD) \cite{Bodwin:1995jh}, an effective field
theory that includes the so-called colour-octet mechanisms.  On the
other hand, the colour-evaporation model
\citer{Fritzsch:1977ay,Gluck:1978zm} of the early days of
quarkonium physics has been revived~\citer{Gavai:1995in,Edin:1997zb}.

However, despite the recent theoretical and experimental developments
the range of applicability of the different approaches is still
subject to debate, as is the quantitative verification of
factorisation. Because the quarkonium mass is still not very large
with respect to the QCD scale, in particular for the charmonium
system, non-factorisable
corrections~\citer{Brodsky:1997tv,Hoyer:1999dr} may not be suppressed
enough, if the quarkonium is not part of an isolated jet, and the
expansions in NRQCD may not converge very well. In this situation, a
global analysis of various processes is mandatory to assess the
importance of different quarkonium production mechanisms, as well as
the limitations of a particular theoretical framework.

\vspace{2mm}

Much of the recent interest and theoretical development in quarkonium
physics has been stimulated by experimental data from high-energy
$p\bar{p}$ and $ep$ colliders. The purpose of this review is to
summarise the theoretical progress in understanding heavy quarkonium
production, and to confront the predictions with recent charmonium and
bottomonium data from the Tevatron and HERA. Forthcoming experimental
results and theoretical advances, for example in the calculation of
higher-order corrections, will sharpen the present picture and allow
more quantitative tests of the theory. Therefore, this review will
mostly focus on the generic features of the various quarkonium
production processes and highlight the prospects for quarkonium
physics at future collider experiments. Quarkonium production has also
been studied in various other processes like $e^+e^-$ annihilation,
$Z$ decays, hadronic collisions at fixed-target experiments and $B$
decays, and several excellent reviews have been presented in the
literature which cover these aspects, see
e.g.~\citer{Braaten:1996pv,Maltoni:2000km}.

\vspace{2mm}

The NRQCD factorisation approach to quarkonium production is
introduced in Section~\ref{sec_qp} and compared to earlier quarkonium
models like the colour-singlet model and the colour-evaporation
model. Section~\ref{sec_tev} summarises the NRQCD predictions for
charmonium production and polarisation at hadron colliders. The
predictions are confronted with recent experimental data from the
Tevatron, and potential shortcomings of the theoretical analysis are
discussed. Charmonium production at HERA is reviewed in
Section~\ref{sec_ep}, and the prospects for quarkonium physics at the
upgraded HERA collider are summarised. Section~\ref{sec_conc}
concludes with an outlook on future work that will be required to test
the theoretical framework more rigorously.

\section{Quarkonium Production Mechanism\label{sec_qp}}

The production of heavy quarkonium at high-energy colliders involves
two distinct scales. The creation of a heavy quark pair is a
short-distance process on scales of the order $1/m_Q$ or smaller and
can be calculated in perturbation theory. The subsequent
non-perturbative transition from the intermediate $Q\overline{Q}$ pair
to a physical quarkonium, on the other hand, involves long-distance
scales of the order of the quarkonium size $1/(m_Q v)$ or
larger. Provided that these two scales are well separated, $1/(m_Q v)
\gg 1/m_Q$, the formation of the bound state should be insensitive to the
details of the heavy quark creation process, which is essentially
local on the scale of the quarkonium size. It is thus intuitive to
expect that long-distance and short-distance physics in quarkonium
production can be separated such that binding effects factorise into
universal non-perturbative parameters.
 
\subsection{Non-Relativistic QCD}
The factorisation approach to quarkonium
production~\cite{Bodwin:1995jh} provides a systematic framework for
separating physics at long-distance and short-distance scales.  It is
based on the use of non-relativistic QCD
(NRQCD)~\cite{Caswell:1986ui}, an effective field theory in which the
heavy quark and antiquark are treated non-relativistically. A
non-relativistic approach to quarkonium physics is motivated by the
success of potential models in describing static properties of
charmonia and bottomonia. The effective field theory framework of
NRQCD allows a rigorous analysis of quarkonium physics which can be
used to systematically improve earlier model approaches.

The most general effective Lagrangian of NRQCD takes the form
\begin{equation}\label{eq_NRQCD}
{\cal L}_{\rm NRQCD} =
\psi^\dagger\left(iD_0+\frac{\vec{D}^2}{2m_Q}\right)\psi +
\chi^\dagger\left(iD_0-\frac{\vec{D}^2}{2m_Q}\right)\chi + {\cal
L}_{\rm light} + \delta {\cal L},
\end{equation}
where $\psi$ and $\chi$ are two-component spinor fields describing the
heavy quark and antiquark, respectively, and ${\cal L}_{\rm light}$ is
the usual relativistic QCD Lagrangian for gluons and light quarks.
The covariant derivative is $D^\mu = \partial^\mu + i g A^\mu$, where
$A^\mu = (A^0,\vec{A})$ is the SU(3) gauge field and $g$ is the QCD
coupling.  The term $\delta{\cal L}$ in (\ref{eq_NRQCD}) includes all
possible operators consistent with the symmetries of QCD and
reproduces the relativistic effects of the full theory. The leading
corrections in $\delta{\cal L}$ are bilinear in the quark (or
antiquark) field:
\begin{eqnarray}\label{eq_dl2}
\delta{\cal L}_{\rm bilinear} & = & 
\frac{c_1}{8m_Q^3}\,\psi^\dagger\, \vec{D}^4\, \psi + 
\frac{c_2}{8m_Q^2}\, \psi^\dagger\, (\vec{D}\cdot g \vec{E}-g\vec{E}\cdot
\vec{D})\, \psi
\nonumber\\
&& \hspace*{-4mm}+\; \frac{c_3}{8m_Q^2}\,\psi^\dagger\, 
(i\vec{D}\times g\vec{E}
- g\vec{E}\times i\vec{D})\cdot\vec{\sigma}\, \psi 
+ \frac{c_4}{2m_Q}\,\psi^\dagger\, g\vec{B} \cdot \vec{\sigma}\, \psi
\nonumber\\[1mm]
&& \hspace*{-4mm} +\; \mbox{charge conjugate terms} , 
\end{eqnarray}
where $E^{i} = G^{0i}$ and $B^{i} = \frac{1}{2}\epsilon^{ijk}G^{jk}$
are the chromoelectric and chromomagnetic components of the gluon
field-strength tensor $G^{\mu\nu}$. The spin symmetry of the minimal
NRQCD Lagrangian is broken by the two terms in (\ref{eq_dl2}) that
contain the Pauli matrix $\vec{\sigma}$.  The creation and
annihilation of heavy quark pairs is taken into account by local
four-fermion operators in $\delta{\cal L}$:
\begin{equation}\label{eq_dl4}
\delta{\cal L}_{\rm 4-fermion} = \sum_i \frac{d_i}{m_Q^2}\,
(\psi^\dagger {\cal K}_i \chi)(\chi^\dagger{\cal K}'_i\psi) .
\end{equation}
The factors ${\cal K}_i,{\cal K}'_i$ are products of a spin and a
colour matrix and may also contain polynomials in the spatial derivate
$\vec{D}$. The operators in $\delta{\cal L}_{\rm 4-fermion}$
annihilate a $Q\overline{Q}$ pair in a colour and angular momentum
state specified by ${\cal K}'_i$ and create a $Q\overline{Q}$ pair at
the same space-time point in a state specified by ${\cal K}_i$.  The
dimensionless coefficients $c_i,d_i$ in the effective Lagrangian are
sensitive only to short distances of order $1/m_Q$ or smaller and can
be determined by matching scattering amplitudes in NRQCD with the
corresponding amplitudes in full QCD.

NRQCD calculations can, in principle, reproduce QCD results for
quarkonium observables to any desired accuracy by adding correction
terms to the effective Lagrangian $\delta{\cal L}$ and tuning the
couplings to appropriate values.

\subsection{NRQCD factorisation}
Within the framework of NRQCD, the cross section for producing a
quarkonium state $H$ can be expressed as a sum of terms, each of which
factors into a short-distance coefficient and a long-distance matrix
element~\cite{Bodwin:1995jh}:
\begin{equation}\label{eq_fac}
d\sigma(H + X) = \sum_n
d\hat{\sigma}(Q\overline{Q}\, [n] + X)\,\langle
{\cal{O}}^{H}\,[n]\rangle .
\end{equation} 
The sum includes all colour and angular momentum states of the
$Q\overline{Q}$ pair, denoted collectively by $n$. The short-distance
coefficients $d\hat{\sigma}$ are proportional to the cross
sections for producing a $Q\overline{Q}$ pair in the state $n$ and
with small relative momentum. They can be calculated perturbatively in
the strong coupling $\alpha_s$.\footnote{In the case of hadronic
production, the short-distance cross sections have to be convoluted
with parton distribution functions.} The non-perturbative transition
probabilities from the $Q\overline{Q}$ state $n$ into the quarkonium
$H$ are given by vacuum expectation values of local four-fermion
operators in NRQCD. Their generic form is
\begin{equation}\label{eq_expvalue}
\langle
{\cal{O}}^{H}\,[n]\rangle \equiv 
\sum_{X,\lambda} \langle 0 | \chi^\dagger {\cal K}_n\psi | H(\lambda) + X 
\rangle \langle H(\lambda) + X | \psi^\dagger {\cal K}'_n \chi | 0 \rangle ,
\end{equation}
where the sum is over the quarkonium polarisation $\lambda$ and any
number of light hadrons $X$ in the final state. The factors ${\cal
K}_n,{\cal K}'_n $ can contain products of colour and spin matrices as
well as covariant derivatives.  The colour and angular momentum
quantum numbers $n$ of the intermediate $Q\overline{Q}$ pair need not
equal those of the physical quarkonium $H$. Soft gluons with energies
of order $m_Q v$ or smaller can be emitted before the bound state is
formed and change the colour and spin of the heavy quark pair. The
effects of these soft gluons are included in the long-distance matrix
elements $\langle {\cal{O}}^{H}\,[n]\rangle$.\footnote{The dependence
of the long-distance matrix elements and short-distance coefficients
on the NRQCD factorisation scale has not been indicated and cancels in
their product (\ref{eq_fac}).}

The derivation of the factorisation formula (\ref{eq_fac}) relies on
the general space-time structure of heavy quark production to separate
the short-distance effects of the hard momentum scale $\sim m_Q$ from
the long-distance effects of the soft momentum scale $\sim
\Lambda_{\rm QCD}$. This step, which is referred to as topological 
factorisation, uses standard factorisation methods of perturbative
QCD~\cite{Collins:1989gx}. The additional step required to derive the
factorisation formula for the production of non-relativistic bound
states involves separating the effect of the $Q\overline{Q}$ relative
momentum $\sim m_Q v$ from the hard momentum scale $m_Q$. This can be
achieved by expanding the hard scattering part of the amplitude as a
Taylor series in terms of the heavy quark relative momentum. The
long-distance factors generated by the Taylor expansion can be
identified with the vacuum expectation values of local four-fermion
operators in NRQCD. In the resulting expression (\ref{eq_fac}) all
effects of the momentum scale $m_Q v$ or smaller are contained in the
long-distance matrix elements $\langle {\cal{O}}^{H}\,[n]\rangle$.
They have to be calculated using non-perturbative methods or determined
from experimental data. The short-distance coefficients
$d\hat{\sigma}$, on the other hand, involve only momentum
scales of order $m_Q$ or larger and therefore have a perturbative
expansion in powers of $\alpha_s(m_Q)$. A general prescription for
calculating the short-distance coefficients $d\hat{\sigma}$ can
be found in~\cite{Braaten:1996jt}.

The factorisation of quarkonium cross sections (\ref{eq_fac}) has not
been proven with complete rigour. Factorisation implies, for example,
that soft gluons connecting the quarkonium and remnant jet parts of
the scattering amplitude cancel for inclusive processes, but no
explicit proof has been presented so far. Although it should be
possible to extend the existing factorisation theorems of perturbative
QCD to inclusive quarkonium processes, the applicability of the NRQCD
factorisation approach has not yet been established theoretically in
all circumstances.

\vspace*{2mm}

The general factorisation formula (\ref{eq_fac}) contains an infinite
number of terms. It provides a viable phenomenological framework only
if the relative magnitude of the various non-perturbative transition
probabilities can be determined systematically. NRQCD predicts a
hierarchy among the matrix elements $\langle
{\cal{O}}^{H}\,[n]\rangle$ in terms of their dependence on the
intrinsic heavy-quark velocity $v$. Using NRQCD power counting rules,
the general expression (\ref{eq_fac}) can be organised into an
expansion in powers of $v$. To any given order in $v$ only a finite
number of terms contribute to the production cross section.

\subsection{Power counting}\label{sec_powercounting}
The importance of the various long-distance matrix elements $\langle
{\cal{O}}^{H}\,[n]\rangle$ can be assessed using power counting
rules. They depend upon the relative size of the three different
low-energy scales in a non-relativistic bound state: the typical heavy
quark three-momentum $m_Q v$ (in the quarkonium rest frame), the
typical kinetic energy $m_Q v^2$ and $\Lambda_{\rm QCD}$.\footnote{The
power counting scheme for effective theories describing heavy-light
mesons is more transparent. Because in a heavy-light meson the
three-momentum and the energy of the heavy quark are both of order
$\Lambda_{\rm QCD}$, the expansion parameter is $\Lambda_{\rm
QCD}/m_Q$ and the importance of the operators in the effective
Lagrangian is determined by their dimensions.}

The standard NRQCD power counting (velocity scaling)
rules~\cite{Lepage:1992tx} have been derived from basic qualitative
properties of non-relativistic bound states and from analysing the
equations of motion for the NRQCD quantum field operators. For
example, the number of heavy quarks in the quarkonium, $\int\! d^3 r\,
\psi^\dagger \psi \approx 1$ implies $\psi \sim (m_Q v)^{3/2}$ since
the quarkonium state can be localised within a region $r\sim 1/(m_Q
v)$. Similarly, the kinetic energy operator $\int\! d^3r\,
\psi^\dagger \,\vec{D}^2/(2m_Q)\, \psi$ has an expectation value of
$m_Q v^2$, and so the spatial part of the covariant derivative has to
scale like $\vec{D}\sim m_Q v$. From the equation of motion for the
vector potential one finds $g\vec{E} \sim m_Q^2 v^3$ and $g\vec{B}
\sim m_Q^2v^4$. According to these rules, the terms 
in $\delta{\cal L}_{\rm bilinear}$ (\ref{eq_dl2}) are suppressed by
${\cal O}(v^2)$ relative to those of the leading NRQCD
Lagrangian.\footnote{The contributions from $\delta{\cal L}_{\rm
4-fermion}$ (\ref{eq_dl4}) contain extra suppression factors ${\cal
O}(\alpha_s^2)$ from the short-distance coefficients $d_i$.}

The power counting rules for the long-distance matrix elements
$\langle {\cal{O}}^{H}\,[n]\rangle$ in (\ref{eq_fac}) can be derived
by considering the Fock state decomposition of a quarkonium state
$|H\rangle$ in Coulomb gauge,
\begin{equation}\label{eq_fock}
| H \rangle = \psi_{Q\overline{Q}}^{H}\, | Q\overline{Q} \rangle +
\psi_{Q\overline{Q}g}^{H}\, | Q\overline{Q} g \rangle +\ldots
\end{equation}
The dominant component $| Q\overline{Q} \rangle $ comprises a heavy
quark pair in a colour-singlet state and with angular momentum quantum
numbers ${}^{2S+1}L_J$ that are consistent with the quantum numbers of
the physical quarkonium. The higher Fock states, such as $|
Q\overline{Q} g \rangle $, contain dynamical gluons or light
$q\bar{q}$ pairs. The heavy quark pair can be in either a
colour-singlet or a colour-octet state with spin $S=0,1$ and angular
momentum $L=0,1,2,\ldots$ All higher Fock states have probabilities
suppressed by powers of $v$ compared to $| Q\overline{Q} \rangle
$. The $| Q\overline{Q} g \rangle $ states with the highest probability
of ${\cal O}(v)$ are those that can be reached from the dominant $|
Q\overline{Q} \rangle $ state through a chromoelectric interaction
induced by the $\psi^\dagger\, (ig\vec{A}\cdot \vec{\partial}) \,
\psi$ term of the NRQCD Lagrangian. The chromoelectric transition
satisfies the selection rules $\Delta L = \pm 1$ and $\Delta S =
0$. Chromomagnetic transitions which proceed through the
$\psi^\dagger\, (g \vec{B}\cdot
\vec{\sigma}) \, \psi$ term in $\delta{\cal L}_{\rm bilinear}$ satisfy 
the selection rules $\Delta L = 0$ and $\Delta S = \pm 1$. Higher Fock
states $| Q\overline{Q} g \rangle $ which can be reached from the
dominant $| Q\overline{Q} \rangle $ state through a chromomagnetic
interaction have probabilities of ${\cal O}(v^2)$.  Both
chromoelectric and chromomagnetic transitions change the colour state
of a $Q\overline{Q}$ pair from colour-singlet to colour-octet, and
from colour-octet to either colour-singlet or colour-octet.  The
chromoelectric transitions preserve the spin of the heavy quarks while
chromomagnetic transitions are spin-symmetry breaking.

The operator ${\cal{O}}^{H}\,[n=1(8),{}^{2S+1}L_J]$ creates and
annihilates a pointlike $Q\overline{Q}$ pair in a colour-singlet
(colour octet) state with angular momentum quantum numbers
${}^{2S+1}L_J$. The magnitude of the corresponding vacuum expectation
value $\langle {\cal{O}}^{H}\,[n]\rangle$ can be determined from the
power counting rules for the operator ${\cal{O}}^{H}\,[n]$ and the
probabilities of the quarkonium's Fock state components that have
non-zero overlap with the quark pair in the state $n$. The general
matrix element $\langle {\cal{O}}^{H} \, [n = 1(8), {}^{2S+1}L_J]
\rangle$ scales as $v^{3+2L+2E+4M}$, where $E$ and $M$ are the minimum
number of chromoelectric and chromomagnetic transitions required for
the $Q\overline{Q}$ pair to reach the state
$Q\overline{Q}\,[n=1(8),{}^{2S+1}L_J]$ from the dominant Fock state of
the quarkonium $H$, and $L$ is the orbital angular momentum quantum
number.

The standard NRQCD power counting rules have been derived assuming
$m_Q v \gg m_Q v^2 \approx \Lambda_{\rm QCD}$ and $\alpha_s(m_Q v)
\sim v$.  Such a hierarchy is likely to be realised for bottomonium states 
where $m_Q \approx 5$~GeV and $v^2 \approx 0.1$. It is, however, not
obvious {\it a priori} that it can be applied to the charmonium system
where $m_Q \approx 1.5$~GeV and $v^2 \approx 0.25$, so that $m_Q v
\approx 750$~MeV. Depending on the relation between the low energy 
scales $m_Q v, m_Q v^2$ and $ \Lambda_{\rm QCD}$, alternative power
counting schemes may be more appropriate for charmonium.  If, for
example, $m_Q v \approx \Lambda_{\rm QCD} \gg m_Q v^2$ the typical
momentum of dynamical gluons would be $\sim \Lambda_{\rm QCD}$, and
chromomagnetic transitions may no longer be suppressed with respect to
chromoelectric transitions~\cite{Beneke:1997av}. Alternative power
counting rules can have important consequences for quarkonium
phenomenology~\citer{Schuler:1997is,Sanchis-Lozano:2001rr}, as
discussed in more detail in the subsequent sections.

\subsection{Colour-octet processes}
The NRQCD formalism implies that so-called colour-octet processes
associated with higher Fock state components of the quarkonium wave
function must contribute to the cross section. Heavy quark pairs that
are produced at short distances in a colour-octet state can evolve
into a physical quarkonium through radiation of soft gluons at late
times in the production process, when the quark pair has already
expanded to the quarkonium size. Such a possibility is ignored in the
{\it colour-singlet model}~\cite{Berger:1981ni,Baier:1981uk}, where
one assumes that only heavy quark pairs produced in the dominant Fock
state form a physical quarkonium.

The most profound theoretical evidence for the incompleteness of the
colour-singlet model comes from the presence of infrared divergences
in the production cross sections and decay rates of $P$-wave
quarkonium. Consider for example the production of $\chi_J$ states.
Within the colour-singlet model, $\chi_J$ is described by a
colour-singlet $Q\overline{Q}$ pair with angular momentum quantum
numbers ${}^{3}P_J$. The cross section is assumed to factorise into a
short-distance coefficient, which describes the production of a
$Q\overline{Q}\,[1,{}^3P_J]$ pair, and a single long-distance factor
which contains all the non-perturbative dynamics of the bound state
formation:
\begin{equation}\label{eq_csm}
d\sigma^{\rm CSM}(\chi_J + X) = 
d\hat{\sigma}(Q\overline{Q}\, [1,{}^3P_J] + X)\,
(2J+1)\,\frac{3N_C}{2\pi}\,|R'(0)|^2 .
\end{equation} 
The long-distance factor is related to the derivative of the radial
wave function at the origin and can be determined from potential model
calculations or from the $\chi_J$ decay width. At next-to-leading
order, the short-distance cross sections for production (or decay) of
$P$-wave quarkonia exhibit a logarithmic infrared divergence due to
soft gluon emission, which can not be factored into $|R'(0)|^2$. In
phenomenological applications of the colour-singlet model, an infrared
cutoff has been introduced and identified with one of the low-energy
scales present in non-relativistic bound states.  It is evident,
however, that the presence of infrared divergencies implies a failure
of the simple factorisation assumption upon which the colour-singlet
model is based.

The NRQCD approach provides a natural solution to this
problem~\cite{Bodwin:1992ye}.  According to the NRQCD power counting
rules, two terms contribute to the production cross section of
$P$-wave states at leading order in the velocity expansion. For
$\chi_J$ production one has
\begin{eqnarray}\label{eq_chioctet}
d\sigma(\chi_J + X) & = &  
d\hat{\sigma}(Q\overline{Q}\, [1,{}^3P_J] + X)\,
\langle {\cal{O}}^{\chi_J}\,[1,{}^3P_J]\rangle 
\nonumber\\
&& \hspace*{-4mm} +\, d\hat{\sigma}(Q\overline{Q}\, 
[8,{}^3S_1] + X)\,\,
\langle {\cal{O}}^{\chi_J}\,[8,{}^3S_1]\rangle
+ {\cal O}(v^2) .
\end{eqnarray}
The matrix element $ \langle {\cal{O}}^{\chi_J}\,[1,{}^3P_J]\rangle $
can be related to the derivative of the radial wave function at the
origin, and the first term in (\ref{eq_chioctet}) corresponds to the
expression of the colour-singlet model. The second term represents a
contribution to the cross section from a colour-octet mechanism. The
short-distance factor is the cross section for producing a
$Q\overline{Q}$ pair in a colour-octet ${}^3S_1$ state, and the
corresponding long-distance matrix element $\langle
{\cal{O}}^{\chi_J}\,[8,{}^3S_1]\rangle$ describes the probability for
such a $Q\overline{Q}\,[8,{}^3S_1]$ pair to form a $\chi_J$
quarkonium. The colour-octet ${}^3S_1$ state can be reached from the
dominant $\chi_J$ Fock state through a single chromoelectric
transition.  According to the NRQCD power counting rules, the $\langle
{\cal{O}}^{\chi_J}\,[1,{}^3P_J]\rangle$ and $\langle
{\cal{O}}^{\chi_J}\,[8,{}^3S_1]\rangle$ matrix elements scale with the
same power of velocity $\sim v^5$ and must both be included in a
consistent theoretical analysis. Other terms in the general
factorisation formula are suppressed by relative order $v^2$ or more.
In the NRQCD analysis, the infrared divergence in the colour-singlet
short-distance cross section $d\hat{\sigma}(Q\overline{Q}\,
[1,{}^3P_J] + X)$ at next-to-leading order is cancelled by a matching
infrared singularity from the radiative corrections to the
colour-octet matrix element $\langle
{\cal{O}}^{\chi_J}\,[8,{}^3S_1]\rangle$. The inclusion of the
colour-octet process proportional to $\langle
{\cal{O}}^{\chi_J}\,[8,{}^3S_1]\rangle$ is crucial to remove the
infrared divergence from the production cross section.\footnote{For
more comprehensive and pedagogical discussions of the treatment of
infrared divergencies in production and decay of $P$-wave quarkonia
see e.g.~\citer{Bodwin:1992ye,Grinstein:2000xb}.}

Colour-octet contributions are needed for a consistent description of
$P$-wave quarkonia, but they can be even more important
phenomenologically for $S$-wave states like $J/\psi$ or
$\psi(2S)$. According to the power counting rules, all colour-octet
matrix elements for the production (or decay) of $S$-wave quarkonia
are suppressed by powers of the velocity compared to the
colour-singlet contribution.  The leading long-distance matrix element
for $J/\psi$ production, for example, involves the operator
${\cal{O}}^{J/\psi}\,[1,{}^3S_1]$, which annihilates and creates a
$c\overline{c}$ pair with the same colour and angular momentum quantum
numbers as in the dominant Fock state of the $J/\psi$. From the power
counting rules it follows that the corresponding vacuum expectation
value $\langle {\cal{O}}^{J/\psi}\,[1,{}^3S_1]\rangle$ scales as
$v^3$.  According to the spin and angular momentum selection rules,
the $c\overline{c}$ pair can reach the colour-octet states
$c\overline{c}\,[8,{}^1S_0]$, $c\overline{c}\,[8,{}^3S_1]$ and
$c\overline{c}\,[8,{}^3P_J]$ through a single chromomagnetic, a double
chromoelectric, and a single chromoelectric transition,
respectively. The corresponding vacuum expectation values $\langle
{\cal O}^{J/\psi}\,[8,{}^1S_0]\rangle$, $\langle {\cal
O}^{J/\psi}\,[8,{}^3S_1]\rangle$ and $\langle {\cal
O}^{J/\psi}\,[8,{}^3P_J]\rangle$ all scale as $v^7$, i.e.\ they are
suppressed by a power $v^4$ compared to the leading $\langle {\cal
O}^{J/\psi}\,[1,{}^3S_1]\rangle$ matrix element.  All other matrix
elements $\langle {\cal O}^{J/\psi}\,[n]\rangle$ scale as $v^{11}$ or
smaller. In the non-relativistic limit $v\to 0$, the NRQCD description
of $S$-wave quarkonium production thus reduces to the colour-singlet
model. However, as discussed in detail in the subsequent sections,
colour-octet processes can become significant, and even dominant, if
the short-distance cross section for producing $Q\overline{Q}$ in a
colour-octet state is enhanced.

NRQCD power counting rules predict a hierarchy among the various
transition probabilities $\langle {\cal{O}}^{H}\,[n]\rangle$.
Truncating the expansion (\ref{eq_fac}) at a given order in the
heavy-quark velocity, only a finite number of terms contribute to the
production cross section. If the operator ${\cal{O}}^{H}\,[n]$ creates
and annihilates a $Q\overline{Q}$ pair in a colour-singlet state with
the angular momentum quantum numbers of the dominant quarkonium Fock
state, its vacuum expectation value $\langle {\cal{O}}^{H}\,[n]
\rangle$ can be related to the quarkonium wave function at the 
origin and determined from the leptonic decay width or potential model
calculations. The number of independent non-perturbative matrix
elements can be reduced further by using the approximate heavy-quark
spin symmetry of the NRQCD Lagrangian~\cite{Bodwin:1995jh}.  The
calculation of the remaining NRQCD matrix elements that contribute at
a given order in velocity expansion requires non-perturbative
methods. In the absence of lattice estimates\footnote{First lattice
results for quarkonium decay matrix elements have been presented
in~\cite{Bodwin:1997mf}. No lattice computations exist, however, for
the non-perturbative matrix elements that describe quarkonium
production.}, the transition probabilities $\langle {\cal{O}}^{H}\,[n]
\rangle$ have to be treated as phenomenological parameters and
determined from experimental data, see Section~\ref{sec_me}.

\vspace*{2mm}

Colour-octet processes are taken into account in the so-called {\it
colour-evaporation model}~\citer{Fritzsch:1977ay,Gluck:1978zm}, where
the cross section for a specific quarkonium state $H$ is given as a
universal fraction $f_H$ of the inclusive $Q\overline{Q}$ production
cross section integrated up to the open heavy quark threshold:
\begin{equation}\label{eq_cem}
d\sigma^{\rm CEM}(H + X) = 
f_H\;\int_{2m_Q}^{ 2M_{Qq}}\! dM_{Q\overline{Q}}\,
\frac{d\sigma(Q\overline{Q}+X)}{dM_{Q\overline{Q}}} .
\end{equation} 
The model assumes that any $Q\overline{Q}$ pair, produced with small
relative momentum, can form a physical quarkonium by emission of soft
gluons. No constraints are imposed on the colour and angular momentum
configuration. The cross section is hence dominated by $Q\overline{Q}$
pairs produced at short distances with vanishing angular momentum and
in either a colour-singlet or colour-octet state.

The colour-evaporation model resembles some features of the NRQCD
factorisation approach, in particular due to the inclusion of
colour-octet processes. However, the colour-evaporation model is not a
rigorous theory, and there is no limit in which it could
systematically reproduce a full QCD calculation. Moreover, it turns
out that the assumption of a single universal long-distance factor is
too restrictive from a phenomenological point of view. It implies, for
example, a universal $\sigma(\chi_c)/\sigma(J/\psi)$ ratio, in
conflict with experimental observations on charmonium production in
hadron-hadron and photon-hadron collisions.

\section{Quarkonium Production at Hadron Colliders\label{sec_tev}}

The first hadron collider measurements of inclusive $J/\psi$,
$\psi(2S)$ and $\chi_c$ cross sections from the \mbox{UA1}
experiment~\cite{Albajar:1991hf} at the \mbox{CERN} $Sp\bar{p}S$ and
from the CDF collaboration~\cite{Abe:1992ww} at the Fermilab Tevatron
could not separate charmonium produced directly in a hard-scattering
reaction from charmonium produced in weak decays of $B$-hadrons. The
early data showed some discrepancies when compared to the
colour-singlet model, which predicts $B$-decays to be the dominant
source of charmonium at hadron colliders~\cite{Glover:1988az}. The
substantial uncertainties in the theoretical calculation, however,
made it difficult to draw definite
conclusions~\cite{Mangano:1993kh}. A rigorous test of the
colour-singlet model became possible when \mbox{CDF} presented a
measurement of the {\it direct} $\psi(2S)$ cross
section~\citer{Sansoni:1996yg,Abe:1997jz}, where the contributions
from $B$-decays had been removed using microvertex-detection. It
turned out that only approximately 25\% of $\psi(2S)$ at the Tevatron
come from the decay of $B$-hadrons.  Surprisingly, the direct
$\psi(2S)$ cross section at $p_t(\psi)\;\simgt\; 5$~GeV proved to be
orders of magnitudes larger than the leading-order colour-singlet
model prediction~\cite{Baier:1983va,Gastmans:1987be}. A similarly
striking discrepancy was observed in the direct $J/\psi$ cross section
after contributions from both $B$-decays and radiative $\chi_c$ decays
had been removed.\footnote{Radiative $\chi$ decays do not contribute
to the $\psi(2S)$ cross section because the transition $\chi_c \to
\psi(2S)$ is kinematically forbidden.}

The cross section for quarkonium hadroproduction is given by 
\begin{eqnarray}
d\sigma(p\bar{p} \to H + X) &\!\!= \!\!& \sum_{i,j} \int dx_1 dx_2
f_{i/p}(x_1) f_{j/\bar{p}}(x_2)\nonumber \\
&&\hspace*{10mm}\times \sum_n
d\hat{\sigma}(i+j \to Q\overline{Q}\, [n] + X)\,\langle
{\cal{O}}^{H}\,[n]\rangle ,
\end{eqnarray}
where $f_{i/p}$ and $f_{j/\bar{p}}$ denote the parton densities, and
$ij = gg,gq,g\bar{q},q\bar{q}$. In the colour-singlet model, the
leading partonic subprocess $g+g \to Q\overline{Q}\, [1,{}^3S_1] + g$
occurs at order $\alpha_s^3$~\cite{Baier:1983va,Gastmans:1987be}.

\begin{figure}[htb]
\epsfig{figure=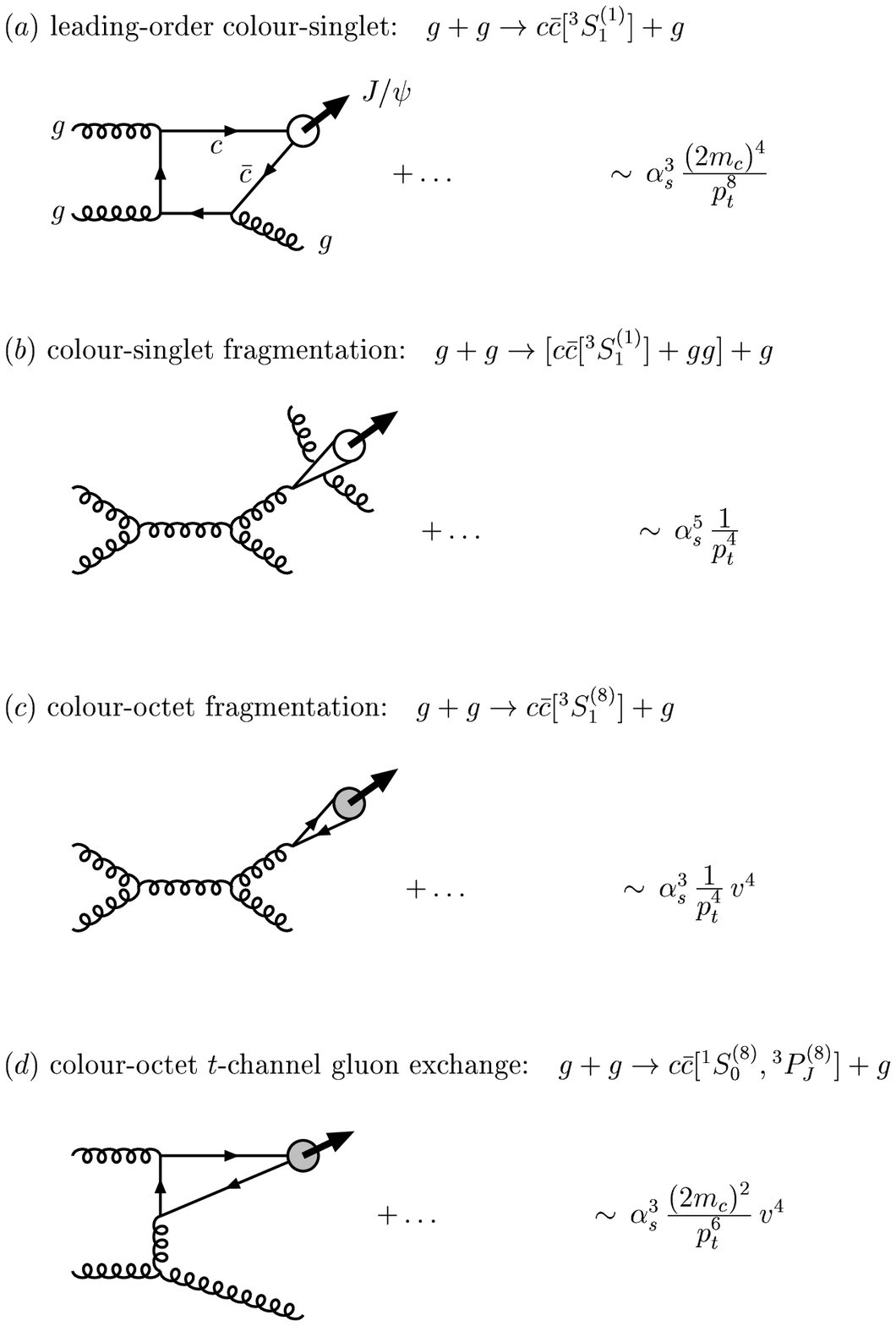,%
        bbllx=65pt,bblly=170pt,bburx=460pt,bbury=760pt,%
        width=0.92\textwidth,clip=}
\caption[Generic Feynman diagrams for $J/\psi$ and $\psi(2S)$ 
 hadroproduction]{\small Generic diagrams for $J/\psi$ and $\psi(2S)$
 production in hadron-hadron collisions through colour-singlet and
 colour-octet channels.}
\label{figure1}
\end{figure}

In retrospect, the shortcoming of the colour-singlet model to describe
direct $J/\psi$ and $\psi(2S)$ hadroproduction can be understood by
examining a typical leading-order Feynman diagram,
Figure~\ref{figure1}(a).
At large transverse momentum, the two internal quark propagators are
off-shell by $\sim p_t^2$ so that the parton differential cross
section scales like $d\sigma/d p_t^2 \sim 1/p_t^8$, as
indicated in the figure. On the other hand, when $p_t \gg 2m_c$ the
quarkonium mass can be considered small and the inclusive charmonium
cross section should scale like any other single-particle inclusive
cross section $\sim 1/p_t^4$. The dominant production mechanism for
charmonium at sufficiently large $p_t$ must thus be via
fragmentation~\cite{Braaten:1993rw}, the production of a parton with
large $p_t$ which subsequently decays into charmonium and other
partons.  A typical fragmentation contribution to colour-singlet
$J/\psi$ production is shown in Figure~\ref{figure1}(b). While
the fragmentation contributions are of higher order in $\alpha_s$
compared to the fusion process Figure~\ref{figure1}(a), they are
enhanced by a power $p_t^4/(2m_c)^4$ at large $p_t$ and can thus
overtake the fusion contribution at $p_t \gg 2m_c$. When
colour-singlet fragmentation is included, the $p_t$ dependence of the
theoretical prediction is in agreement with the Tevatron data but the
normalisation is still underestimated by about an order of magnitude
\citer{Cacciari:1994dr,Roy:1994ie}, indicating that an additional 
fragmentation contribution is still missing. It is now generally
believed that gluon fragmentation into colour-octet ${}^3S_1$ charm
quark pairs \cite{Braaten:1995vv,Cacciari:1995yt}, as shown in
Figure~\ref{figure1}(c), is the dominant source of $J/\psi$ and
$\psi(2S)$ at large $p_t$ at the Tevatron. The probability of forming
a $J/\psi$ particle from a pointlike $c\bar{c}$ pair in a colour-octet
${}^3S_1$ state is given by the NRQCD matrix element $\langle {\cal
O}^{J/\psi}[8,{}^3S_1] \rangle$ which is suppressed by $v^4$ relative
to the non-perturbative factor of the leading colour-singlet
term. However, this suppression is more than compensated by the gain
in two powers of $\alpha_s/\pi$ in the short-distance cross section
for producing colour-octet ${}^3S_1$ charm quark pairs, as compared to
colour-singlet fragmentation. At ${\cal{O}}(v^4)$ in the velocity
expansion, two additional colour-octet channels have to be included,
Figure~\ref{figure1}(d), which do not have a fragmentation
interpretation at order $\alpha_s^3$ but which become significant at
moderate $p_t\sim 2m_c$~\cite{Cho:1996vh,Cho:1996ce}. The importance
of the $c\bar{c}[8,{}^1S_0]$ and $c\bar{c}[8,{}^3P_J]$ contributions
can not be estimated from naive power counting in $\alpha_s$ and $v$
alone, but rather follows from the dominance of $t$-channel gluon
exchange, forbidden in the leading-order colour-singlet cross
section. The leading-order subchannels which contribute to $J/\psi$
and $\psi(2S)$ hadroproduction at order $v^4$ are thus
\begin{eqnarray}
g + g      & \rightarrow & c\bar c
\left[1,\!{}^3\!S_1; \,8,\!{}^3\!S_1; \,8,\!{}^1\!S_0; \,8,\!{}^3\!P_J
 \right] + g, \label{subgg} \\
g + q/\bar{q}      & \rightarrow & c\bar c
\left[8,\!{}^3\!S_1; \,8,\!{}^1\!S_0; \,8,\!{}^3\!P_J \right] + q/\bar{q}, \\
q + \bar q & \rightarrow & c\bar c
\left[8,\!{}^3\!S_1; \,8,\!{}^1\!S_0; \,8,\!{}^3\!P_J \right] + g,
\end{eqnarray}
Explicit expressions for the parton cross sections can be found
in~\cite{Cho:1996ce} and \cite{Beneke:1998re}.

\subsection{Charmonium cross sections}
\label{sec_cxs}
\begin{figure}[htb]
\vspace*{-8mm}
\hspace*{8mm}
\includegraphics[width=0.9\textwidth,clip]{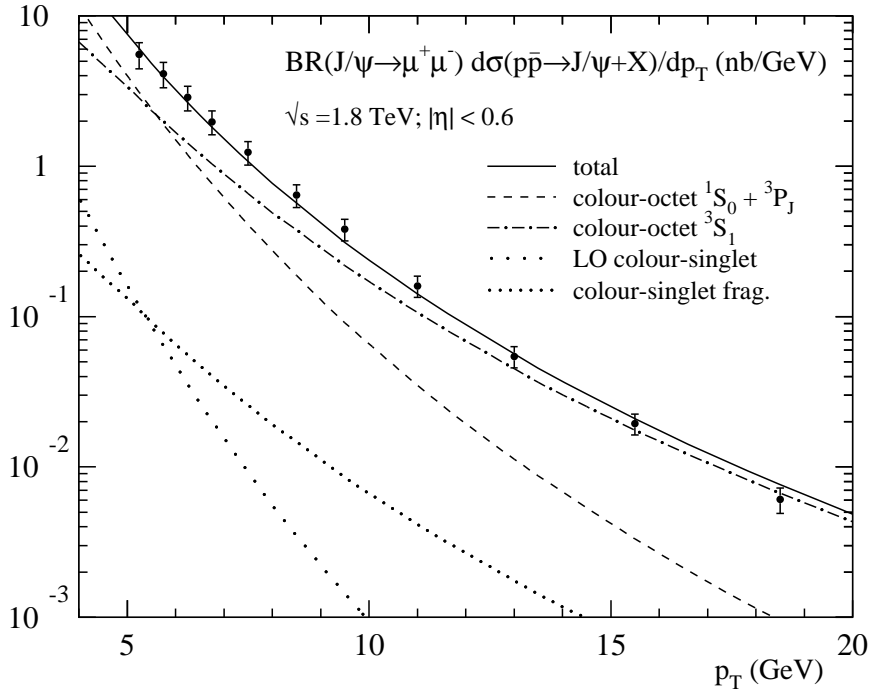}
\vspace*{-5mm}
\caption[Colour-singlet and colour-octet contributions to direct
 $J/\psi$ production at the Tevatron compared to experimental data]
 {\small Colour-singlet and colour-octet contributions to direct
 $J/\psi$ production in $p\bar{p} \to J/\psi+X$ at the Tevatron
 ($\sqrt{s}=1.8$~TeV, pseudorapidity cut $|\eta|<0.6$)) compared to
 experimental data from CDF~\cite{Abe:1997jz}. Parameters: CTEQ5L
 parton distribution functions~\cite{Lai:1999wy}; factorisation and
 renormalisation scale $\mu = \sqrt{p_t^2+4m_c^2}$; $m_c =
 1.5$~GeV. The leading logarithms $(\alpha_s\ln p_t^2/(2m_c)^2)^n$
 have been summed by solving the Altarelli-Parisi evolution equations
 for the gluon fragmentation function. NRQCD matrix elements as
 specified in Table~\ref{table1}.}
\label{figure2} 
\end{figure}

\begin{figure}[htb]
\vspace*{-8mm}
\hspace*{8mm}
\includegraphics[width=0.9\textwidth,clip]{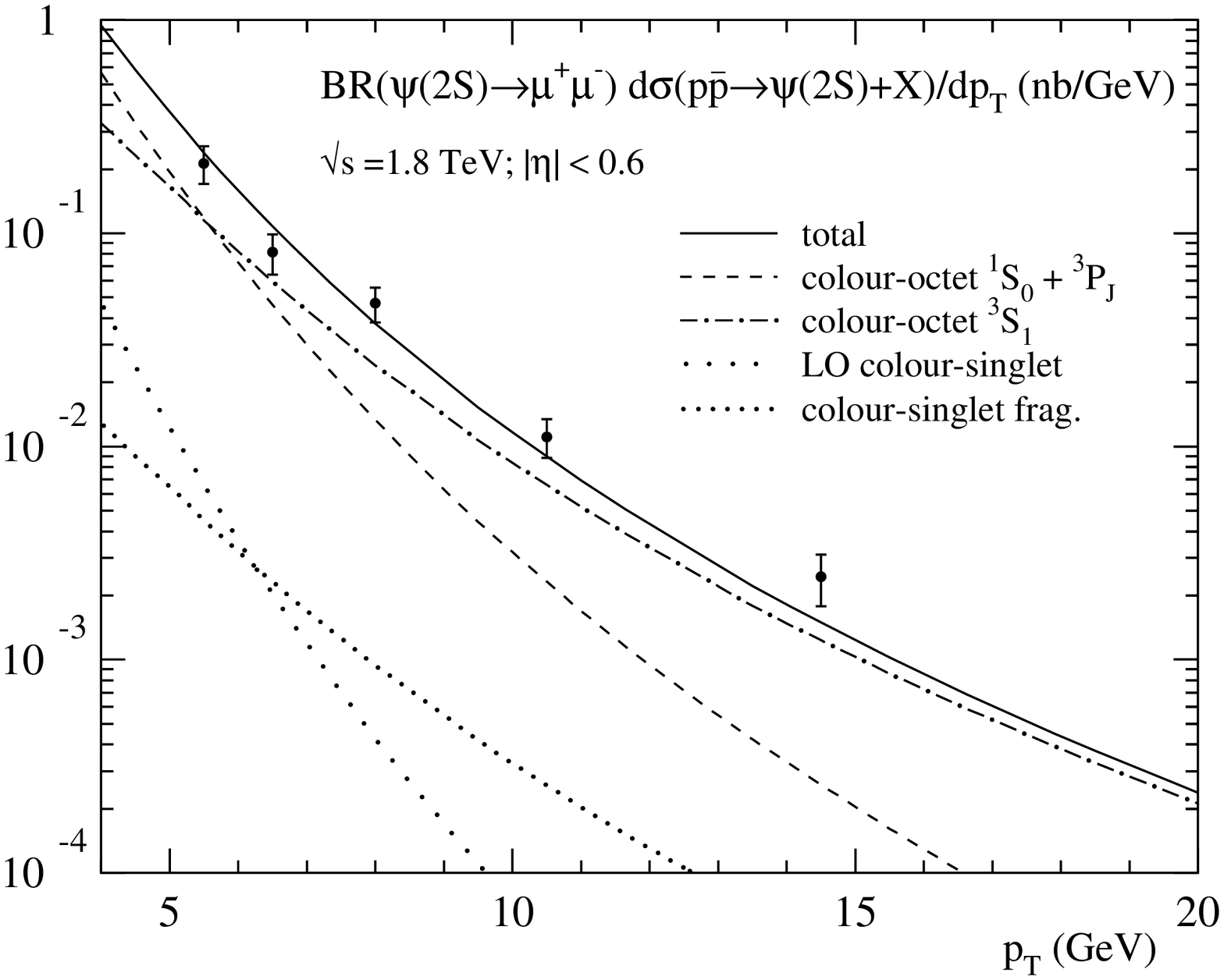}
\vspace*{-8mm}
\caption[Same as Figure~\ref{figure2} for $\psi(2S)$
 production]
 {\small Same as Figure~\ref{figure2} for $\psi(2S)$
 production.}
\label{figure3} 
\end{figure}

The different contributions to the $J/\psi$ transverse momentum
distribution are compared to the CDF data \cite{Abe:1997jz} in
Figure~\ref{figure2}. As mentioned above, the colour-singlet
model at lowest order in $\alpha_s$ fails dramatically when confronted
with the experimental results. When colour-singlet fragmentation is
included, the prediction increases by more than an order of magnitude
at large $p_t$, but it still falls below the data by a factor of $\sim
30$.  The CDF results on $J/\psi$ production can be explained by
including the leading colour-octet contributions
$c\overline{c}\,[8,{}^1S_0]$, $c\overline{c}\,[8,{}^3S_1]$ and
$c\overline{c}\,[8,{}^3P_J]$, and adjusting the corresponding
non-perturbative parameters to fit the data.\footnote{The colour-octet
${}^1S_0$ and ${}^3P_J$ channels have a similar $p_t$ dependence, and
the transverse momentum distribution is sensitive only to a linear
combination of the $\langle {\cal O}^{J/\psi}\,[8,{}^1S_0]\rangle$ and
$\langle {\cal O}^{J/\psi}\,[8,{}^3P_J]\rangle$ matrix elements.}
Numerically one finds the non-perturbative matrix elements to be of
${\cal{O}}(10^{-2}~\mbox{GeV}^3)$, see Table~\ref{table1},
consistent with the $v^4$ suppression expected from the velocity
scaling rules. Similar conclusions can be drawn for $\psi(2S)$
production, see Figure~\ref{figure3}.
\begin{table}[bht]
\vspace*{5mm}
\begin{center}
\renewcommand{\arraystretch}{1.5}
$$
\begin{array}{|c|ccc|}
\hline\hline
 H & \langle {\cal{O}}_1^{H} \rangle  & \langle
 {\cal{O}}^{H}[8,{}^3S_1] \rangle  &
 M_{3.5}^{H}(8,{}^1S_0,{}^3P_0)\\ \hline
 J/\psi   & 1.16~{\rm GeV^3} & (1.19 \pm 0.14)\times 10^{-2}~{\rm GeV}^3 &  
 (4.54 \pm 1.11)\times 10^{-2}~{\rm GeV}^3 \\[-1mm] 
 \psi(2S) & 0.76~{\rm GeV^3} & (0.50 \pm 0.06)\times 10^{-2}~{\rm GeV}^3 & 
 (1.89 \pm 0.46)\times 10^{-2}~{\rm GeV}^3 
 \\[-1mm]
 \chi_{c0} & 0.11~{\rm GeV^5} & (0.31 \pm 0.04)\times 10^{-2}~{\rm GeV}^3 & 
 \\[1mm] \hline \hline
\end{array}
$$
\renewcommand{\arraystretch}{1.0}
\caption[NRQCD matrix elements for charmonium production obtained from 
 the transverse momentum distribution at the Tevatron] {\small NRQCD
 matrix elements for charmonium production.  The colour-singlet matrix
 elements are taken from the potential model calculation of
 \cite{Buchmuller:1981su,Eichten:1995ch}. The colour-octet matrix
 elements have been extracted from the CDF data \cite{Abe:1997jz},
 where $M_k^H(8,{}^1\!S_0,{}^3\!P_0)\equiv \langle {\cal
 O}^{H}[8,{}^1S_0]\rangle + k\,\langle {\cal O}^{H}
 [8,{}^3P_0]\rangle/m_c^2$. The errors quoted are statistical
 only. Parameters: CTEQ5L parton distribution functions
 \cite{Lai:1999wy}, renormalisation and factorisation scale
 $\mu=\sqrt{p_t^2+4 m_c^2}$ and $m_c=1.5\,$GeV. The Altarelli-Parisi
 evolution has been included for the $c\bar{c}[8,{}^3S_1]$
 fragmentation contribution. See \cite{Beneke:1997yw} for further
 details.}
\label{table1}
\end{center}
\end{table}

\begin{figure}[htb]
\vspace*{-8mm}
\hspace*{8mm}
\includegraphics[width=0.9\textwidth,clip]{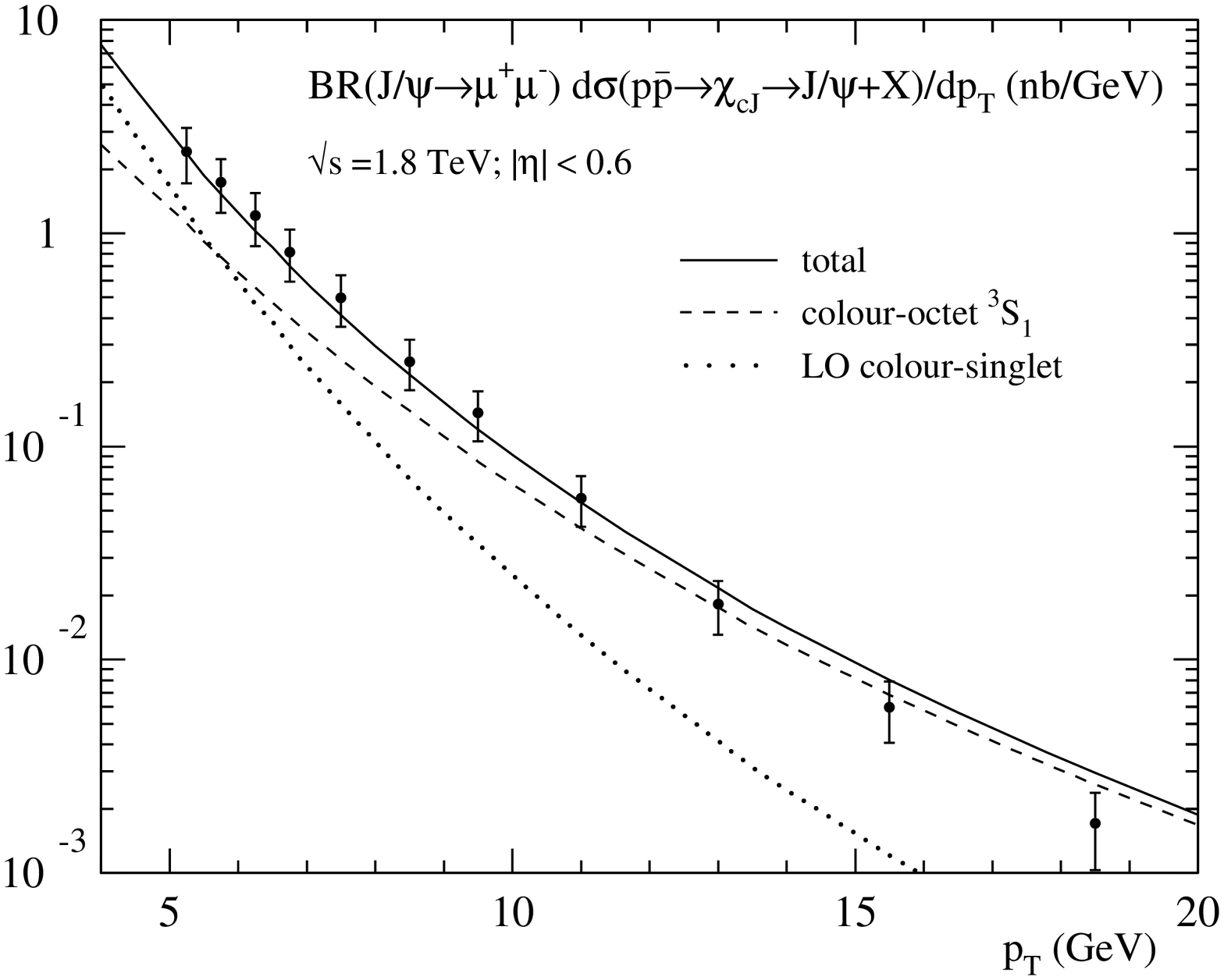}
\vspace*{-8mm}
\caption[Same as Figure~\ref{figure2} for $\chi_c$
 production] {\small Same as Figure~\ref{figure2} for $\chi_c$
 production. Experimental data from~\cite{Abe:1997yz}.}
\label{figure4} 
\end{figure}

Colour-octet processes need to be included in a theoretically
consistent description of $\chi_c$ production. According to
Eq.(\ref{eq_chioctet}) both $c\overline{c}\,[1,{}^3P_J]$ and
$c\overline{c}\,[8,{}^3S_1]$ intermediate states contribute to the
$\chi_c$ cross section at leading order in the velocity
expansion. While the discrepancy between the colour-singlet prediction
and the experimental results~\cite{Abe:1997yz} is less dramatic than
in the case of $S$-wave charmonia, the inclusion of the ${}^3S_1$
colour-octet process (with $\langle {\cal O}^{\chi_c} \, [8,{}^3S_1]
\rangle$ adjusted to the value listed in Table~\ref{table1})
significantly improves the theoretical description. The comparison
with recent data is presented in Figure~\ref{figure4}. The
measured ratio of production cross sections
$\sigma(\chi_{c2})/\sigma(\chi_{c1}) = 0.96 \pm 0.27 (\mbox{stat}) \pm
0.11 (\mbox{sys})$~\cite{Papadimitriou:2001bb} is also in good
agreement with the NRQCD expectation of $1.1\pm0.2$~\cite{fabio}.

The analysis of the CDF cross section data alone, although very
encouraging, does not provide a conclusive test of NRQCD factorisation
because free parameters have to be introduced to fit the data.
However, if factorisation holds the non-perturbative matrix elements,
Table~\ref{table1}, are universal and can be used to make
predictions for various processes and observables. The polarisation
signature of $J/\psi$ and $\psi(2S)$ at large transverse momentum is
one of the single most crucial tests of the NRQCD approach and will be
discussed in some detail in Section~\ref{sec_tevpol}. Ultimately, a
global analysis of different production processes will be needed to
assess the universality of the colour-octet contributions. Such a
program, however, requires a careful discussion of the theoretical
uncertainties associated with the determination of the
non-perturbative matrix elements from experimental data. For
charmonium hadroproduction the uncertainties are mainly associated
with the calculation of the short-distance cross sections and, as
argued in Section~\ref{sec_me}, have not yet been reliably quantified.

The $J/\psi$ and $\psi(2S)$ transverse momentum distributions at the
Tevatron can also be described by the colour-evaporation
model~\citer{Gavai:1995in,Amundson:1997qr} or the related soft-colour
interaction model~\cite{Edin:1997zb}. Both approaches allow
colour-octet $c\overline{c}$ pairs produced in gluon fragmentation to
hadronise into charmonium and thus resemble the NRQCD cross section
prediction at large $p_t$. The colour-evaporation model, however,
assumes unsuppressed gluon emission from the $c\bar{c}$ pair during
hadronisation, which randomises spin and colour, and consequently
predicts unpolarised quarkonium. The NRQCD approach, in contrast,
incorporates the spin-symmetry of QCD in the heavy quark limit, which
implies transverse polarisation of direct $J/\psi$ and $\psi(2S)$ at
large $p_t$. The polarisation prediction of NRQCD will be discussed in
detail in Section~\ref{sec_tevpol}.

\subsection{Determination of NRQCD matrix elements}
\label{sec_me}
The numerical values for the non-perturbative matrix elements obtained
from the fit to the Tevatron data are sensitive to any effects that
modify the shape of the charmonium $p_t$ distribution. This is true in
particular for the $\langle {\cal O}^{\psi}[8,{}^1S_0]\rangle$ and
$\langle {\cal O}^{\psi}[8,{}^3P_0]\rangle$ matrix elements which are
determined from data at relatively small $p_t\;\simlt\; 8$~GeV.  The
theoretical uncertainties include the small-$x$ behaviour of the gluon
distribution \cite{Beneke:1997yw}, the evolution of the strong
coupling \cite{Beneke:1997yw}, as well as systematic effects inherent
in NRQCD, such as the inaccurate treatment of the energy conservation
in the hadronisation of the colour-octet $c\bar{c}$
pairs~\cite{Beneke:1997qw}.  Moreover, higher-order QCD corrections
are expected to play an important role, as discussed in more detail
below.

The inclusion of higher-order QCD corrections is required to reduce
the theoretical uncertainty and to allow a more precise prediction of
the charmonium cross sections. Next-to-leading order (NLO)
calculations for quarkonium production at hadron colliders are
presently available only for total cross
sections~\cite{Kuhn:1993qw,Petrelli:1998ge}.  Significant higher-order
corrections to differential distributions are expected from the strong
renormalisation and factorisation scale dependence of the
leading-order results~\cite{Beneke:1997yw}. Moreover, the NLO
colour-singlet cross section includes processes like $g + g \to
Q\overline{Q}[1,{}^3S_1] + g + g $ which are dominated by $t$-channel
gluon exchange and scale as $\sim \alpha_s^4 (2m_Q)^2/p_t^6$. At
$p_t\gg 2m_Q$ their contribution is enhanced with respect to the the
leading-order cross section, Figure~\ref{figure1}(a), which scales as
$\sim \alpha_s^3 (2m_Q)^4/p_t^8$. This is born out by preliminary
studies~\cite{Petrelli:2000rh} which include part of the NLO
hadroproduction cross section and by the complete calculation of NLO
corrections to the related process of quarkonium photoproduction
\cite{Kramer:1996nb}. The NLO colour-singlet cross section may be
comparable in size to the colour-octet ${}^1S_0 $ and ${}^3P_J$
processes, which scale as $ \sim \alpha_s^3 v^4 (2m_Q)^2/p_t^6 $ (see
Figure~\ref{figure1}(d)), and affect the determination of the
corresponding NRQCD matrix elements from the Tevatron data.  A full
NLO analysis is however needed before quantitative conclusions can be
drawn.
 
Another source of potentially large higher-order corrections is the
multiple emission of soft or almost collinear gluons from the initial
state partons. These corrections, as well as effects related to
intrinsic transverse momentum, are expected to modify the shape of the
transverse momentum distribution predominantly at relatively low
values of $p_t\;\simlt\; 2m_Q$. Initial state radiation can be
partially summed in perturbation theory \cite{Collins:1985kg}, but so
far only total cross sections have been considered in the literature
\cite{Cacciari:2000sy}.  An estimate of the effect on the transverse
momentum distribution should be provided by phenomenological models
where a Gaussian $k_t$-smearing is added to the initial state
partons. The result of these calculations, however, not only depends
on the average $\langle k_t \rangle$, which enters as a free
parameter, but also on the details of how the smearing is
implemented. Moreover, a lower cut-off has to be provided which
regulates the divergences at $p_t = 0$.\footnote{Using the NLO
calculation for the total cross section~\cite{Petrelli:1998ge}, one
can obtain the rough estimate that perturbative Sudakov effects should
be confined below $p_t \sim 1-2$ GeV for charmonium production at
Tevatron energies.} Qualitatively, the inclusion of a Gaussian
$k_t$-smearing leads to an enhancement of the short distance cross
section at small $p_t$, which results in smaller values for the fits
of the $\langle {\cal O}^{\psi}[8,{}^1S_0]\rangle$ and $\langle {\cal
O}^{\psi}[8,{}^3P_0] \rangle$ NRQCD matrix elements
\cite{Sridhar:1998rt,Petrelli:2000rh}. The actual size of the effect,
however, turns out to be different for the two models studied in the
literature.

An alternative approach to treat the effect of initial state radiation
is by means of Monte Carlo event generators which include multiple
gluon emission in the parton shower approximation. Comprehensive
phenomenological analyses have been carried out for charmonium
production at the Tevatron and at the
LHC~\citer{Cano-Coloma:1997rn,Kniehl:1999qy} using the event generator
Pythia \cite{Sjostrand:1994yb} supplemented by the leading
colour-octet processes \cite{Cano-Coloma:1997rn}. The inclusion of
initial state radiation as implemented in Pythia leads to an
enhancement of the short-distance cross section. The size of the
effect is significantly larger than for the Gaussian $k_t$-smearing
mentioned above, and it extents out to large $p_t$. Consequently, the
$\langle {\cal O}^{\psi}[8,{}^1S_0]\rangle$ and $\langle {\cal
O}^{\psi}[8,{}^3P_0] \rangle$ NRQCD matrix elements estimated from the
Monte Carlo analysis of the Tevatron cross sections are significantly
lower than those listed in Table~\ref{table1}.

Recently, the impact of a non-vanishing initial state transverse
momentum has been analysed in the context of the $k_t$-factorisation
formalism. This approach has been developed to sum large logarithmic
corrections in the high-energy limit $s\gg 4 m_c^2$. The summation
generates initial state gluons with transverse momenta comparable to
the hard scale $m_Q$. They are described by so-called unintegrated
gluon distribution functions. The $k_t$-factorisation
calculations~\cite{Hagler:2001eu,Yuan:2001cp} indicate that the
colour-singlet contribution to the direct $J/\psi$ cross section could
be substantially larger than in the conventional collinear
factorisation approach. Moreover, the non-zero transverse momentum of
the initial state partons allows for the production of high-$p_t$
colour-octet $c\overline{c}$ pairs without emission of hard
gluons. The shape of the transverse momentum distribution for the
various colour-octet processes in the $k_t$-factorisation formalism is
thus different from the curves displayed in Figure~\ref{figure2}. The
analyses~\cite{Hagler:2001eu,Yuan:2001cp} suggest that the direct
$J/\psi$ cross section might be dominated by the
$c\overline{c}\,[8,{}^1S_0]$ and $c\overline{c}\,[8,{}^3P_J]$ channels
up to large transverse momenta of the order $p_t\;\simlt\; 20$~GeV.
However, more work is needed to establish the quantitative predictions
of the $k_t$-factorisation approach in general.  At present there is
no reliable estimate of the corresponding theoretical uncertainties
due to higher-order corrections and, in particular, due to the poorly
known unintegrated gluon distribution.  The impact of non-vanishing
initial state transverse momentum has to be quantified in various
other high-energy QCD processes before the importance of the different
quarkonium production mechanisms can be assessed conclusively in this
framework.

\vspace{1mm}

To illustrate the uncertainty in extracting the NRQCD colour-octet
matrix elements from the $J/\psi$, $\psi(2S)$ and $\chi_c$ transverse
momentum distributions at the Tevatron, the results of various
numerical analyses are collected in Tables~\ref{table2},
\ref{table3} and \ref{table4}. It is evident from
the comparison that the theoretical error is not very well under
control.  The discrepancies between the analyses that rely on the
leading-order collinear factorisation calculation can be attributed to
different choices for the input parameters and different prescriptions
for the interpolation between the fusion cross section at low $p_t$
and the fragmentation cross section including summation of
$\ln(p_t^2/(4m_c^2)$ terms at large $p_t$. More accurate data on the
$J/\psi$ and $\psi(2S)$ cross sections could help to reduce some of
the ambiguities. Ultimately, the calculation of next-to-leading order
corrections to both colour-singlet and colour-octet processes is
needed to improve the analysis and reduce the large factorisation and
renormalisation scale uncertainty of the leading-order prediction.
Effects beyond fixed order in perturbation theory, like the summation
of soft gluon radiation, have to be addressed in a more rigorous and
systematic way, and a partial summation of the NRQCD expansion might
be needed to account for the inaccurate treatment of the kinematics in
the hadronisation of the colour-octet $c\bar{c}$ pairs.

\begin{table}
\begin{center}
\vskip0.2cm
\renewcommand{\arraystretch}{1.3}
$$
\begin{array}{|c|cc|ccc|}
\hline\hline
 \mbox{Reference} & \multicolumn{2}{c|}{\mbox{PDF}} & \langle
 {\cal{O}}^{J/\psi}[8,{}^3S_1] \rangle &
 M_{k}^{J/\psi}(8,{}^1S_0,{}^3P_0) & k \\ \hline\hline
 \multicolumn{6}{|c|}{\mbox{LO collinear factorisation}} \\ \hline
 \cite{Cho:1996ce} &
 \multicolumn{2}{c|}{\mbox{MRS(D0)}~\cite{Martin:1993zi}} & 0.66 \pm
 0.21 & 6.6 \pm 1.5 & 3 \\ \hline &
 \multicolumn{2}{c|}{\mbox{CTEQ4L}~\cite{Lai:1997mg}} & 1.06 \pm
 0.14^{+1.05}_{-0.59} & 4.38 \pm 1.15^{+1.52}_{-0.74} & \\
 \cite{Beneke:1997yw} &
 \multicolumn{2}{c|}{\mbox{GRV-LO(94)}~\cite{Gluck:1995uf}} & 1.12 \pm
 0.14^{+0.99}_{-0.56} & 3.90 \pm 1.14^{+1.46}_{-1.07} & 3.5 \\ &
 \multicolumn{2}{c|}{\mbox{MRS(R2)}~\cite{Martin:1996as}} & 1.40 \pm
 0.22^{+1.35}_{-0.79} & 10.9 \pm 2.07^{+2.79}_{-1.26} & \\ \hline &
 \multicolumn{2}{c|}{\mbox{MRST-LO(98)}~\cite{Martin:1998sq}} & 0.44
 \pm 0.07 & 8.7 \pm 0.9 & \\
 \raisebox{2ex}[-2ex]{\cite{Braaten:2000qk}} &
 \multicolumn{2}{c|}{\mbox{CTEQ5L}~\cite{Lai:1999wy}} & 0.39 \pm 0.07
 & 6.6 \pm 0.7 & \raisebox{2ex}[-2ex]{3.4} \\[0.5mm] \hline\hline
 \multicolumn{6}{|c|}{\mbox{Parton shower radiation}} \\ \hline &
 \multicolumn{2}{c|}{\mbox{CTEQ2L}~\cite{Tung:1994ua}} & 0.96 \pm 0.15
 & 1.32 \pm 0.21 & \\ \cite{Sanchis-Lozano:2000um} &
 \multicolumn{2}{c|}{\mbox{MRS(D0)}~\cite{Martin:1993zi}} & 0.68 \pm
 0.16 & 1.32 \pm 0.21 & 3 \\ &
 \multicolumn{2}{c|}{\mbox{GRV-HO(94)}~\cite{Gluck:1995uf}} & 0.92 \pm
 0.11 & 0.45 \pm 0.09 & \\ \hline \cite{Kniehl:1999qy} &
 \multicolumn{2}{c|}{\mbox{CTEQ4M}~\cite{Lai:1997mg}} & 0.27 \pm 0.05
 & 0.57 \pm 0.18 & 3.5 \\[0.5mm] \hline \hline
 \multicolumn{6}{|c|}{\mbox{$k_t$-smearing}} \\ \hline & & \langle k_t
 \rangle \mbox{[GeV]} & & & \\ & & 1 & 1.5\pm 0.22 & 8.6\pm 2.1 & \\
 \raisebox{2ex}[-2ex]{\cite{Petrelli:2000rh}} &
 \raisebox{2ex}[-2ex]{CTEQ4M~\cite{Lai:1997mg}} & 1.5 & 1.7 \pm 0.19 &
 4.5 \pm 1.5 & \raisebox{2ex}[-2ex]{3.5}\\ \hline & & 0.7 & 1.35 \pm
 0.30 & 8.46 \pm 1.41 & \\ \raisebox{2ex}[-2ex]{\cite{Sridhar:1998rt}}
 & \raisebox{2ex}[-2ex]{MRS(D$'_-$)~\cite{Martin:1993zi}} & 1 & 1.5
 \pm 0.29 & 7.05 \pm 1.17 & \raisebox{2ex}[-2ex]{3} \\[0.5mm]
 \hline\hline \multicolumn{6}{|c|}{\mbox{$k_t$-factorisation}} \\
 \hline \cite{Hagler:2001eu} &
 \multicolumn{2}{c|}{\mbox{KMS}~\cite{Kwiecinski:1997ee}} & \approx
 0.04 \pm 0.01 & \approx 6.5 \pm 0.5 & 5\\ \hline \hline
\end{array}
$$
\renewcommand{\arraystretch}{1.0}
\caption[NRQCD matrix elements for $J/\psi$ production obtained from
 various other analyses of the transverse momentum distribution]
 {\small NRQCD matrix elements for $J/\psi$ production obtained from
 various analyses of the $J/\psi$ transverse momentum distribution at
 the Tevatron. Values are given in units $10^{-2}$~GeV${}^3$.
 $M_k^H(8,{}^1\!S_0,^3\!P_0)\equiv \langle {\cal O}^{H}
 [8,{}^1S_0]\rangle + k\,\langle {\cal O}^{H}
 [8,{}^3P_0]\rangle/m_c^2$.  The first error quoted is statistical,
 the second error, when listed, due to variation of the renormalisation
 and factorisation scales between $\mu=1/2\,\sqrt{p_t^2+4 m_c^2}$
 and $2\,\sqrt{p_t^2+4 m_c^2}$.}
\label{table2}
\end{center}
\end{table}
\begin{table}
\begin{center}
\vskip0.2cm
\renewcommand{\arraystretch}{1.3}
$$
\begin{array}{|c|cc|ccc|}
\hline\hline
 \mbox{Reference} & \multicolumn{2}{c|}{\mbox{PDF}} & \langle
 {\cal{O}}^{\psi(2S)}[8,{}^3S_1] \rangle &
 M_{k}^{\psi(2S)}(8,{}^1S_0,{}^3P_0) & k \\ \hline\hline
 \multicolumn{6}{|c|}{\mbox{LO collinear factorisation}} \\ \hline
 \cite{Cho:1996ce} &
 \multicolumn{2}{c|}{\mbox{MRS(D0)}~\cite{Martin:1993zi}} & 0.46 \pm
 0.1 & 1.77 \pm 0.57 & 3 \\ \hline &
 \multicolumn{2}{c|}{\mbox{CTEQ4L}~\cite{Lai:1997mg}} & 0.44 \pm
 0.08^{+0.43}_{-0.24} & 1.80 \pm 0.56^{+0.62}_{-0.30} & \\
 \cite{Beneke:1997yw} &
 \multicolumn{2}{c|}{\mbox{GRV-LO(94)}~\cite{Gluck:1995uf}} & 0.46 \pm
 0.08^{+0.41}_{-0.23} & 1.60 \pm 0.51^{+0.60}_{-0.44} & 3.5 \\ &
 \multicolumn{2}{c|}{\mbox{MRS(R2)}~\cite{Martin:1996as}} & 0.56 \pm
 0.11^{+0.54}_{-0.32} & 4.36 \pm 0.96^{+1.11}_{-0.50} & \\ \hline &
 \multicolumn{2}{c|}{\mbox{MRST-LO(98)}~\cite{Martin:1998sq}} & 0.42
 \pm 0.1 & 1.3 \pm 0.5 & \\
 \raisebox{2ex}[-2ex]{\cite{Braaten:2000qk}} &
 \multicolumn{2}{c|}{\mbox{CTEQ5L}~\cite{Lai:1999wy}} & 0.37 \pm 0.09
 & 0.78 \pm 0.36 & \raisebox{2ex}[-2ex]{3.4} \\[0.5mm] \hline\hline
 \multicolumn{6}{|c|}{\mbox{Parton shower radiation}} \\ \hline &
 \multicolumn{2}{c|}{\mbox{CTEQ2L}~\cite{Tung:1994ua}} & 0.14 \pm 0.03
 & 0.33 \pm 0.09 & \\ \cite{Cano-Coloma:1997rn,Sanchis-Lozano:2000um} &
 \multicolumn{2}{c|}{\mbox{MRS(D0)}~\cite{Martin:1993zi}} & 0.11 \pm
 0.03 & 0.28 \pm 0.07 & 3 \\ &
 \multicolumn{2}{c|}{\mbox{GRV-HO(94)}~\cite{Gluck:1995uf}} & 0.13 \pm
 0.02 & 0.04 \pm 0.05 & \\ \hline \hline
\end{array}
$$
\renewcommand{\arraystretch}{1.0}
\caption[Same as Table~\ref{table2} for $\psi(2S)$ production]
 {\small Same as Table~\ref{table2} for $\psi(2S)$ production.  Values
 are given in units $10^{-2}$~GeV${}^3$.}
\label{table3}
\end{center}
\end{table}

\begin{table}
\begin{center}
\vskip0.2cm
\renewcommand{\arraystretch}{1.3}
$$
\begin{array}{|c|cc|cc|}
\hline\hline
 \mbox{Reference} & \multicolumn{2}{c|}{\mbox{PDF}} & \langle
 {\cal{O}}^{\chi_{c0}}[1,{}^3P_0] \rangle~[\mbox{GeV}^5] & \langle
 {\cal{O}}^{\chi_{c0}}[8,{}^3S_1] \rangle~[10^{-2}~\mbox{GeV}^3] \\
 \hline\hline \multicolumn{5}{|c|}{\mbox{LO collinear factorisation}}
 \\ \hline \cite{Cho:1996ce} &
 \multicolumn{2}{c|}{\mbox{MRS(D0)}~\cite{Martin:1993zi}} &
 0.11~(\mbox{input}) & 0.33 \pm 0.04 \\ \hline \cite{Kniehl:1999qy} &
 \multicolumn{2}{c|}{\mbox{CTEQ4L}~\cite{Lai:1997mg}} & 0.23 \pm 0.03
 & 0.068 \pm 0.018\\ \hline &
 \multicolumn{2}{c|}{\mbox{MRST-LO(98)}~\cite{Martin:1998sq}} & 0.09
 \pm 0.01~(\mbox{input}) & 0.23 \pm 0.03 \\
 \raisebox{2ex}[-2ex]{\cite{Braaten:2000qk}} &
 \multicolumn{2}{c|}{\mbox{CTEQ5L}~\cite{Lai:1999wy}} & 0.09 \pm
 0.01~(\mbox{input}) & 0.19 \pm 0.02 \\[0.5mm] \hline\hline
 \multicolumn{5}{|c|}{\mbox{$k_t$-factorisation}} \\ \hline
 \cite{Hagler:2001dd} &
 \multicolumn{2}{c|}{\mbox{KMS}~\cite{Kwiecinski:1997ee}} &
 0.11~(\mbox{input}) & 0.03 \pm 0.01 \\ \hline \hline
\end{array}
$$
\renewcommand{\arraystretch}{1.0}
\caption[Same as Table~\ref{table2} for $\chi_c$ production]
 {\small Same as Table~\ref{table2} for $\chi_c$ production. }
\label{table4}
\end{center}
\end{table}

Taking into account the large theoretical uncertainties of present
analyses, most results appear to be consistent with the NRQCD power
counting rules. The colour-octet matrix elements extracted from the
$k_t$-factorisation calculation, however, do not follow the expected
pattern.  Such extreme values for the NRQCD matrix elements would
indicate a serious flaw in our understanding of the power counting.

It is instructive to compare the determination of the colour-octet
matrix elements from the Tevatron data with analyses of other
production processes. NRQCD matrix elements have been extracted from
charmonium hadroproduction at fixed-target experiments and from
$B$-decays. The calculations include ${\cal O}(\alpha_s)$ corrections
to the short-distance coefficients and, in the case of $B$-decays,
power corrections $\sim \Lambda_{\rm QCD}/m_b$.  Representative values
obtained for specific linear combinations of $\langle {\cal
  O}^{\psi}[8,{}^1S_0]\rangle$ and $\langle {\cal O}^{\psi}
[8,{}^3P_0]\rangle$ are\\[-2mm] 
$$
\begin{array}{cl}
\mbox{fixed-target hadroproduction} & \left\{
\begin{array}{lc}
M_{6.4}^{J/\psi}\;\,\, \approx 1.9 \times 10^{-2}~\mbox{GeV}^3 & \\[2mm]
M_{6.4}^{\psi(2S)} \approx 0.28 \times 10^{-2}~\mbox{GeV}^3 &
\raisebox{2ex}[-2ex]{\cite{Maltoni:2000km}}
\end{array}
\right.\\[8mm]
\mbox{$B$ decays} & \left\{
\begin{array}{lc}
M_{3.1}^{J/\psi} \;\,\,\approx  (1.5^{+0.8}_{-1.1}) 
\times 10^{-2}~\mbox{GeV}^3 & \\[1mm]
M_{3.1}^{\psi(2S)} \approx (0.5 \pm 0.5)\times 10^{-2}~\mbox{GeV}^3 &
\raisebox{2ex}[-2ex]{\cite{Beneke:1999ks}} \\[2mm]
M_{3.4}^{J/\psi}\;\,\, \approx 2.4 \times 10^{-2}~\mbox{GeV}^3  & \\[1mm]
M_{3.4}^{\psi(2S)} \approx 1.0 \times 10^{-2}~\mbox{GeV}^3  & 
\raisebox{2ex}[-2ex]{\cite{Ma:2000bz}}
\end{array}
\right.
\end{array}
$$\\ Potentially large higher-twist effects at small $p_t$ may render
the application of the NRQCD factorisation formula to quarkonium
hadroproduction at fixed-target experiments unreliable. Nevertheless,
the values obtained for the non-perturbative matrix elements are in
qualitative agreement with the determination at the Tevatron. $B$-decays
provide a more accurate testing ground for NRQCD factorisation. The
linear combination of $\langle {\cal O}^{\psi}[8,{}^1S_0]\rangle$ and
$\langle {\cal O}^{\psi} [8,{}^3P_0]\rangle$ extracted from the
branching ratios BR($B\to \psi + X$) is consistent with Tevatron
analyses that include soft-gluon effects in the parton-shower
approximation.

In subsequent numerical analysis the following ranges of values 
for the non-per\-tur\-ba\-tive matrix elements will be considered:
\begin{equation}\label{eq_mevar}
\left. \begin{array}{ccc}
\langle
 {\cal{O}}^{J/\psi}[8,{}^3S_1] \rangle & = & (0.3 - 2.0)\times 
10^{-2}~\mbox{GeV}^3 \\[1mm]
\langle {\cal O}^{J/\psi}[8,{}^1S_0]\rangle 
+ 3.5\,\langle {\cal O}^{J/\psi} [8,{}^3P_0]\rangle/m_c^2 
& = & (1.0 - 10) \times 10^{-2}~\mbox{GeV}^3\\[4mm]
\langle {\cal{O}}^{\psi(2S)}[8,{}^3S_1] \rangle & = & 
(0.1 - 1.0) \times 10^{-2}~\mbox{GeV}^3\\[1mm]
\langle {\cal O}^{\psi(2S)}[8,{}^1S_0]\rangle 
+ 3.5\,\langle {\cal O}^{\psi(2S)} [8,{}^3P_0]\rangle/m_c^2 
& = & (0.1 - 2.0) \times 10^{-2}~\mbox{GeV}^3\\[4mm]
\langle {\cal{O}}^{\chi_{c0}}[8,{}^3S_1] \rangle & = & 
(0.1 - 0.5) \times 10^{-2}~\mbox{GeV}^3
\end{array}
\quad \right\}
\end{equation}\\ 
Note that although the theoretical errors seem conservative, they do
not accommodate the non-perturbative matrix elements obtained from
calculations in the $k_t$ factorisation approach.\footnote{As alluded
to above, the matrix elements obtained from calculations in the $k_t$
factorisation approach are not consistent with the NRQCD power
counting rules.}

\subsection{$J/\psi$ and $\psi(2S)$  polarisation\label{sec_tevpol}}
The analysis of $J/\psi$ and $\psi(2S)$ polarisation at large
transverse momentum is one of the most decisive tests of the NRQCD
factorisation approach. At large $p_t$, the direct $J/\psi$ and
$\psi(2S)$ cross sections should be dominated by gluon fragmentation
into colour-octet ${}^3S_1$ charm quark pairs,
Figure~\ref{figure1}(c).  When $p_t \gg 2m_c$ the fragmenting
gluons are effectively on-shell and the intermediate heavy quark pairs 
transversely polarised. According to the standard NRQCD power counting
rules, Section~\ref{sec_powercounting}, $c\bar{c}\,[8,{}^3S_1]$
evolves into $J/\psi$ or $\psi(2S)$ predominantly through a double
chromoelectric transition which preserves the heavy quark spin.  The
dominance of gluon fragmentation into colour-octet ${}^3S_1$ charm
quark pairs at large $p_t$ and NRQCD spin-symmetry thus imply
transversely polarised $J/\psi$ and $\psi(2S)$ at large transverse
momentum~\cite{Cho:1995ih}.

The polarisation can be measured through the angular distribution in
the leptonic decay,
\begin{equation}
\frac{d\Gamma(\psi\to l^+l^-)}{d\!\cos\theta} \propto 
1+\alpha\,\cos^2\theta ,
\end{equation}
where $\theta$ denotes the angle between the lepton three-momentum in
the $\psi$ rest frame and the $\psi$ three-momentum in the laboratory
frame. The polar-angle asymmetry $\alpha$ can take values in the range
$-1 \le \alpha \le 1$, where $\alpha=1$ corresponds to transverse
polarisation.

Corrections to the asymptotic NRQCD prediction of pure transverse
polarisation can arise from spin-symmetry breaking chromomagnetic
interactions in the non-perturbative evolution of the colour-octet
${}^3S_1$ charm quark pair. NRQCD power counting rules imply that
spin-symmetry breaking transitions, which lead to random polarisation,
are suppressed by a factor $v^4$. Hence their effect on $\alpha$
should not exceed $\approx$~15\%~\cite{Beneke:1996yb}.  Additional
corrections to transverse polarisation result from higher-order terms
in the short-distance part of the gluon fragmentation function. After
hard gluon radiation, the intermediate heavy quark pair can be in
various spin and angular momentum states and hadronise into
unpolarised or longitudinal charmonium. The explicit
calculation~\cite{Beneke:1996yb} reveals that the effect is
numerically quite small and that radiative ${\cal O}(\alpha_s)$
corrections can decrease the transverse polarisation by only about
10\%. The dominant source of depolarisation comes from the
colour-octet ${}^1S_0$ and ${}^3P_J$ fusion processes,
Figure~\ref{figure1}(d). They are important for moderate
transverse momentum and need to be included to predict quarkonium
polarisation in the $p_t$-range probed by present
data~\cite{Beneke:1997yw,Leibovich:1997pa}.

\begin{figure}[htb]
\vspace*{-8mm}
\hspace*{10mm}
\includegraphics[width=0.85\textwidth,clip]{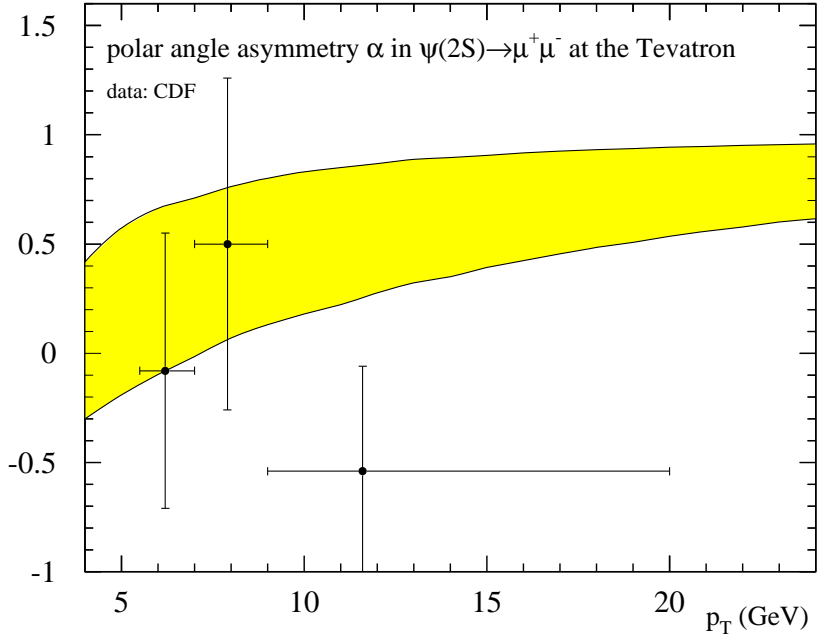}
\vspace*{-5mm}
\caption[Polar angle asymmetry $\alpha$ for $\psi(2S)$ production 
 at the Tevatron compared to experimental data] {\small Polar angle
 asymmetry $\alpha$ for $\psi(2S)$ production in $p\bar{p} \to
 \psi(2S)(\to \mu^+\mu^-)+X$ at the Tevatron as a function of $p_t$
 compared to experimental data from CDF~\cite{Affolder:2000nn}.
 Parameter specifications as in Figure~\ref{figure2}.  NLO corrections
 to the fragmentation contribution \cite{Beneke:1996yb,Beneke:1997yw}
 have been included. The error band is obtained as a combination of
 the uncertainty (statistical only) in the extraction of the NRQCD
 matrix elements [Table~\ref{table1}] and the limiting cases that
 either $\langle {\cal O}^{\psi}[8,{}^1S_0]\rangle$ or $\langle {\cal
 O}^{\psi}[8,{}^3P_0]\rangle$ is set to zero in the linear combination
 extracted from the data.}
\label{figure5}
\vspace*{-2mm} 
\end{figure}

At ${\cal{O}}(v^4)$ in the velocity expansion, the polarisation yield
from all production channels can be calculated unambiguously in NRQCD
in terms of the non-perturbative matrix elements that have been
determined from the unpolarised cross section.\footnote{The
calculation of the polarised cross section requires special care as
interference contributions from intermediate $c\overline{c}$ pairs in
different angular momentum states have to be
included~\cite{Beneke:1996yb,Braaten:1996jt}.}  Figure~\ref{figure5}
displays the theoretical prediction for the asymmetry $\alpha$ in
$\psi(2S)$ production at the Tevatron as a function of the $\psi(2S)$
transverse momentum.  Both the ${\cal O}(\alpha_s)$ corrections to the
fragmentation function and the ${\cal O}(4m_c^2/p_t^2)$ corrections
due to non-fragmentation contributions are included, but no attempt
has been made to quantify spin-symmetry breaking effects.  The shaded
error band reflects the uncertainty in the determination of the NRQCD
matrix elements from the unpolarised cross section. The matrix
elements have been varied in the range specified in
Equation~\ref{eq_mevar}, and the limiting cases that either $\langle
{\cal O}^{\psi(2S)}\,[8,{}^1S_0]\rangle$ or $\langle {\cal
O}^{\psi(2S)}\,[8,{}^3P_J]\rangle$ is set to zero in the combination
$M_{3.5}^{\psi(2S)}(8,{}^1S_0,{}^3P_0) $ have been
considered.  No transverse polarisation is expected at $p_t\sim 5\,$
GeV, but the angular distribution is predicted to change drastically
as $p_t$ increases. A first measurement from
CDF~\cite{Affolder:2000nn} does not support this prediction, but the
experimental errors are too large to draw definite conclusions.

The analysis of $J/\psi$ polarisation is complicated by the fact that
the experimental data sample~\cite{Affolder:2000nn} includes $J/\psi$
that have not been produced directly but come from decays of $\chi_c$
and $\psi(2S)$ mesons. The contribution from radiative $\chi$ decays
decreases the transverse $J/\psi$ polarisation at large $p_t$, but
tends to be cancelled out by the contribution from $\psi(2S)$
decays~\cite{Braaten:2000qk,Kniehl:2000nn}. The overall polarisation
yield for $J/\psi$ at large transverse momentum is thus not strongly
diluted by decays of higher charmonium states. Again, the first
experimental data do not confirm the NRQCD
prediction~\cite{Braaten:2000qk,Kniehl:2000nn}, see
Figure~\ref{figure6}.

The absence of transversely polarized $J/\psi$ and $\psi(2S)$ at large
transverse momentum, if confirmed with higher-statistics data at the
Tevatron Run~II, represents a serious challenge for the application of
NRQCD to charmonium production. In the following, some uncertainties
and potential shortcomings of the present theoretical
analysis will be addressed:

\begin{figure}[htb]
\vspace*{7mm}
\hspace*{25mm}
\includegraphics[width=0.65\textwidth,clip]{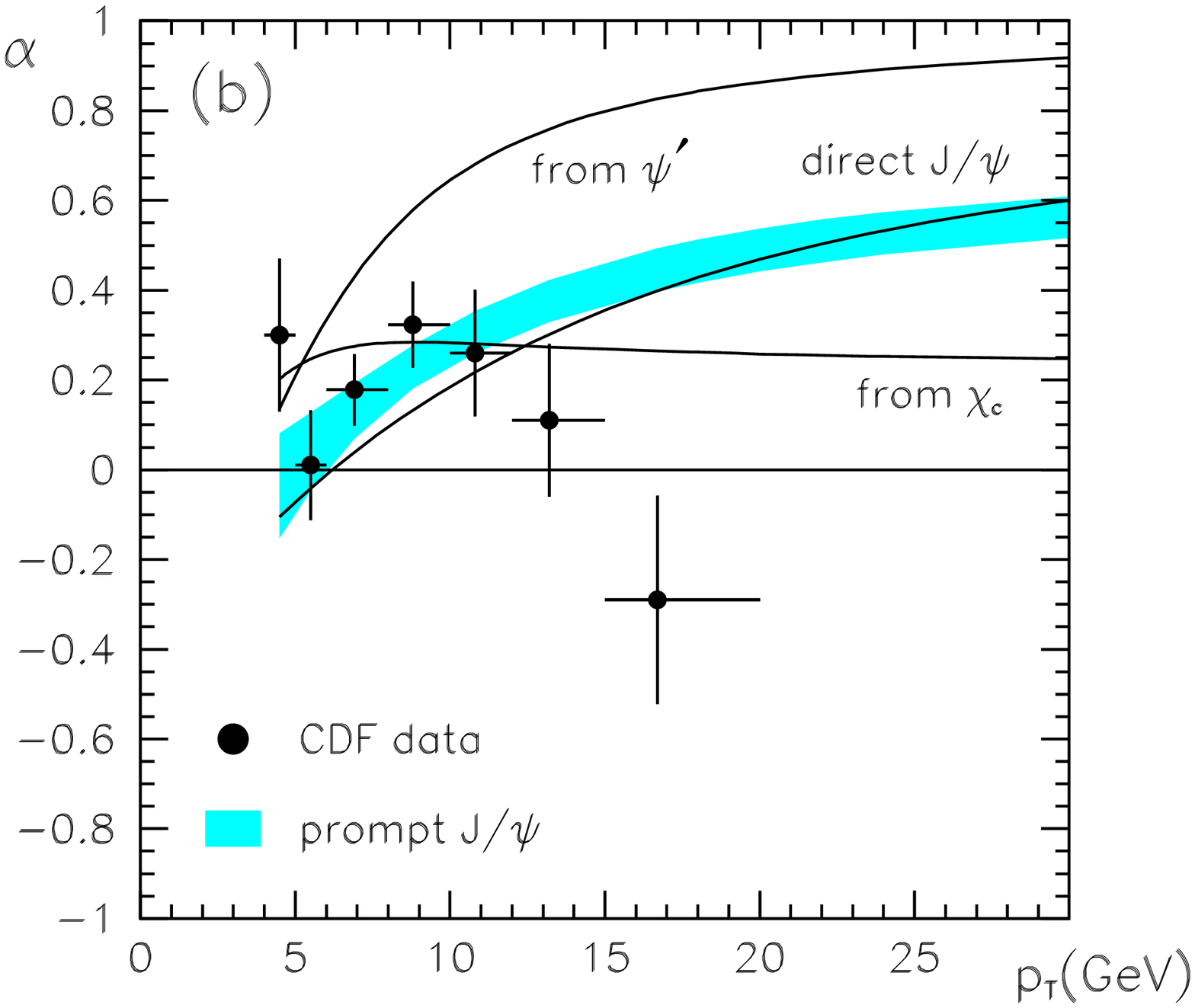}
\vspace*{0mm}
\caption[Same as Figure~\ref{figure5} for prompt $J/\psi$
 production]{\small Polar angle asymmetry $\alpha$ for prompt $J/\psi$
 production (i.e.\ not from B-decay) in $p\bar{p} \to J/\psi(\to
 \mu^+\mu^-)+X$ at the Tevatron as a function of $p_t$ compared to
 experimental data from CDF~\cite{Affolder:2000nn}. From
 \cite{Braaten:2000qk}.}  \label{figure6}
\end{figure}

 The {\it uncertainty in the determination of NRQCD matrix elements}
 $\langle {\cal O}^{\psi}\,[8,{}^1S_0]\rangle$, $\langle {\cal
 O}^{\psi}\,[8,{}^3S_1]\rangle$ and $\langle {\cal
 O}^{\psi}\,[8,{}^3P_J]\rangle$ translates into an uncertainty in the
 predicted yield of transversely polarized $J/\psi$ and $\psi(2S)$.
 In the large-$p_t$ limit, the ${}^3S_1$ production channel yields
 transverse polarisation, while the ${}^1S_0$ and ${}^3P_J$ channels
 both yield unpolarised quarkonia in this limit.  A decreased value of
 the $\langle {\cal O}^{\psi}\,[8,{}^3S_1]\rangle$ matrix element,
 relative to $\langle {\cal O}^{\psi}\,[8,{}^1S_0,{}^3P_J]\rangle$,
 would thus delay the onset of transverse polarisation and reduce the
 discrepancy between NRQCD predictions and data.  However, extreme
 values for the non-perturbative matrix elements outside the range
 specified in Equation~\ref{eq_mevar} would be in conflict with most
 of the analyses listed in Tables~\ref{table2} and \ref{table3}, and
 moreover be inconsistent with conventional NRQCD power counting.

 {\it Higher-order QCD effects} in the short-distance cross section
 should be included to improve the accuracy of the theoretical
 prediction, in particular in the intermediate $p_t$ region.
 Preliminary studies indicate large NLO effects for the unpolarized
 $p_t$-distributions, see Section~\ref{sec_cxs}, but no attempt has
 been made so far to estimate the impact of higher-order corrections
 on the spin-dependence of the cross section. First studies within the
 $k_t$-factorisation approach~\cite{Yuan:2000qe} suggest that the
 $J/\psi$ and $\psi(2S)$ cross section might not be dominated by gluon
 fragmentation but rather by the colour-octet ${}^1S_0$ process up to
 $p_t\;\simlt \; 20$~GeV. Since the ${}^1S_0$ production channel
 yields unpolarised quarkonia, the overall $J/\psi$ and $\psi(2S)$
 polarisation might be greatly reduced. However, as mentioned
 previously, the theoretical uncertainties in the $k_t$-factorisation
 prediction need yet to be quantified.\footnote{The polarisation of 
 $\chi_{cJ}$ has also been studied recently in the context of the 
 $k_t$-factorisation approach~\cite{Yuan:2001xw}.}

 {\it Heavy quark spin-symmetry} is violated by higher-order terms in
 the NRQCD Lagrangian, and longitudinal polarisation can arise if the
 binding of the charm quark pair into $J/\psi$ and $\psi(2S)$ proceeds
 through two chromomagnetic transitions.  The suppression of the
 spin-symmetry breaking chromomagnetic interactions follows from the
 conventional NRQCD power counting rules. These rules, however, have
 not been firmly established for the case of charmonium.  Alternative
 power counting schemes~\cite{Beneke:1997av,Fleming:2000ib} imply that
 the chromomagnetic transitions might not be suppressed with respect
 to chromoelectric transitions. As a consequence, the $\langle {\cal
 O}^{\psi}\,[8,{}^1S_0]\rangle$ matrix element would be enhanced
 relative to $\langle {\cal O}^{\psi}\,[8,{}^3S_1]\rangle$ and
 $\langle {\cal O}^{\psi}\,[8,{}^3P_J]\rangle$. Such a hierarchy is
 not inconsistent with the analyses of the unpolarised $J/\psi$ and
 $\psi(2S)$ cross sections, if one assumes that the linear combination
 $M_{k}^{\psi}(8,{}^1S_0,{}^3P_0) $ is dominated by $\langle
 {\cal O}^{\psi}\,[8,{}^1S_0]\rangle$.  Since chromomagnetic
 transitions do not preserve heavy quark spin-symmetry the
 polarisation of $J/\psi$ and $\psi(2S)$ could be significantly
 reduced in the alternative power counting schemes.

 Finally, charmonium production mechanisms {\it beyond NRQCD} could be
 responsible for $J/\psi$ and $\psi(2S)$ depolarisation. For example,
 the leading-twist formalism of NRQCD factorisation does not include
 possible rescattering interactions between the intermediate heavy
 quark pair and a comoving colour field.  It has been shown that such
 rescattering corrections could yield unpolarized
 quarkonium~\cite{Marchal:2000wd,Maul:2001fw}. The
 analysis~\cite{Marchal:2000wd,Maul:2001fw}, however, relies on
 several simplifying assumptions, and further work is needed to
 establish the importance of comover interactions for charmonium
 production at large $p_t$.

\vspace*{2mm}

The distinctive NRQCD prediction of transverse $J/\psi$ and $\psi(2S)$
polarisation at large $p_t$ is not supported by first experimental
data. However, the measurements still have rather large errors and
no conclusions should be drawn until the statistics improves. If, in
Run~II of the Tevatron, the experimental results continue to disagree
with NRQCD predictions, one might have to conclude that either
higher-order QCD effects could be crucial to describe the the
spin-dependence of the charm cross section, or that alternative power
counting schemes might be needed for the case of charmonium.

\subsection{Bottomonium production}
The application of NRQCD should be on safer grounds for the
bottomonium system. As $v^2 \sim 0.1$ for bottomonium, higher-order
terms in the velocity expansion (in particular colour-octet
contributions) are expected to be less significant than in the case of
charmonium. Cross sections for the production of $\Upsilon$ and
$\chi_b$ states have been measured at the Tevatron in the region $p_t
\;\simlt\;20\,\mbox{GeV}$ \citer{Abe:1995an,CDF-ups}.
The leading-order colour-singlet model predictions underestimate the
data, but the discrepancy is much less significant than in the case of
charmonium. Given the large theoretical uncertainties in the cross
section calculation, in particular at small
$p_t\;\simlt\;M_{\Upsilon}$, the need for colour-octet contributions
is not yet as firmly established as for charmonium production.  The
inclusion of both next-to-leading order corrections and the summation
of soft gluon radiation is required to obtain a realistic description
of the $\Upsilon$ cross section in the $p_t$-range probed by present
data. Such calculations have not yet been performed.

\begin{figure}[htb]
\vspace*{-8mm}
\hspace*{8mm}
\includegraphics[width=0.9\textwidth,clip]{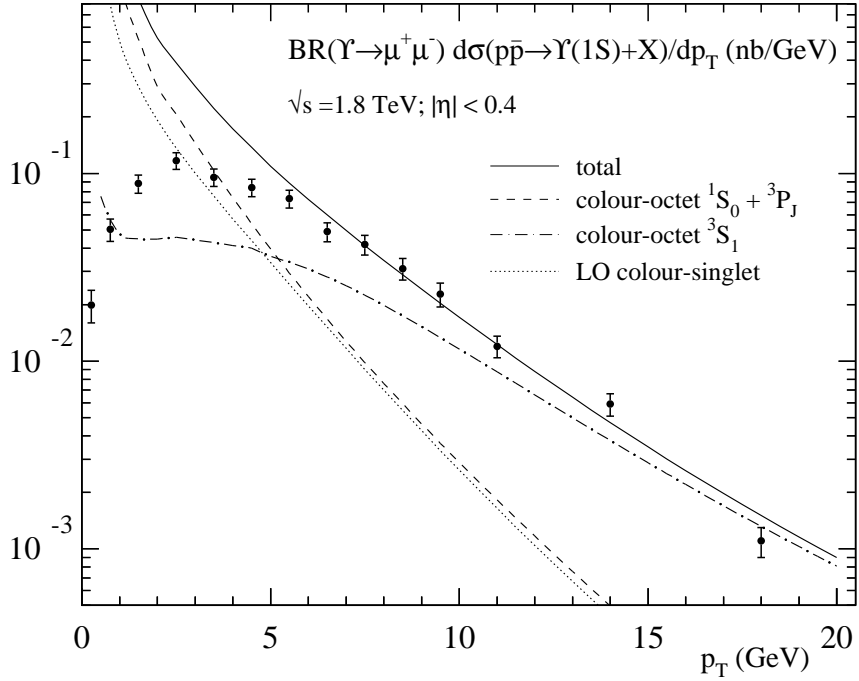}
\vspace*{-8mm}
\caption[Colour-singlet and colour-octet contributions to inclusive
 $\Upsilon(1S)$ production at the Tevatron compared to experimental
 data] {\small Colour-singlet and colour-octet contributions
 to inclusive $\Upsilon(1S)$ production in $p\bar{p} \to
 \Upsilon(1S)+X$ at the Tevatron ($\sqrt{s}=1.8$~TeV, pseudorapidity
 cut $|\eta|<0.4$)) compared to experimental data from CDF
 \cite{Abe:1997jz}. Parameters: CTEQ5L parton distribution functions
 \cite{Lai:1999wy}; factorisation and renormalisation scale $\mu =
 \sqrt{p_t^2+4m_b^2}$; $m_b = 4.88$~GeV. NRQCD matrix elements as
 specified in Table~\ref{table5}.}
\label{figure7} 
\end{figure}

The inclusive cross section for $\Upsilon(1S)$ production at the
Tevatron is compared to the leading-order NRQCD calculation in
Figure~\ref{figure7}. 
The theoretical curves are based on the simple choice of
non-perturbative parameters listed in
Table~\ref{table5}~\cite{LHC-workshop}.
\begin{table}
\begin{center}
\vskip0.2cm
\renewcommand{\arraystretch}{1.5}
$$
\begin{array}{|c|ccc|}
\hline\hline
 H & \langle {\cal{O}}_1^{H} \rangle  & \langle
 {\cal{O}}^{H}[8,{}^3S_1] \rangle  
 &\langle {\cal{O}}^{H}[8,{}^1S_0] \rangle\\ \hline\hline
 \Upsilon(1S)   & 9.28  ~{\rm GeV^3} & 15 \times 10^{-2}~{\rm GeV}^3 &
 2.0 \times 10^{-2}~{\rm GeV}^3 \\[-1mm] 
 \Upsilon(2S)   & 4.63  ~{\rm GeV^3} & 4.5 \times 10^{-2}~{\rm GeV}^3 &
 0.6 \times 10^{-2}~{\rm GeV}^3 \\[-1mm]
 \Upsilon(3S)   & 3.54  ~{\rm GeV^3} & 7.5 \times 10^{-2}~{\rm GeV}^3 &
 1.0 \times 10^{-2}~{\rm GeV}^3 \\[-1mm]
 \chi_{0}(1P)   & 2.03  ~{\rm GeV^5} & 4.0 \times 10^{-2}~{\rm GeV}^3 &
 \\[-1mm]
 \chi_{0}(2P)   & 2.57  ~{\rm GeV^5} & 6.5 \times 10^{-2}~{\rm GeV}^3 &
 \\ 
\hline \hline
\end{array}
$$
\renewcommand{\arraystretch}{1.0}
\caption[NRQCD matrix elements for bottomonium production obtained 
 from the transverse momentum distribution at the Tevatron] {\small
 NRQCD matrix elements for bottomonium production.  The colour-singlet
 matrix elements are taken from the potential model calculation of
 \cite{Buchmuller:1981su,Eichten:1995ch}. The colour-octet matrix
 elements have been determined from the CDF data for $p_t > 8~{\rm
 GeV} $ \cite{Affolder:1999wm}, where $\langle {\cal
 O}^{H}[8,{}^1S_0]\rangle = \langle {\cal
 O}^{H}[8,{}^3P_0]\rangle/m_b^2$ has been assumed for
 simplicity. Parameters: CTEQ5L parton distribution functions
 \cite{Lai:1999wy}, renormalisation and factorisation scale
 $\mu=\sqrt{p_t^2+4 m_b^2}$ and $m_b=4.88\,$GeV.}
\label{table5}
\end{center}
\end{table}
The colour-octet matrix elements have been determined from the CDF
data for $p_t > 8$~GeV, where the contributions from direct
$\Upsilon(1S)$ production and from radiative $\chi_b$ decays are
measured separately.  Given the large uncertainty in the leading-order
calculation of the short distance cross section, no attempt has been
made to quantify the error of the non-perturbative parameters listed
in Table~\ref{table5}. The ambiguities in the determination of the
non-perturbative matrix elements for bottomonium production are best
demonstrated by comparing the results in Table~\ref{table5} with other
numerical analyses~\cite{Cho:1996ce,Braaten:2001cm}, collected in
Table~\ref{table6} for the case of direct $\Upsilon(1S)$ production.

\begin{table}[thb]
\begin{center}
\vskip0.2cm
\renewcommand{\arraystretch}{1.3}
$$
\begin{array}{|c|cc|}
\hline\hline
 \mbox{Reference} &  {\cal{O}}^{\Upsilon(1S)}[8,{}^3S_1] \rangle & 
 \langle {\cal{O}}^{\Upsilon(1S)}[8,{}^1S_0] \rangle \\
\hline\hline
 \cite{Cho:1996ce} & 0.59 \pm 0.19 & 3.6\pm 5 \\ \hline
 \cite{Braaten:2001cm} & 2.0 \pm 4.1^{-0.6}_{+0.5} & 13.6 \pm 6.8
 ^{+10.8}_{-7.5} \\ \hline \hline
\end{array}
$$
\renewcommand{\arraystretch}{1.0}
\caption[NRQCD matrix elements for $\Upsilon(1S)$ production obtained
 from other analyses of the transverse momentum distribution]{\small
 NRQCD colour-octet matrix elements for $\Upsilon(1S)$ production
 obtained from various analyses of the transverse momentum
 distribution at the Tevatron.  Values are given in units
 $10^{-2}$~GeV${}^3$. The first error quoted is statistical, the
 second error, when listed, due to variation of the renormalisation
 and factorisation scales between $\mu=1/2\,\sqrt{p_t^2+m_b^2}$ 
 and $2\,\sqrt{p_t^2+m_b^2}$.}
\label{table6}
\end{center}
\end{table}

The impact of initial state gluon radiation on the $\Upsilon$ cross
sections at the Tevatron has been estimated by adding a Gaussian
$k_t$-smearing as discussed previously in the context of charmonium
production. An average $\langle k_t \rangle \sim 3$~GeV and a
$K$-factor $\sim 3$ are found to bring the leading-order
colour-singlet cross section in line with the inclusive
$\Upsilon(1S,2S)$ data~\cite{Mangano:1995yd}. Similar results have
been obtained within a Monte Carlo analysis \cite{Domenech:2000qg},
leading to significantly lower fit values for the colour-octet NRQCD
matrix elements than those determined from a leading-order
calculation. Moreover, the Monte Carlo results imply that no feeddown
from $\chi$ states produced through colour-octet ${}^3S_1$ $b\bar{b}$
states is needed to describe the inclusive $\Upsilon$ cross section,
in contrast to what is found at leading-order. The calculation of
next-to-leading order corrections and a systematic treatment of soft
gluon radiation within perturbation theory are required to resolve
these issues.

A preliminary measurement shows that inclusive $\Upsilon(1S)$ is
produced unpolarised within the experimentally accessible range of
transverse momentum. This result is consistent with NRQCD
factorisation which predicts the onset of transverse $\Upsilon$
polarisation at $p_t(\Upsilon)\;\simgt\;
15$~GeV~\cite{Braaten:2001gw}, beyond what is probed by present
data.\footnote{Transverse polarisation of $\Upsilon$ states produced
in $p$-Cu collisions at $\sqrt{s} = 38.8$~GeV has been reported
recently~\cite{Brown:2001bz}. The observation of a non-zero
polarisation disproves the assumption that gluon emission from the
$b\bar{b}$ pair during hadronisation randomises spin and colour, as
implicit in the colour-evaporation model. On the other hand, the
degree of transverse polarisation observed experimentally is smaller
than predicted by a NRQCD
calculation~\cite{Kharchilava:1999wa,Tkabladze:1999mb}.}

\subsection{Prospects for the Tevatron~Run~II and the LHC\label{sec_tev2}}
Run~II at the Tevatron will provide a substantial increase in
luminosity and will allow to determine the $J/\psi$, $\psi(2S)$ and
$\chi_c$ cross sections more precisely and up to larger values of
$p_t$. A accurate measurement of the $J/\psi$ and $\psi(2S)$
polarisation at large transverse momentum will be the most crucial
test of NRQCD factorisation. In addition, improved data on the
$J/\psi$ and $\psi(2S)$ cross sections will help to reduce some of the
ambiguities in extracting the colour-octet matrix elements.

With increased statistics it might be possible to access other
charmonium states like the $\eta_c(nS)$ or the $h_c(nP)$.  Heavy-quark
spin symmetry provides approximate relations between the
non-perturbative matrix elements that describe spin-singlet and
spin-triplet states~\cite{Bodwin:1995jh}. The matrix elements for
$\eta_c(nS)$ are related to those for $\psi(nS)$, while the leading
matrix elements for $h_c(nP)$ can be obtained from those for
$\chi_c(nP)$. Within NRQCD the rates for $\eta(nS)$ and $h(nP)$ can
thus be predicted unambiguously in terms of the non-perturbative
matrix elements that describe the $J/\psi$, $\psi(2S)$ and $\chi_c$
cross sections. Comparing the various charmonium production rates
would hence provide a stringent test of NRQCD factorisation and
heavy-quark spin symmetry.  The cross sections for producing $\eta_c$
and $h_c$ at Run~II of the Tevatron are
large~\cite{Mathews:1998nk,Sridhar:1996vd}, but the acceptances and
efficiencies for observing the decay modes that can be triggered upon
are in general small, and detailed experimental studies are needed to
quantify the prospects.  Other charmonium processes that have been
studied in the literature include the production of $D$ wave
states~\cite{Qiao:1997wb}, $J/\psi$ production in association with
photons~\cite{Kim:1997bb,Mathews:1999ye} and double gluon
fragmentation to $J/\psi$ pairs~\cite{Barger:1996vx}.

The larger statistics expected at Run~II of the Tevatron will also
allow to improve the measurement of the bottomonium cross
sections. Yet undiscovered states like $\eta_b(1S)$ could be detected,
for example in the decay $\eta_b \to J/\psi +
J/\psi$~\cite{Braaten:2001cm}, and the associated production of
$\Upsilon$ and electroweak bosons might be
accessible~\cite{Braaten:1999th}. If sufficient statistics can be
accumulated, the onset of transverse $\Upsilon(nS)$ polarisation may
be visible at $p_t(\Upsilon)\;\simgt\; 15$~GeV.

\vspace*{2mm}

\begin{figure}[h,t,b]
\vspace*{-13mm}
\hspace*{10mm}
\includegraphics[width=0.85\textwidth,clip]{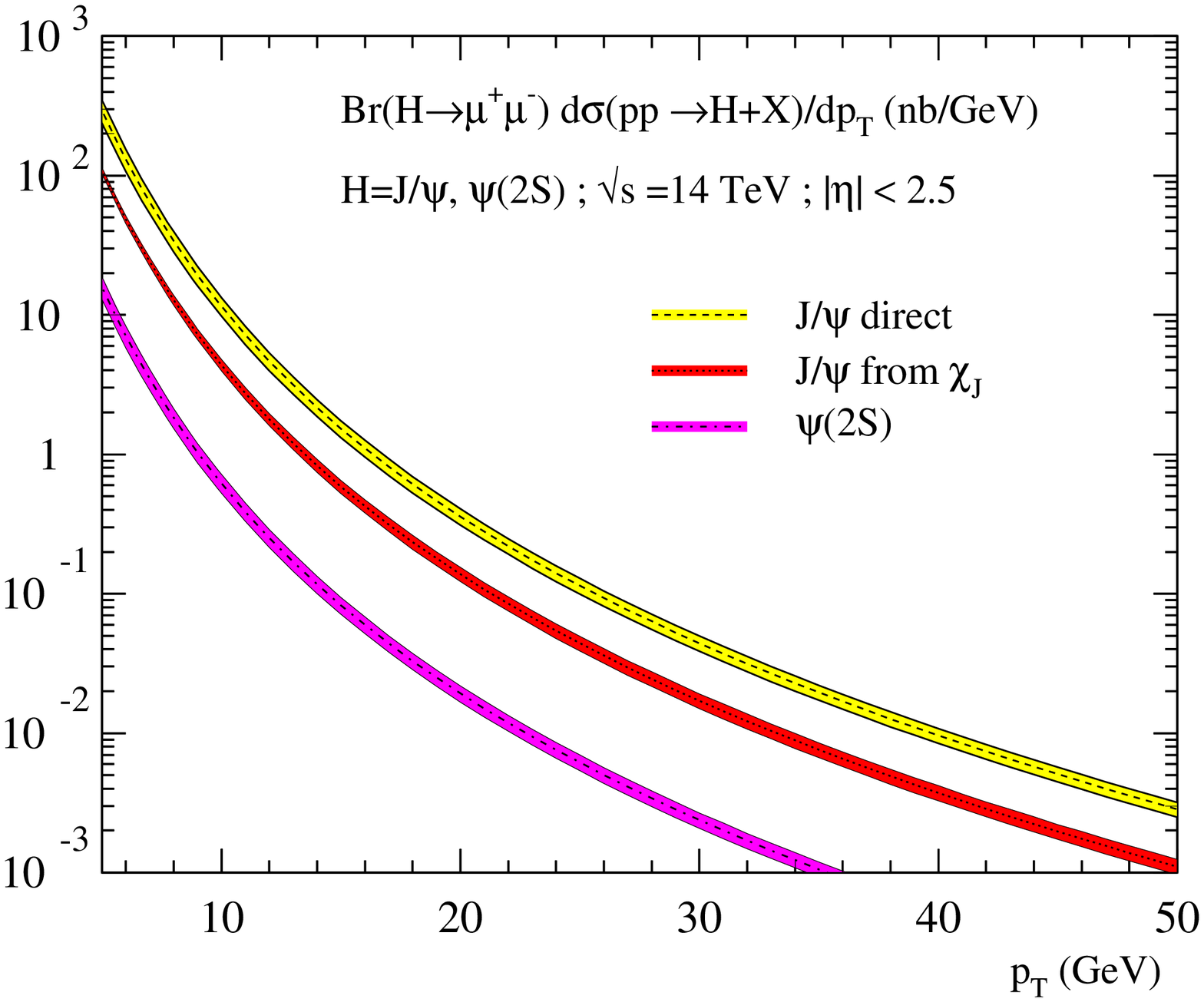}

\vspace*{-19mm}
\hspace*{10mm}
\includegraphics[width=0.85\textwidth,clip]{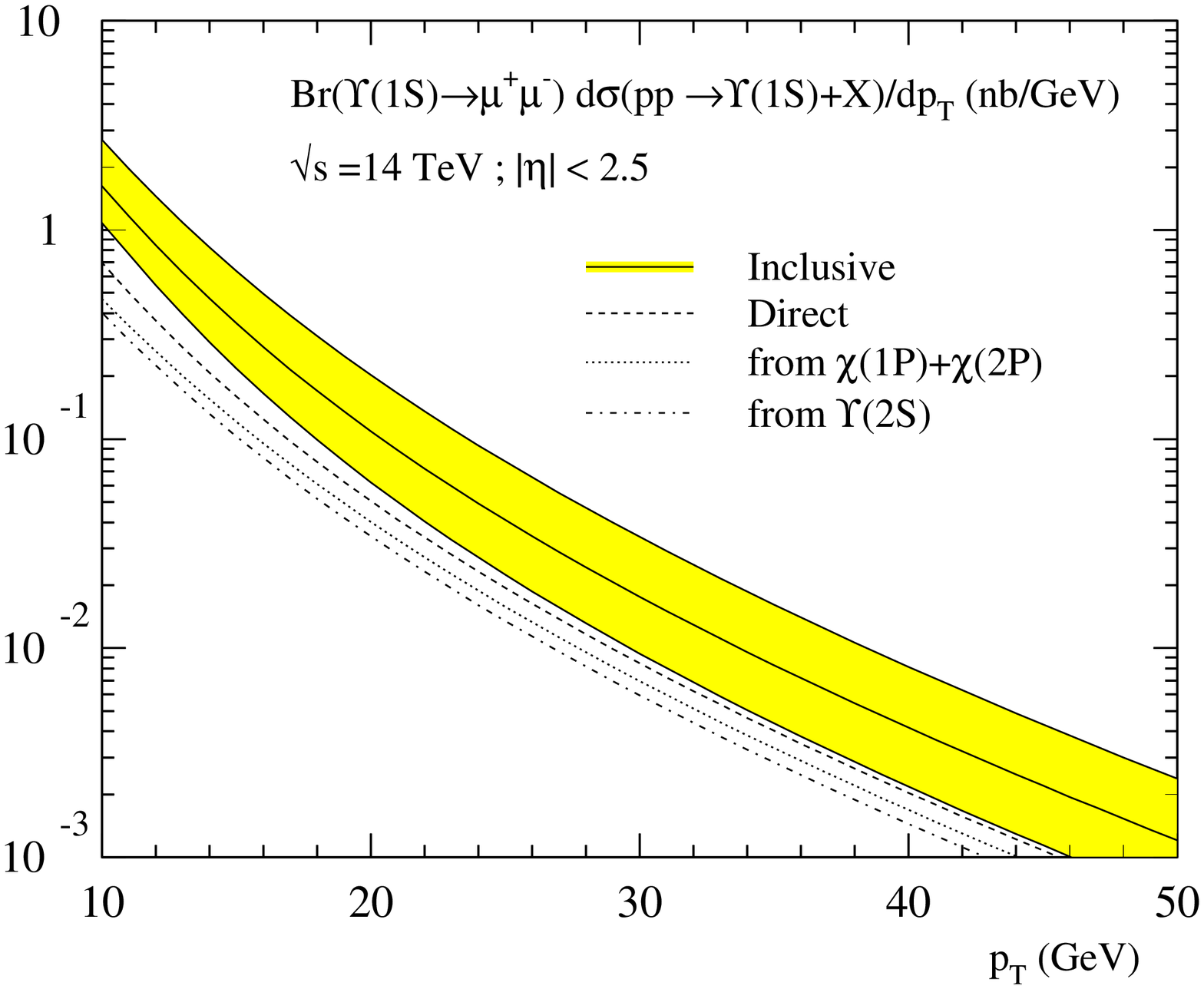}
\vspace*{-8mm}
\caption[Charmonium and bottomonium cross sections at the LHC] 
 {\small Cross sections for charmonium and bottomonium production $pp
 \to H+X$ at the LHC ($\sqrt{s}=14$~TeV, pseudorapidity cut
 $|\eta|<2.5$). Parameters: CTEQ5L parton distribution
 functions~\cite{Lai:1999wy}, factorisation and renormalisation scale
 $\mu = \sqrt{p_t^2+4m_Q^2}$, $m_c = 1.5$~GeV$, m_b = 4.88$~GeV. The
 leading logarithms $(\alpha_s\ln p_t^2/(2m_b)^2)^n$ have been summed
 by solving the Altarelli-Parisi evolution equations for the gluon
 fragmentation function. NRQCD matrix elements as in
 Tables~\ref{table1},\ref{table5}. For bottomonium production the
 error band is obtained by varying the colour-octet matrix elements
 between half and twice their central value for illustration.}
\label{figure8} 
\end{figure}

A comprehensive test of the NRQCD factorisation approach in the
bottomonium sector and a measurement of the $\Upsilon$ polarisation at
$p_t\gg M_{\Upsilon}$ will have to wait until the LHC starts
operating. By that time, new experimental data from the Tevatron and
HERA, as well as theoretical progress, e.g.\ in the calculation of
higher-order corrections, will have significantly improved the present
picture. Still, in order to provide some benchmark cross sections for
quarkonium production at the LHC, the NRQCD predictions for selected
charmonium and bottomonium states have been collected in
Figure~\ref{figure8}~\cite{LHC-workshop}. Given the substantial
uncertainty in the determination of the non-perturbative matrix
elements from present data, the predictions should be regarded as
order-of-magnitude estimates.\footnote{For a collection of LHC cross
section predictions obtained from a Monte Carlo event generator
see~\cite{Domenech:2001ri}.}

\begin{figure}[h,t,b]
\vspace*{-8mm}
\hspace*{10mm}
\includegraphics[width=0.85\textwidth,clip]{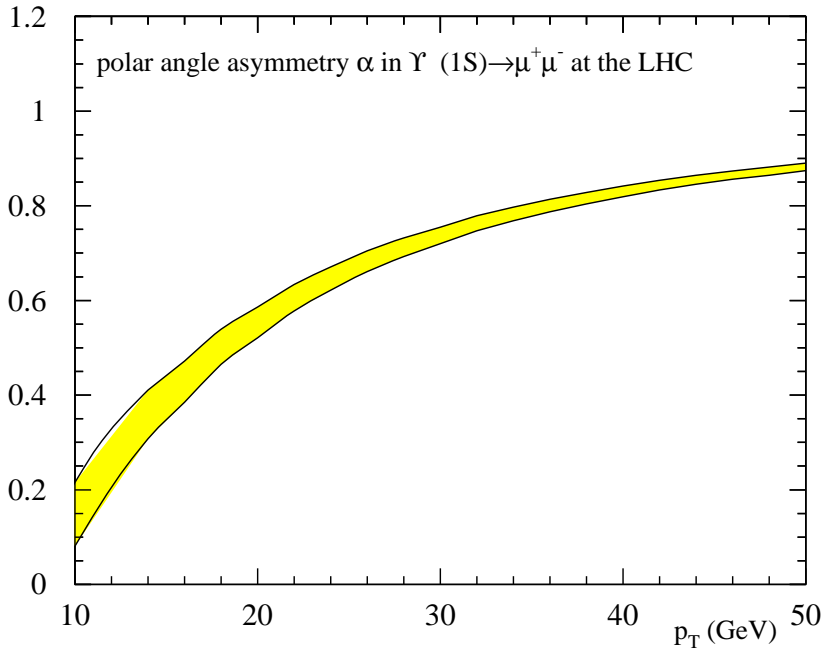}
\vspace*{-8mm}
\caption[Polar angle asymmetry $\alpha$ for direct $\Upsilon(1S)$
 production at the LHC]{\small Polar angle asymmetry $\alpha$ for
 direct $\Upsilon(1S)$ production in $pp \to \Upsilon(1S)(\to
 \mu^+\mu^-)+X$ at the LHC as a function of $p_t$. NRQCD matrix
 elements as specified in Table~\ref{table5}, other parameters as in
 Figure~\ref{figure7}. The error band reflects the limiting cases that
 either $\langle {\cal O}^{\psi}[8,{}^1S_0]\rangle$ or $\langle {\cal
 O}^{\psi}[8,{}^3P_0]\rangle$ is set to zero in the linear combination
 extracted from the data.}
\label{figure9}
\vspace*{10mm}
\end{figure}

The polarisation pattern predicted for direct $\Upsilon(1S)$
production at the LHC is presented in Figure~\ref{figure9}, based on
the NRQCD matrix elements of Table~\ref{table5}.
Higher-order corrections to the gluon fragmentation function
\cite{Beneke:1996yb} will lead to a small reduction of
the transverse polarisation at large $p_t$ and should be included once
data become available. If the charmonium mass is indeed not large
enough for a nonrelativistic expansion to be reliable, the onset of
transverse $\Upsilon$ polarisation at $p_t \gg M_{\Upsilon}$ may
become the most decisive test of the NRQCD factorisation approach.

\section{Quarkonium production at $ep$ colliders\label{sec_ep}}
The analysis of charmonium and bottomonium cross sections at the
high-energy $ep$ collider \mbox{HERA} provides a powerful tool to
assess the importance of the different quarkonium production
mechanisms and to test the general picture developed in the context of
NRQCD factorisation. Section~\ref{sec_gp} focusses on $J/\psi$ and
$\psi(2S)$ photoproduction, for which most of the experimental data
have been collected. The impact of colour-octet processes and
higher-order QCD corrections on cross sections and various
differential distributions is reviewed, and the theoretical
predictions are confronted with recent experimental
results. Quarkonium production in deep-inelastic scattering is
discussed in Section~\ref{sec_dis}. The high-statistics data expected
at the upgraded \mbox{HERA} collider will allow detailed studies of
more exclusive observables like $J/\psi$ polarisation, see
Section~\ref{sec_eppol}. Moreover, various other processes may be
accessible, including $\chi_c$ production, associated $J/\psi +
\gamma$ production and the production of bottomonium states. The
prospects for quarkonium physics at the upgraded HERA collider will be
discussed in some detail in Section~\ref{sec_heraup}.

\subsection{$J/\psi$ and $\psi(2S)$ photoproduction\label{sec_gp}}
Quarkonium production in high-energy $ep$ collisions at \mbox{HERA} is
dominated by photoproduction events where the electron is scattered by
a small angle producing photons of almost zero
virtuality. Experimental results have been reported for $J/\psi$ and
$\psi(2S)$ photoproduction in a wide kinematical region,
$30~\mbox{GeV}\;\simlt\; \sqrt{\vphantom{l}s_{\gamma p}} \;\simlt\;
200$~GeV. The cross section is dominated by photon-gluon fusion, where
the photon interacts as a pointlike particle. Besides the direct
production channel, charmonium photoproduction at \mbox{HERA} can also
take place through resolved reactions, in which case the photon
participates in the hard scattering through its parton content.
Resolved processes contribute significantly to the lower endpoint of
the charmonium energy spectrum and have been probed at \mbox{HERA} for
the first time recently.  Higher-twist phenomena like
elastic/diffractive quarkonium
production~\citer{Donnachie:1987pu,Brodsky:1994kf} are not included in
the leading-twist calculation of NRQCD (see~\cite{Beneke:1998re} for a
detailed discussion) and have to be eliminated by either a cut in the
$J/\psi$ transverse momentum $p_t\;\simgt\; 1$~GeV, or by a cut in the
$J/\psi$ energy variable $z\equiv p_P \cdot k_{\psi}/p_P \cdot
k_{\gamma}
\;\simlt \; 0.9$ (in the proton rest frame, $z$ is the ratio of the
$J/\psi$ to $\gamma$ energy, $z = E_{\psi}/E_{\gamma}$).

\vspace*{2mm}

\begin{figure}[htb]
\vspace*{5mm}
\epsfig{figure=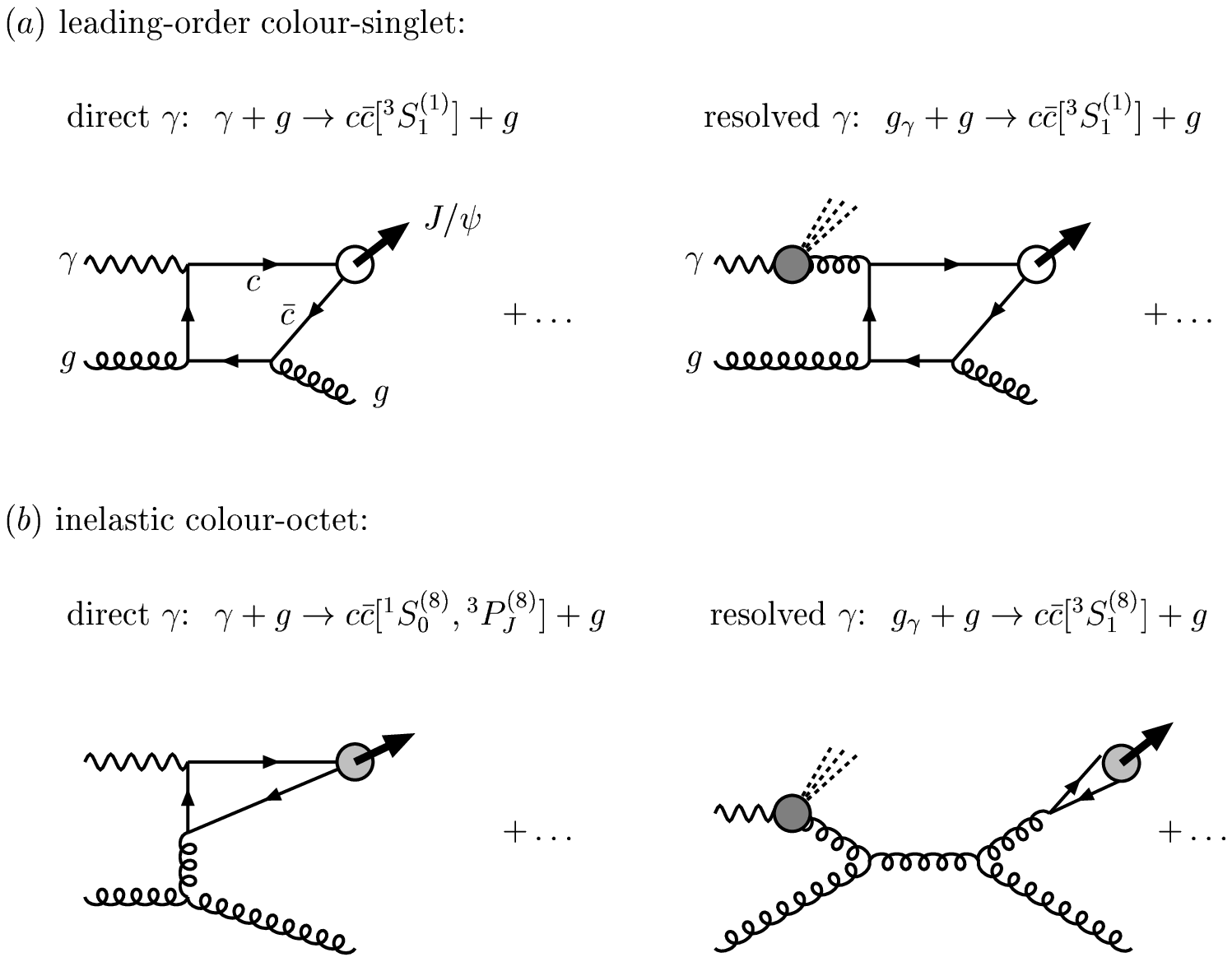,%
        bbllx=85pt,bblly=360pt,bburx=540pt,bbury=745pt,%
        width=0.92\textwidth,clip=}
\vspace*{-3mm}
\caption[Generic diagrams for inelastic 
 $J/\psi$ and $\psi(2S)$ photoproduction]{\small Generic diagrams for
 inelastic $J/\psi$ and $\psi(2S)$ photoproduction through
 colour-singlet and colour-octet channels.}
\label{figure10}
\end{figure}

\noindent
{\bf Photoproduction mechanism$\;$} 
The cross section for quarkonium photoproduction can be written as
\begin{eqnarray}
d\sigma(\gamma p \to H + X) &\!\!= \!\!& \sum_{i,j\in\{\gamma,g,q\}} 
\int dx_1 dx_2 f_{i/\gamma}(x_1) f_{j/p}(x_2)\nonumber \\ 
&&\hspace*{10mm}\times \sum_n
d\hat{\sigma}(i+j \to Q\overline{Q}\, [n] + X)\,\langle
{\cal{O}}^{H}\,[n]\rangle ,
\end{eqnarray}
where $f_{i/\gamma}$ and $f_{j/p}$ denote the parton distributions in
the photon and the proton, respectively. For the direct photon
processes $i=\gamma$ and $f_{\gamma/\gamma}(x) = \delta(1-x)$.  In the
following, it is assumed that the $J/\psi$ transverse momentum is
restricted to $p_t > 1$~GeV, in order to suppress the elastic
contribution and higher-twist corrections in general. Generic Feynman
diagrams which contribute to the parton cross section in the inelastic
region away from $p_t=0$ (or $z=1$) are shown in
Figure~\ref{figure10}. At leading order in the velocity expansion,
$J/\psi$ is produced through the colour-singlet
channel~\cite{Berger:1981ni}, Figure~\ref{figure10}(a). Relativistic
corrections at ${\cal{O}}(v^2)$ modify the large-$z$ and small-$p_t$
region~\citer{Keung:1983jb,Paranavitane:2000if} but can be neglected
for $p_t\;\simgt\; 1\,$~GeV. The next-to-leading order QCD corrections
to the direct photon process~\cite{Kramer:1995zi,Kramer:1996nb}
increase the cross section prediction significantly, in particular at
$p_t \;\simgt\; m_c$ as discussed in detail below.

At ${\cal{O}}(v^4)$ relative to the colour-singlet contribution, the
$J/\psi$ can also be produced through intermediate colour-octet
$^3\!S_1$, $^1\!S_0$ and $^3\!P_J$
configurations~\citer{Cacciari:1996dg,Ko:1996xw}, see
Figure~\ref{figure10}(b).  Next-to-leading order corrections to the
colour-octet channels are only known for the total photoproduction
cross section (integrated over all $z$ and
$p_t$)~\cite{Maltoni:1998pt}.

The importance of the colour-octet production processes follows from a
kinematical enhancement of the short-distance cross section, similar
to the case of $J/\psi$ hadroproduction.  The $^1\!S_0$ and $^3\!P_J$
configurations can be produced through $t$-channel exchange of a gluon
already at lowest order in $\alpha_s$. (For the $^3\!S_1$ octet
channel this is true for the resolved process.) This leads to a
potentially significant colour-octet contribution, in particular in
the large-$z$ region.

The leading-order subchannels which contribute to inelastic $J/\psi$
photoproduction through direct photon processes are
\begin{eqnarray}
\gamma + g & \rightarrow & c\bar c 
\left[1,{}^3\!S_1; \,8,\!{}^3\!S_1; \,8,\!{}^1\!S_0; \,8,\!{}^3\!P_J
 \right] + g, \label{subpg}\\
\gamma + q/\bar{q} & \rightarrow & c\bar c
\left[8,\!{}^3\!S_1; \,8,\!{}^1\!S_0; \,8,\!{}^3\!P_J \right] + q/\bar{q},
\end{eqnarray}
where the initial-state parton originates from the target proton. For 
resolved photon processes, the subchannels are
\begin{eqnarray}
g + g      & \rightarrow & c\bar c
\left[1,{}^3\!S_1; \,8,\!{}^3\!S_1; \,8,\!{}^1\!S_0; \,8,\!{}^3\!P_J
 \right] + g, \label{subgg2} \\
g + q/\bar{q}      & \rightarrow & c\bar c
\left[ 8,\!{}^3\!S_1; \,8,\!{}^1\!S_0; \,8,\!{}^3\!P_J \right] + q/\bar{q}, \\
q + \bar q & \rightarrow & c\bar c
\left[ 8,\!{}^3\!S_1; \,8,\!{}^1\!S_0; \,8,\!{}^3\!P_J \right] + g,
\end{eqnarray}
where one of the initial-state partons originates from the photon and
the other originates from the proton. Explicit expressions for the
parton differential cross sections are given in~\cite{Beneke:1998re}.
The direct-photon processes dominate in the region $z \;\simgt\; 0.2$,
whereas resolved-photon processes become important in the region $z
\;\simlt\; 0.2$. At HERA energies, photon-quark fusion can contribute 
about 10\%--15\% to the cross section at large $z$. Quark-gluon fusion
constitutes about 20\%--40\% of the resolved cross section at $z
\;\simlt\; 0.2$ and becomes more important than gluon-gluon fusion at
larger $z$.  Quark-antiquark fusion is always completely negligible.

Fragmentation contributions in photon-gluon
fusion~\citer{Godbole:1996ie,Kniehl:1997gh} exist only at the next
order in $\alpha_s$ and are suppressed for $p_t\;\simlt\; 10$~GeV.
The $c\bar{c}[8,{}^3\!S_1]$-channel is insignificant for the direct-photon
contribution, but dominates in resolved-photon interactions at
$p_t\;\simgt\; 5$~GeV because it includes a gluon fragmentation
component.  The resolved photon amplitudes are identical to those
relevant to $J/\psi$ production in hadron-hadron collisions, and at
HERA energies the relative importance of the various contributions as
functions of $p_t$ is nearly the same as at Tevatron energies.

\begin{figure}[t,h,b]
\vspace*{-8mm}
\hspace*{8mm}
\includegraphics[width=0.9\textwidth,clip]{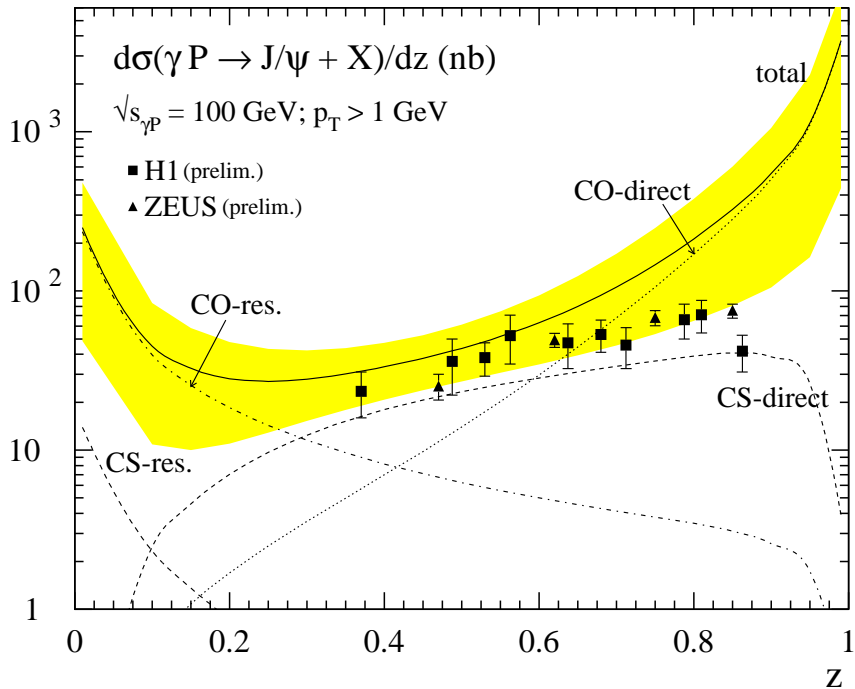}
\vspace*{-8mm}
\caption[Colour-singlet and colour-octet contributions to the $J/\psi$ 
 energy distribution in photoproduction at HERA compared to
 experimental data]{\small Colour-singlet (CS) and colour-octet (CO)
 contributions due to direct and resolved photons to the $J/\psi$
 energy distribution $d\sigma/dz$ at the photon-proton centre-of-mass
 energy $\sqrt{s_{\gamma p}} =100\,$GeV compared to experimental data
 from H1 and ZEUS~\cite{H1inelastic,ZEUSinelastic}.  The shaded area
 represents the sum of all contributions with colour-octet matrix
 elements according to Equation~\ref{eq_mevar}. The lines
 corresponding to separate colour-octet channels are plotted for
 $\langle {\cal O}^{J/\psi}[8,{}^1\!S_0]\rangle= \langle {\cal
 O}^{J/\psi}[8,{}^3\!P_0]\rangle/m_c^2=0.012\,\mbox{GeV}^3$.  The
 colour-singlet cross section is evaluated in leading order in
 $\alpha_s$.  Other parameters: GRV(LO) proton and photon parton
 distribution functions~\cite{Gluck:1995uf,Gluck:1992jc};
 $\Lambda^{(4)}_{LO} = 200$~MeV; factorisation and renormalisation
 scale $\mu=2 m_c$.}
\label{figure11}
\end{figure}

\vspace*{2mm}

\noindent
{\bf The \begin{boldmath}$J/\psi$\end{boldmath} energy
distribution$\;$} The $J/\psi$ energy spectrum is a particularly
sensitive probe of the different production mechanisms.  The
colour-singlet and colour-octet contributions to the differential
cross section through direct and resolved photon processes are
presented in
Figure~\ref{figure11}~\cite{Cacciari:1996dg,Beneke:1998re}. The results
are shown as a function of the scaling variable $z=p_\psi \cdot p_p /
p_\gamma \cdot p_p$ at a typical
\mbox{HERA} energy of $\sqrt{s\hphantom{tk}} \!\!\!\!\!  _{\gamma
p}\,\, = 100$~GeV in the inelastic region $p_t > 1$~GeV. Adopting the
NRQCD matrix elements as determined from the Tevatron $J/\psi$ data
(see Section~\ref{sec_me}, Equation~\ref{eq_mevar}), colour-octet
processes are predicted to contribute significantly near the upper and
lower endpoint of the energy spectrum. As mentioned previously, the
colour-octet enhancement is a consequence of $t$-channel gluon
exchange in the $c\bar{c}[8,{}^1S_0]$ and $c\bar{c}[8,{}^3P_J]$
processes. The actual size of the effect, however, is difficult to
quantify given the substantial uncertainties in the values of the
$\langle {\cal O}^{J/\psi}[8,{}^1S_0]\rangle$ and $\langle {\cal
O}^{J/\psi} [8,{}^3P_0]\rangle$ matrix elements, as indicated by the
shaded error band.

The theoretical predictions are compared to experimental results
obtained by the \mbox{H1}~\cite{H1inelastic} and
\mbox{ZEUS}~\cite{ZEUSinelastic} collaborations at~\mbox{HERA} in
Figure~\ref{figure11}. The colour-octet enhancement in the region
$z\;\simgt\;0.6$ is not supported by the experimental data. At first
sight this could lead to rather stringent constraints on the
colour-octet matrix elements $\langle {\cal
O}^{J/\psi}[8,{}^1S_0]\rangle$ and $\langle {\cal O}^{J/\psi}
[8,{}^3P_0]\rangle$ and might point towards an inconsistency with the
values obtained for these matrix elements from other
processes. However, the peaked shape of the $z$-distribution is
derived neglecting the energy transfer of soft-parton emission in the
non-perturbative transition of the colour-octet charm quark
pairs. Such kinematic effects become important near the endpoint of
the $J/\psi$ energy spectrum, and a summation of singular higher-order
terms in the velocity expansion is necessary to predict the shape of
the $z$-distribution in the region $z\;\simgt\;
0.7$~\cite{Beneke:1997qw}.\footnote{The situation is analogous to the
well-known problem of extracting the CKM matrix element $|V_{ub}|$
from the endpoint region of the lepton energy distribution in
semileptonic $B$ decay.} A quantitative study~\cite{Beneke:2000gq}
reveals that the kinematic impact of soft-gluon emission in the
hadronisation of colour-octet $c\bar{c}$ pairs at large $z$ is not
well under control, even after the summation of higher-order NRQCD
corrections. A more stringent cut on the $J/\psi$ transverse momentum
is needed to remove the sensitivity of the prediction to processes
which contribute in the elastic region at $z\approx 1$. At present, no
reliable constraints on the size of colour-octet contributions in
$J/\psi$ photoproduction can be derived from the $J/\psi$ energy
spectrum near the upper endpoint.

The low-$z$ region is not expected to be sensitive to higher-order
terms in the velocity expansion. Therefore, if the data could be
extended to $z\;\simlt\; 0.3$, an important resolved photon
contribution should be
visible~\cite{Cacciari:1996dy,Kniehl:1997fv,Kniehl:1997gh}, if the
colour-octet matrix elements are not significantly smaller than
extracted from the Tevatron data and expected from power counting
rules. A preliminary measurement~\cite{H1lowz} indeed shows evidence
for resolved processes, but the statistical error is still too large
to establish the importance of colour-octet channels in the low-$z$
region.

\vspace*{2mm}

\begin{figure}[h,t,b]
\vspace*{-10mm}
\hspace*{10mm}
\includegraphics[width=0.8\textwidth,clip]{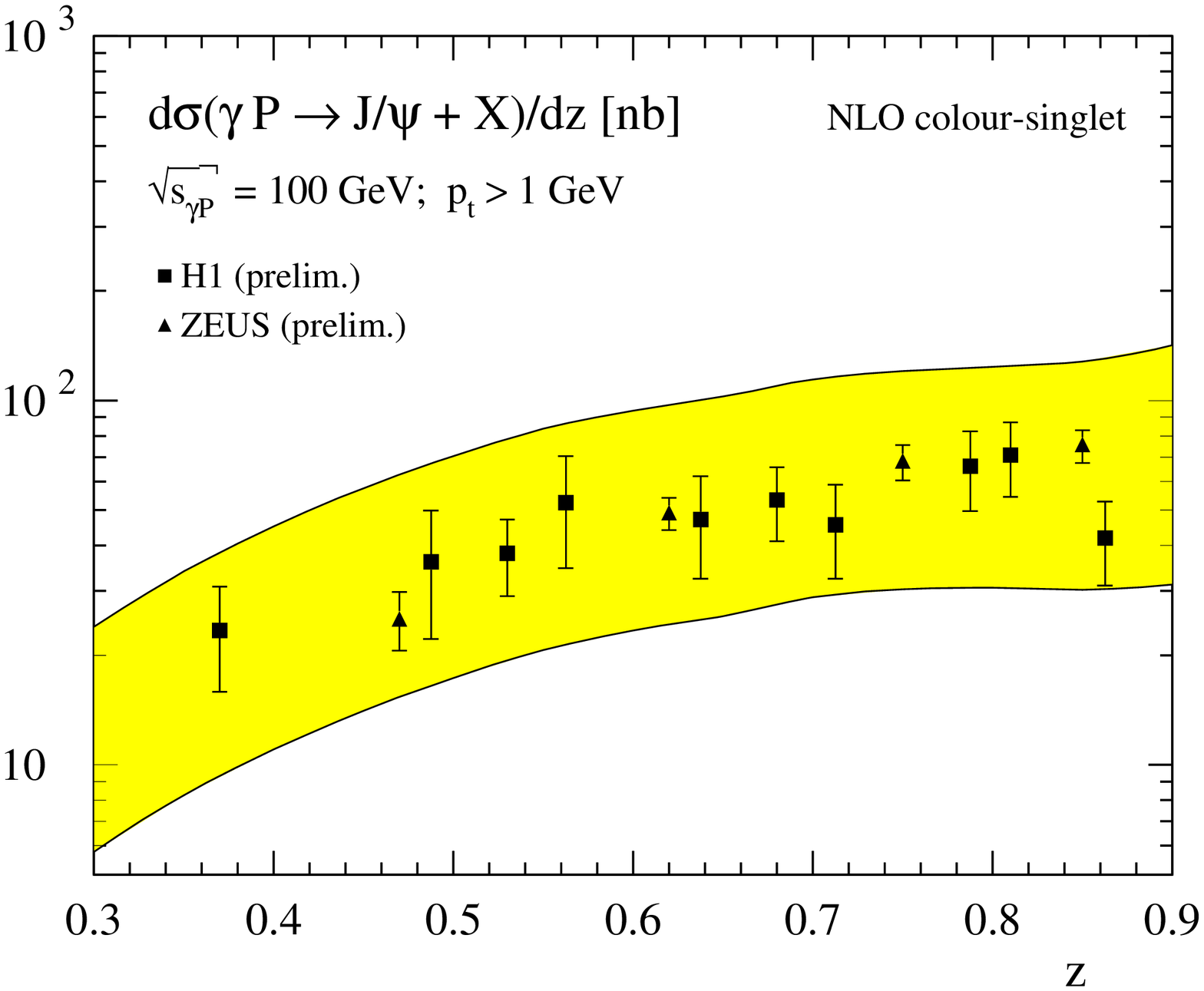}

\vspace*{-15mm}
\hspace*{10mm}
\includegraphics[width=0.8\textwidth,clip]{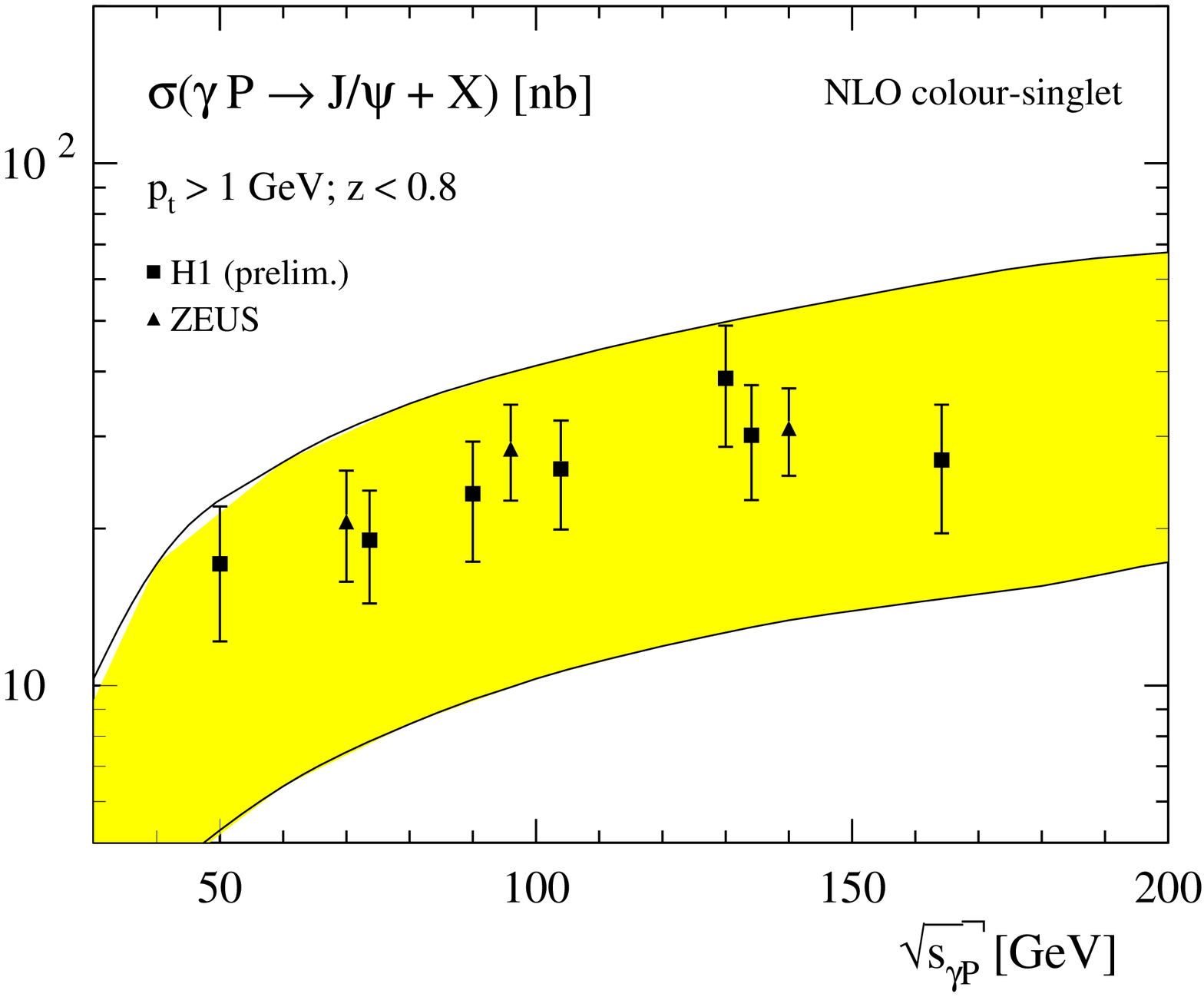}
\vspace*{-5mm}
\caption[Next-to-leading order colour-singlet prediction for
 inelastic $J/\psi$ photoproduction at HERA compared to experimental
 data] {\small Next-to-leading order colour-singlet prediction for
 inelastic $J/\psi$ photoproduction compared to experimental data from
 H1 and ZEUS~\cite{H1inelastic,ZEUSinelastic,Breitweg:1997we}. Upper
 Figure: $J/\psi$ energy distribution $d\sigma/dz$ at the
 photon-proton centre-of-mass energy $\sqrt{s_{\gamma p}}
 =100\,$GeV. Lower Figure: inelastic $J/\psi$ photoproduction cross
 section as a function of the photon-proton centre-of-mass energy
 $\sqrt{s_{\gamma p}}$.  Parameters: MRST parton distribution
 functions~\cite{Martin:2000ww}; factorisation and renormalisation
 scale $\mu = \sqrt{2}m_c$~\cite{Kramer:1996nb}. The error band is
 obtained by varying the charm quark mass and $\alpha_s$ in the range
 1.3~GeV $\le m_c\le$ 1.5~GeV and 0.1225 $\ge \alpha_s(M_Z)\ge$
 0.1175, respectively.}
\label{figure12}
\end{figure}

\noindent
{\bf NLO colour-singlet contributions$\;$} The current experimental
results on $J/\psi$ photoproduction are adequately described by the
colour-singlet channel alone, once higher-order QCD corrections are
included~\cite{Kramer:1995zi,Kramer:1996nb} and once the normalisation
uncertainty due to the choice for the charm quark mass and the QCD
coupling is taken into account.  The NLO calculation for the $J/\psi$
energy distribution is confronted with the experimental data in
Figure~\ref{figure12}. The shaded error band illustrates the large
normalisation uncertainty of the theoretical prediction: the upper
boundary corresponds to a charm quark mass $m_c = 1.3$~GeV and
$\alpha_s(M_Z) = 0.1225$, while the lower boundary has been obtained
using $m_c = 1.5$~GeV and $\alpha_s(M_Z) = 0.1175$.\footnote{The
variation of the cross section with $\alpha_s$ has been studied in a
consistent way by adopting the corresponding set of parton
distribution functions~\cite{Martin:2000ww}.} From
Figure~\ref{figure12} one can infer that the shape of the energy
distribution in the colour-singlet channel is not very sensitive to
the choice of input parameters. The theoretical prediction is in good
agreement with the experimental results.  A similar conclusion can be
drawn from the analysis of the inelastic cross section as a function
of the photon-proton energy, Figure~\ref{figure12}.

\begin{figure}[h,t,b]
\vspace*{-10mm}
\hspace*{8mm}
\includegraphics[width=0.9\textwidth,clip]{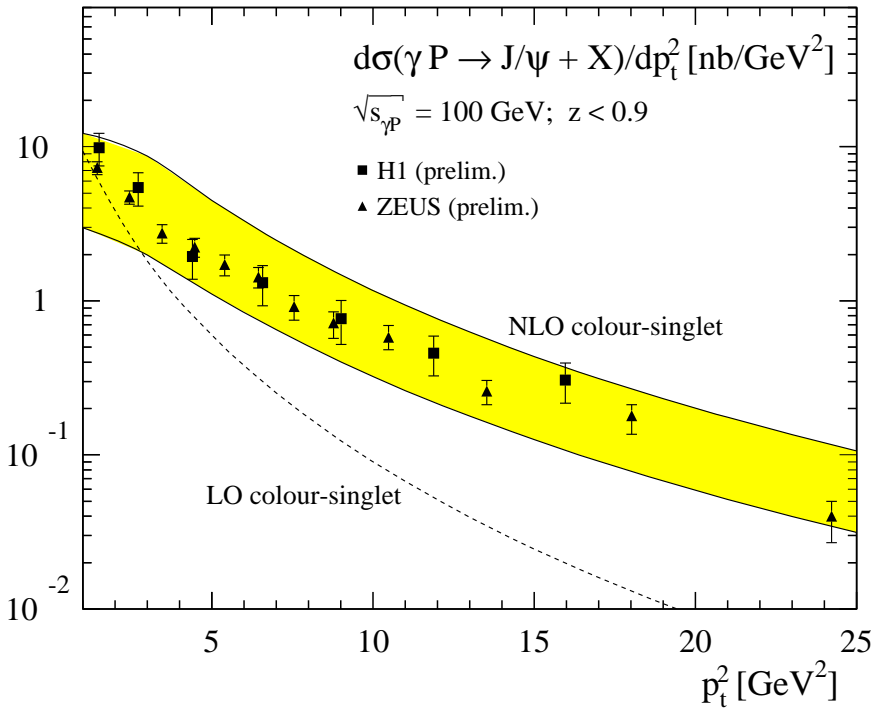}
\vspace*{-4mm}
\caption[Leading order and next-to-leading order colour-singlet prediction
 for the $J/\psi$ transverse momentum distribution in photoproduction
 at HERA compared experimental data]{\small Leading order and
 next-to-leading order colour-singlet prediction for the $J/\psi$
 transverse momentum spectrum $d\sigma/dp_t^2$ at the photon-proton
 centre-of-mass energy $\sqrt{s_{\gamma p}} =100\,$GeV compared to
 experimental data from H1 and
 ZEUS~\cite{H1inelastic,ZEUSinelastic}. Parameters as in
 Figure~\ref{figure12}.}
\label{figure13}
\end{figure}

The next-to-leading order QCD corrections are crucial to describe the
shape of the $J/\psi$ transverse momentum distribution. The NLO
colour-singlet cross section includes processes like $\gamma + g \to
Q\overline{Q}[1,{}^3S_1] + g + g $ which are dominated by
$t$-channel gluon exchange and scale as $\sim \alpha_s^3 m_Q^2 /
p_t^6$. At $p_t\;\simgt\;m_Q$ their contribution is enhanced with
respect to the leading-order cross section, which scales as $\sim
\alpha_s^2 m_Q^4/p_t^8$. The comparison with the experimental data,
Figure~\ref{figure13}, confirms the importance of the higher-order corrections.

The colour-singlet channel provides a successful description of
$J/\psi$ photoproduction as far as present data on energy and
transverse momentum distributions is concerned. The substantial
normalisation uncertainties of the theoretical prediction, as well as
the inaccurate kinematical treatment of soft gluon emission in the
hadronisation of colour-octet $c\bar{c}$ pairs, however, do not allow
to place stringent constraints on the role of colour-octet
processes.  With the expected increase in luminosity at
\mbox{HERA}, observables like $J/\psi$ polarisation and various other
quarkonium final states will provide more information on the
importance of the different production mechanisms, see
Sections~\ref{sec_eppol} and \ref{sec_heraup}.

\vspace*{2mm}

\noindent
{\bf \begin{boldmath}$\psi(2S)$\end{boldmath} photoproduction$\;$} 
While a non-relativistic treatment may be appropriate to describe
$J/\psi$ photoproduction in many circumstances, significant
relativistic corrections are expected for radial excitations like
$\psi(2S)$. In the static limit $v \to 0$, the
$\sigma(\psi(2S))/\sigma(J/\psi)$ cross section ratio is given by the
ratio of the NRQCD long-distance factors $\langle {\cal
O}^{\psi(2S)}[1,{}^3S_1]\rangle /$ $\langle {\cal
O}^{J/\psi}[1,{}^3S_1]\rangle$ to all orders in perturbation theory.
The colour-singlet matrix elements can be related to the quarkonium
decay widths, so that $\sigma(\psi(2S)) / \sigma(J/\psi)$ $=
\Gamma^{\psi(2S)}_{ee}/$ $\Gamma^{J/\psi}_{ee} =$ $0.41\pm 
0.05$~\cite{Groom:2000in} in the static approximation. The
non-relativistic estimate is not confirmed by recent HERA measurements
which yield $\sigma(\psi(2S))/\sigma(J/\psi) = 0.210 \pm
0.058$~\cite{H1inelastic} and $\sigma(\psi(2S))/\sigma(J/\psi) = 0.242
\pm 0.065$~\cite{ZEUSpsii}, respectively. Higher-order terms in the
velocity expansion will modify the theoretical prediction, their
impact is, however, hard to quantify.  Colour-octet processes might
increase or decrease the cross section ratio, depending on the value
of the non-perturbative matrix elements.  No quantitative prediction
is possible given the present theoretical error on $\langle {\cal
O}^{\psi} [8,{}^1S_0]
\rangle$ and $\langle {\cal O}^{\psi} [8,{}^3P_0] \rangle$. Another
source of relativistic corrections originates from NRQCD operators
which involve total derivatives on fermion
bilinears~\cite{Beneke:1997qw,Beneke:1997av}. Although these operators
are formally suppressed by powers of $v^2$ relative to the leading
colour-singlet contribution, they are crucial to reproduce the
hadronic phase space constraints of the quarkonium production cross
section. According to factorisation, all dependence on the specific
quarkonium state resides within the NRQCD matrix elements. Since the
short-distance coefficients are consequently expressed in terms of the
heavy quark mass $2 m_Q$ rather than the quarkonium mass $m_H$, the
physical phase space restrictions for producing a specific bound state
are not manifest at fixed order in the velocity expansion. While $2
m_Q - m_H \sim m_Q v^2$ is formally suppressed and numerically small
for $J/\psi$, the difference between parton and hadron kinematics can
lead to a large ambiguity in predicting the cross section of radial
excitations like $\psi(2S)$ where $2 m_c - m_{\psi} \approx
1~\mbox{GeV}$. The ambiguity can be resolved by summing certain
higher-order terms in the velocity expansion~\cite{Beneke:1997qw}. The
summation, however, leads to a non-perturbative distribution function,
which can not be calculated except within models. A rough
estimate~\cite{Kramer:1996nb}, obtained by assuming that the effective
charm masses in the short-distance cross sections scale like the
corresponding $\psi(2S)$ and $J/\psi$ masses, implies a significant
phase space suppression factor of $(m_{J/\psi} / m_{\psi})^3 \approx
0.6$, consistent with the experimental observation.

\subsection{$J/\psi$ production in deep-inelastic scattering\label{sec_dis}}
Leptoproduction of quarkonium has been extensively studied in the
context of the colour-singlet
model~\citer{Baier:1982zz,Merabet:1994sm} and in the framework of the
NRQCD factorisation approach including colour-octet
processes~\cite{Fleming:1998fq,Yuan:2001cn}. At large photon
virtualities $Q^2$, the theoretical prediction is under better control
than in the case of photoproduction because higher-order corrections
and higher-twist contributions should become less important as $Q^2$
increases. Moreover, diffractive processes~\cite{Brodsky:1994kf} are
expected to be suppressed at $Q^2
\gg 4m_Q^2$, and the total quarkonium leptoproduction cross section
can be sensibly compared with the NRQCD prediction, contrary to
photoproduction.  The leading-order $J/\psi$ production process is
${\cal O}(\alpha\alpha_s)$ and proceeds purely through intermediate
${}^1S_0$ and ${}^3P_J$ colour-octet configurations. The
leptoproduction cross section in the high-energy limit $Q^2,s\gg
4m_c^2$~\cite{Fleming:1998fq},
\begin{eqnarray}
\lefteqn{\hspace*{-10mm}
\lim_{m^2_c/Q^2,m^2_c/s\to 0}\sigma(eP\to e + J/\psi+X)\to 
\int\!{dQ^2 \over Q^2} \int\!\! dy \int\!\! dx \, f_{g/p}(x) 
\, \delta(xys-Q^2) }
\nonumber \\
& & \;\;\;\;\;\; \times
{2 \alpha_s(Q^2)\alpha^2 e^2_c \pi^2 \over m_c
Q^2}\,{1+(1-y)^2\over y}
\left( \langle{\cal O}^{J/\psi}[8,{}^1S_0]\rangle + 3
{\langle{\cal O}^{J/\psi}[8,{}^3P_0]\rangle \over m^2_c} \right)
\label{helim}
\end{eqnarray}
($y$ is the momentum fraction of the $J/\psi$ relative to the incoming
electron) thus directly probes the NRQCD matrix elements $\langle
{\cal O}^{J/\psi}[8,{}^1\!S_0]\rangle$ and $\langle {\cal
O}^{J/\psi}[8,{}^3\!P_0]\rangle$ and provides a sensitive test of
NRQCD factorisation. A more precise knowledge of $\langle {\cal
O}^{J/\psi}[8,{}^1\!S_0]\rangle$ and $\langle {\cal
O}^{J/\psi}[8,{}^3\!P_0]\rangle$ would also help to clarify the role
of colour-octet processes in $J/\psi$ photoproduction. At ${\cal
O}(\alpha\alpha_s^2)$ $J/\psi$ is produced through ${}^3S_1$
colour-singlet and ${}^1S_0$, ${}^3S_1$ and ${}^3P_J$ colour-octet
configurations. Results for the parton cross sections can be found
in~\cite{Yuan:2001cn}. The ${\cal O}(\alpha\alpha_s^2)$ colour-octet
processes are particularly important at large values of the
inelasticity $z$. Note, however, that the upper endpoint region of the
$z$-distribution in $J/\psi$ leptoproduction can not be predicted
reliably without summing large higher-order corrections in the NRQCD
velocity expansion, similar to photoproduction.

\begin{figure}[htb]
\vspace*{-5mm}
\hspace*{10mm}
\includegraphics[width=0.85\textwidth,clip=]{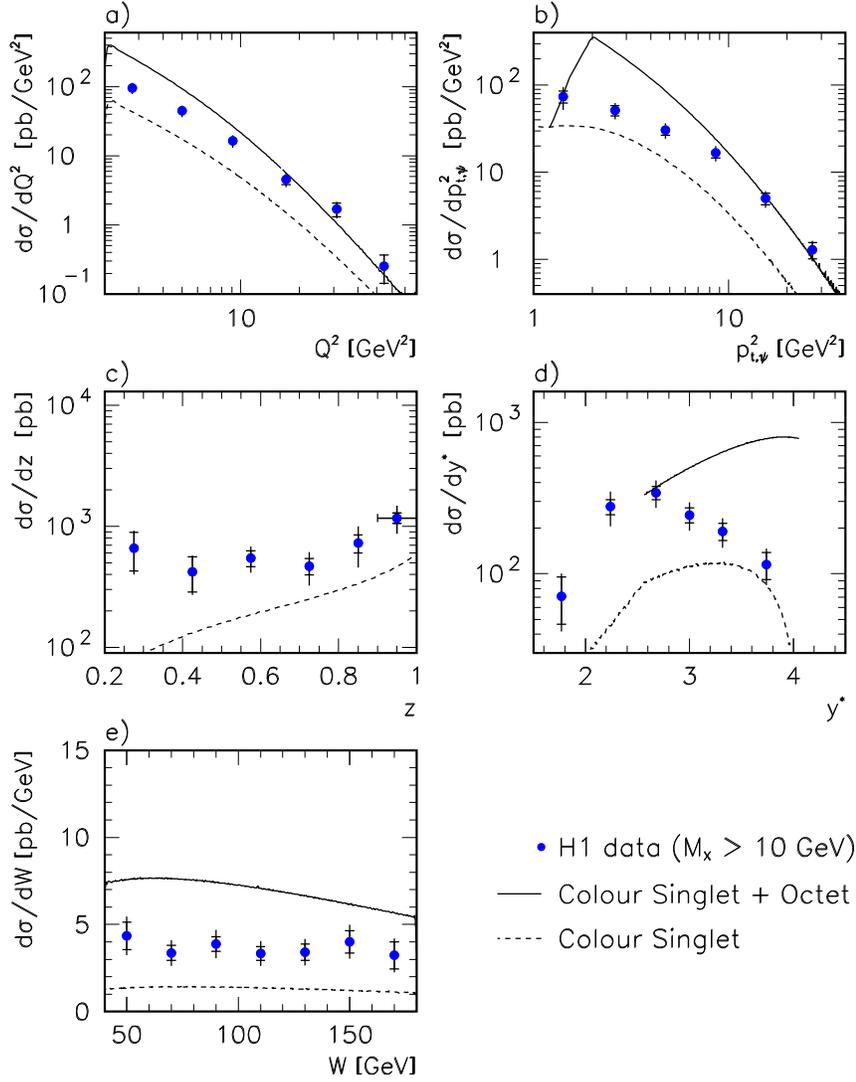}
\vspace*{-5mm}
\caption[Colour-singlet and colour-octet contributions to inelastic $J/\psi$ 
 leptoproduction at HERA compared to experimental data]{\small
 Colour-singlet and colour-octet contributions to inelastic $J/\psi$
 leptoproduction $ep \to e\, J/\psi\, X$~\cite{Fleming:1998fq}
 compared to experimental data from H1~\cite{Adloff:1999zs}.  The
 kinematic region is 2~GeV${}^2 \le Q^2\le 80$~GeV${}^2$, 40~GeV~$\le
 W \le 180$~GeV. The variable $y^*$ denotes the rapidity of the
 $J/\psi$ in the $\gamma^*p$ centre-of-mass system. The invariant mass
 of the hadronic final state has been restricted to $M_X > 10$~GeV to
 suppress diffractive contributions. Parameters: GRV(LO) parton
 distribution functions~\cite{Gluck:1995uf,Gluck:1992jc};
 factorisation and renormalisation scales $\mu = \sqrt{Q^2+4m_c^2}$;
 $m_c = 1.5$~GeV. NRQCD matrix elements $\langle {\cal
 O}^{J/\psi}[1,^3\!S_1]\rangle = 1.1$~GeV${}^3$, $\langle {\cal
 O}^{J/\psi}[8,^1\!S_0]\rangle = 0.01$~GeV${}^3$, and $\langle {\cal
 O}^{J/\psi}[8,^3\!P_0]\rangle/m_c^2= 0.005$~GeV${}^3$.}
\label{figure14}
\end{figure}

Experimental results on $J/\psi$ leptoproduction at
HERA~\cite{Adloff:1999zs} are shown in Figure~\ref{figure14} compared
to the leading-order NRQCD predictions. The theoretical cross sections
include the ${\cal O}(\alpha\alpha_s^2)$ colour-singlet process and
the ${\cal O}(\alpha\alpha_s)$ colour-octet contributions. Taken at
face value, the colour-singlet channel underestimates the data by a
factor $\sim 2-3$. There is, however, a considerable amount of
uncertainty in the cross section normalisation, in particular from the
value of the charm quark mass, the strong coupling, and the choice of
renormalisation and factorisation scales. On the other hand, adopting
$\langle {\cal O}^{J/\psi}[8,{}^1\!S_0]\rangle = 0.01$~GeV${}^3$ and
$\langle {\cal O}^{J/\psi}[8,{}^3\!P_0]\rangle/m_c^2= 0.005$~GeV${}^3$
the sum of colour-singlet and colour-octet contributions exceeds the
data by up to a factor of three. The shape of the distributions favour
the prediction of the colour-singlet channel, but more experimental
information extending to larger $Q^2$ values is needed before firm
conclusions can be drawn.

\subsection{$J/\psi$ and $\psi(2S)$ polarisation\label{sec_eppol}}

Polarisation measurements are powerful to assess the relative
importance of colour-octet processes in different kinematic
regions. First preliminary results~\cite{ZEUSpol} indicate a non-zero
polarisation signature in $J/\psi$ photoproduction at HERA. With the
expected increase in luminosity, the analysis of decay angular
distributions could provide information on the relevance of the
various quarkonium production processes, complementing the previously
discussed polarisation measurement at hadron colliders.

\begin{figure}[htb]
\vspace*{-10mm}
\hspace*{8mm}
\epsfig{figure=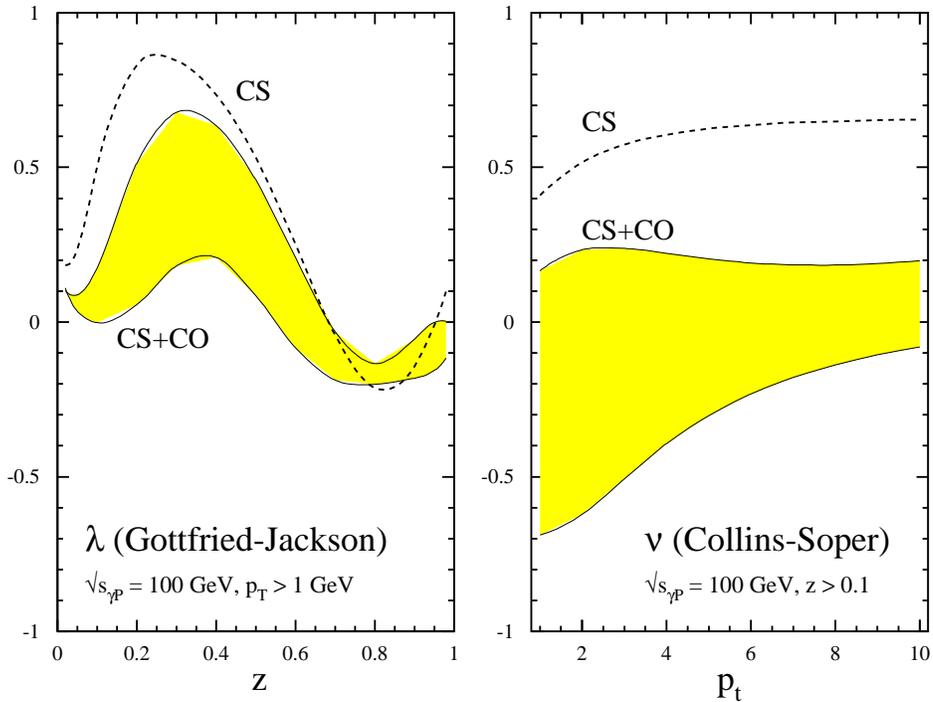,%
        bbllx=20pt,bblly=200pt,bburx=550pt,bbury=635pt,%
        width=0.85\textwidth,clip=}
\vspace*{-7mm}
\caption[Colour-singlet and colour-octet contributions to the polar and
 azimuthal angular parameters in the leptonic decay of $J/\psi$
 produced in photoproduction]{\small Colour-singlet and colour-octet
 contributions to the polar and azimuthal angular parameters in the
 leptonic decay of $J/\psi$ produced in
 photoproduction. (See~\cite{Beneke:1998re} for the definition of the
 Gottfried-Jackson and Collins-Soper reference frame.)  The dashed
 line is the colour-singlet channel prediction. The shaded area
 bounded by the solid lines represents the sum of all contributions
 with colour-octet matrix elements according to
 Equation~\ref{eq_mevar}.  Parameters: GRV(LO) parton distribution
 functions~\cite{Gluck:1995uf,Gluck:1992jc}; factorisation and
 renormalisation scale $\mu = 2m_c^2$; $m_c = 1.5$~GeV.}
\label{figure15}
\vspace*{-2mm}
\end{figure}

The general decay angular distribution can be parametrised as
\begin{equation}
  \frac{d\Gamma(J/\psi\to l^+l^-)}{d\Omega}
  \propto  
  1 + \lambda \cos^2\theta + \mu \sin 2\theta \cos\phi
  + \frac{\nu}{2} \sin^2\theta \cos 2\phi,
  \label{paramdef}
\end{equation}
where $\theta$ and $\phi$ refer to the polar and azimuthal angle of
the $l^+$ three-momentum with respect to a coordinate system defined
in the $J/\psi$ rest frame. The parameters $\lambda, \mu, \nu$ can be
calculated within NRQCD as a function of the kinematic variables like
$z$ and $p_t$~\cite{Beneke:1998re}. Two examples are shown in
Figure~\ref{figure15}, the polar angle parameter $\lambda$ and the
azimuthal angle parameter $\nu$ as a function of the energy fraction
$z$ and the transverse momentum $p_t$, respectively. Each plot
exhibits the result from the colour-singlet channel alone (dashed
line) and the result after including colour-octet contributions. Since
the decay angular distribution parameters involve ratios of cross
sections, the dependence on parameters that affect the absolute
normalisation, such as the charm quark mass, strong coupling, the
renormalisation/factorisation scale and parton distribution, cancels
to a large extent and does not constitute a significant uncertainty.

The $z$-dependence of the polar angle asymmetry
(Figure~\ref{figure15}, left plot) is distinctive and can be used to
discriminate between NRQCD and the colour-evaporation model, which
always predicts unpolarised $J/\psi$.\footnote{The predictions for the
angular parameters in the endpoint region of the $z$ distribution
depend on the non-perturbative distribution functions that result from
a summation of higher-oder corrections in the velocity
expansion. Note, however, that the dependence is significant only if
the angular parameter varies strongly in the endpoint region and
disappears entirely if its distribution is flat~\cite{Beneke:1998re}.}
The transverse momentum distribution, on the other hand, can prove
very useful to determine the relative magnitude of colour-singlet and
colour-octet contributions: While colour-octet processes tend to
predict unpolarised quarkonium in the $p_t$ region considered here,
large and positive values of the azimuthal parameter $\nu$ are
expected for $p_t\;\simgt\;5$~GeV if the cross section is dominated by
the colour-singlet channel, see Figure~\ref{figure15}, right plot.  To
observe the unique transverse polarisation signature of gluon
fragmentation into colour-octet $c\bar{c}$ pairs one would have to
isolate the resolved photon contribution at small $z\;\simlt\; 0.3$
and measure the polar angular distribution for large transverse
momenta $p_t\;\simgt\; 10$~GeV. A comprehensive analysis of the $p_t$
and $z$ dependence of the angular parameters $\lambda, \mu, \nu$ for
different reference frames can be found in~\cite{Beneke:1998re}.

Polarisation in $J/\psi$ photoproduction through the colour-singlet
channel has also been studied in the context of the
$k_t$-factorisation approach~\cite{Baranov:1998af}.  The virtuality of
the initial state gluon increases the fraction of longitudinally
polarised $J/\psi$, in particular at large transverse momentum. The
effect becomes significant at $p_t\;\simgt\; 3$~GeV, and could be
probed at the upgraded HERA collider by analysing the polar angle
asymmetry. The impact of colour-octet processes on the polarisation
prediction in the $k_t$-factorisation approach has not been estimated
so far.

Information on the quarkonium production mechanism can also be
obtained from the angular distribution in inelastic $J/\psi$
leptoproduction~\cite{Korner:1982fm,Fleming:1998fq,Yuan:2001cn}.  The
theoretical prediction at leading order only depends on the ratio of
the NRQCD colour-octet matrix elements $\langle {\cal
O}^{J/\psi}[8,^1\!S_0]\rangle$ and $\langle {\cal
O}^{J/\psi}[8,^3\!P_0]\rangle$. Longitudinally polarised $J/\psi$ is
expected if the $^3\!P_J$ channel dominates, while the $^1\!S_0$
channel leads to unpolarised quarkonium. The experimental result,
$\lambda = 0.77^{+0.44}_{-0.38}$~\cite{Adloff:1999zs}, is consistent
with colour-singlet dominance, which implies $\lambda \approx
0.5$~\cite{Adloff:1999zs}. The experimental uncertainties are,
however, still too large to put stringent limits on the size of
colour-octet contributions.

\subsection{Prospects for the upgraded HERA collider\label{sec_heraup}}
With the HERA luminosity upgrade, a wealth of new quarkonium data will
become available. The existing measurements can be extended into new
kinematic regions, and observables like $J/\psi$ polarisation and
various other quarkonium final states will be accessible.  The future
analyses of quarkonium production at HERA offer unique possibilities
to test the theoretical framework of NRQCD factorisation and to
clarify the role of colour-octet processes. Some of the most interesting 
and promising reactions will be discussed in the following.


\noindent
{\bf Resolved photon processes$\;$} As emphasised previously, the
low-$z$ region of the $J/\psi$ energy spectrum in photoproduction is
dominated by resolved photon processes, which resemble the $J/\psi$
production mechanisms relevant at hadron colliders. Away from the
upper endpoint region, the energy spectrum is not expected to be
sensitive to higher-order terms in the velocity expansion, and a
measurement of the resolved cross section can provide a sensible test
of the importance of colour-octet processes.

\begin{figure}[htb]
\vspace*{-2mm}
\hspace*{0mm}
\epsfig{figure=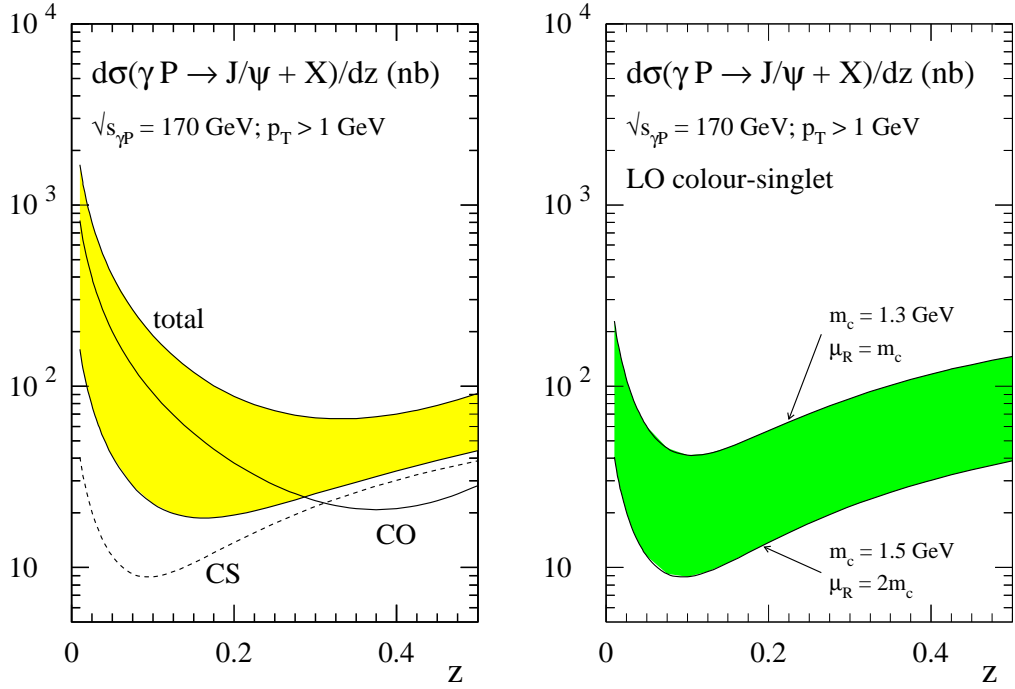,%
        bbllx=25pt,bblly=25pt,bburx=750pt,bbury=520pt,%
        width=0.95\textwidth,clip=}
\vspace*{-5mm}
\caption[Direct and resolved photon contributions to the $J/\psi$ 
 energy distribution in photoproduction]{\small Direct and
 resolved photon contributions to the $J/\psi$ energy distribution
 $d\sigma/dz$ at the photon-proton centre-of-mass energy
 $\sqrt{s_{\gamma p}} =170\,$GeV. Left: Colour-singlet (CS) and
 colour-octet (CO) contributions. The shaded area bounded by the solid
 lines represents the sum of all contributions with colour-octet
 matrix elements according to Equation~\ref{eq_mevar}. The separate
 colour-octet channel is plotted for $\langle {\cal
 O}^{J/\psi}[8,^1\!S_0]\rangle= \langle {\cal
 O}^{J/\psi}[8,^3\!P_0]\rangle/m_c^2=0.012\,\mbox{GeV}^3$.  Other
 parameters as in Figure~\ref{figure11}. Right: Colour-singlet
 contribution only. The error band is obtained by varying the charm
 quark mass and the renormalisation scale in the range 1.3~GeV $\le
 m_c\le$ 1.5~GeV and $m_c \le \mu_R \le 2 m_c$, respectively.}
\label{figure16}
\vspace*{0mm}
\end{figure}

Colour-singlet and colour-octet contributions to the low-$z$ region of
the energy spectrum are displayed in Figure~\ref{figure16}, left, at a
photon-proton energy of $W_{\gamma p} = 170$~GeV. Note that, because
only a small fraction of the incoming photon energy participates in the
hard interaction, the typical $W_{\gamma p}$ values for resolved
photon events within the detector acceptance are higher than for the
medium-$z$ analyses~\cite{H1lowz}. The right plot of
Figure~\ref{figure16} shows the uncertainty of the leading-order
colour-singlet cross section due to the variation of the charm quark
mass and the renormalisation scale.  Next-to-leading order corrections
to inelastic $J/\psi$ photoproduction through resolved photons are not
yet available.

A preliminary measurement~\cite{H1lowz} clearly establishes the
existence of resolved processes at small $z$, but experimental errors
and theoretical normalisation uncertainties do not allow to constrain
the size of colour-octet contributions. To confirm the significant
excess over the colour-singlet process observed at the Tevatron it
will be necessary to analyse resolved photon processes at larger
values of $p_t$.

\vspace*{2mm}

\noindent
{\bf \begin{boldmath}$\chi_c$\end{boldmath} photoproduction$\;$} The
measurement of inelastic $\chi_c$ photoproduction is a particularly
powerful way to discriminate between NRQCD and the colour-evaporation
model. The assumption of a single universal long-distance factor in
the colour-evaporation model implies a universal
$\sigma(\chi_c)/\sigma(J/\psi)$ ratio, and a large $\chi_c$ cross
section is predicted in photon-proton collisions, similar to hadron
colliders where $\sigma(\chi_c) / \sigma(J/\psi)
\approx 0.5$~\cite{Abe:1997yz}. In NRQCD, on the other hand, the 
$\sigma(\chi_c)/\sigma(J/\psi)$ ratio is process-dependent and
strongly suppressed in photoproduction. While $\chi_c$ can be produced
copiously in hadron-hadron collisions through $gg$, $gq$ and
$q\bar{q}$ interactions, the direct photon process $\gamma + g \to
c\bar c [1,{}^3\!P_J] + g$ is forbidden at leading order in
$\alpha_s$ due to the specific colour structure and charge conjugation
invariance. Apart from higher-order corrections and resolved photon
interactions, $\chi_c$ can thus only be produced through the
colour-octet process $\gamma + g \to c\bar c [8,{}^3\!S_1] +
g$~\cite{Ma:1996fd,Cacciari:1996dy}.  According to factorisation, and
up to corrections of ${\cal O}(\alpha_s,v^2)$, the cross section ratio
$\sigma(\chi_c)/\sigma(J/\psi)$ in direct photon interactions can be
expressed in terms of the corresponding long-distance factors
\begin{eqnarray}
\frac{\sigma(\gamma p \to \chi_J\,X)}
     {\sigma(\gamma p \to J/\psi \,X)} 
& \approx & \frac{\sigma(\gamma p \to c\bar c [8,{}^3\!S_1]\,X)\,
   \langle {\cal O}^{\chi_J}[8,^3\!S_1]\rangle }
        {\sigma(\gamma p \to c\bar c [1,{}^3\!S_1]\,X)
   \langle {\cal O}^{J/\psi}[1,{}^3\!S_1]\rangle }\nonumber\\[2mm]
& =& \frac{15}{8}\, (2J+1)\, \frac{\langle {\cal O}^{\chi_0} 
   [8,^3\!S_1]\rangle}
   {\langle {\cal O}^{J/\psi}[1,^3\!S_1]\rangle}
 \approx (2J+1)\,0.005 ,
\end{eqnarray}
with NRQCD matrix elements as listed in Table~\ref{table1}. A search
for $\chi_c$ production at HERA resulting in a cross section
measurement or upper cross section limit would directly probe the
colour-octet matrix element $\langle {\cal
O}^{\chi_J}[8,^3\!S_1]\rangle$ and test the assumption of a single
universal long-distance factor implicit in the colour-evaporation
model.

\vspace*{2mm}

\noindent
{\bf Photoproduction of spin-singlet states$\;$} The inclusion of
colour-octet processes is crucial to describe $\eta_c(nS)$ and
$h_c(nP)$ photoproduction. With regard to the $P$-wave state $h_c$,
the colour-octet contribution is required to cancel the infrared
divergence present in the colour-singlet cross
section~\cite{Fleming:1998md}. The production of $\eta_c$, on the
other hand, is dominated by colour-octet processes since the
colour-singlet cross section vanishes at leading-order due to charge
conjugation invariance~\cite{Hao:1999kq,Hao:2000ci}, similar to
$\chi_c$ photoproduction.  As discussed previously in the context of
future quarkonium physics at the Tevatron, Section~\ref{sec_tev2},
heavy-quark spin-symmetry allows to predict the production of
spin-singlet states like $\eta_c(nS)$ and $h_c(nP)$ in terms of the
NRQCD matrix elements that describe $J/\psi$ and $\chi_c$
production. The cross sections for $\eta_c(nS)$ and $h_c(nP)$
photoproduction are sizeable~\cite{Fleming:1998md,Hao:1999kq}, but it
is not obvious that these particles can be detected experimentally, 
even with high-statistics data.

\begin{figure}[t,h,b]
\vspace*{-13mm}
\hspace*{12mm}
\includegraphics[width=0.82\textwidth,clip]{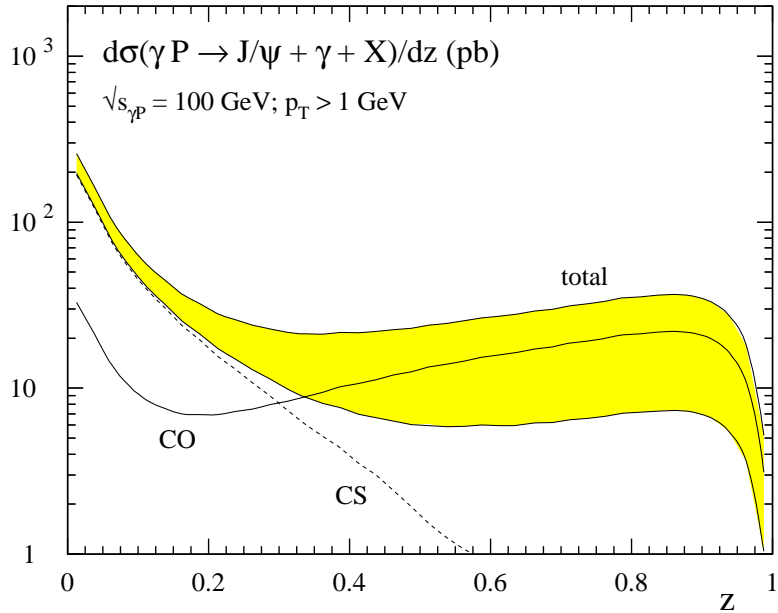}
\vspace*{-10mm}
\caption[Colour-singlet and colour-octet contributions  to the $J/\psi$
 energy distribution in associated $J/\psi + \gamma$ photoproduction]
 {\small Colour-singlet (CS) and colour-octet (CO) contributions due
 to direct and resolved photons to the $J/\psi$ energy distribution
 $d\sigma/dz$ in associated $J/\psi + \gamma$ production at the
 photon-proton centre-of-mass energy $\sqrt{s_{\gamma p}}
 =100\,$GeV. The shaded area bounded by the solid lines represents the
 sum of all contributions with colour-octet matrix elements according
 to Equation~\ref{eq_mevar}. The separate colour-octet channel is
 plotted for $\langle {\cal O}^{J/\psi}[8,^1\!S_0]\rangle= \langle
 {\cal O}^{J/\psi}[8,^3\!P_0]\rangle/m_c^2=0.012\,\mbox{GeV}^3$.
 Other parameters as in Figure~\ref{figure11}.}  \label{figure17}
\vspace*{-1mm}
\end{figure}

\noindent
{\bf Associated \begin{boldmath}$J/\psi+\gamma$\end{boldmath}
production$\;$} The energy spectrum of $J/\psi$ produced in
association with a photon, $\gamma p \to J/\psi +
\gamma$\,X, is a distinctive probe of colour-octet 
processes~\cite{Kim:1993at,Cacciari:1996dy,Mehen:1997vx,Cacciari:1997zu}.
In the colour-singlet channel and at leading-order in $\alpha_s$,
$J/\psi + \gamma$ can only be produced through resolved photon
interactions. The corresponding energy distribution is thus peaked at
low values of $z$. The intermediate and large-$z$ region of the energy
spectrum is expected to be dominated by the colour-octet process
$\gamma g \to c\bar c [8,{}^3\!S_1]\,\gamma$. Colour-singlet and
colour-octet contributions to associated $J/\psi + \gamma$
photoproduction through direct and resolved photon processes are shown
in Figure~\ref{figure17} as a function of the $J/\psi$
energy. Observation of
a substantial fraction of $J/\psi + \gamma$ events at $z\;\simgt\;
0.5$ would provide clear evidence for the presence of colour-octet
processes in quarkonium photoproduction. For a more comprehensive 
study of associated $J/\psi + \gamma$ production including various 
other differential distributions see~\cite{Cacciari:1997zu}.

\newpage
\vspace*{2mm}

\begin{figure}[t,h,b]
\vspace*{-10mm}
\hspace*{12mm}
\includegraphics[width=0.82\textwidth,clip]{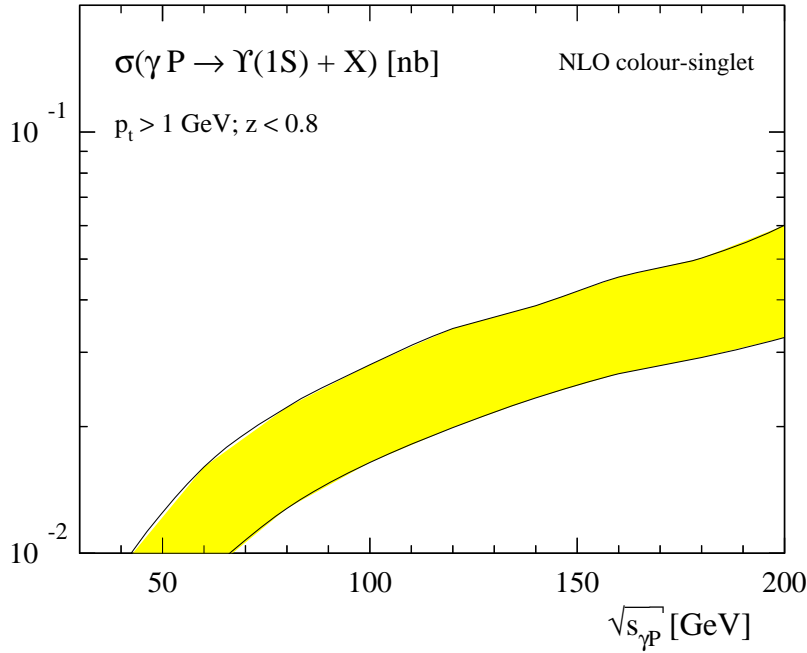}
\vspace*{-5mm}
\caption[Next-to-leading order colour-singlet prediction for inelastic
 $\Upsilon(1S)$ photoproduction]{\small Next-to-leading order
 colour-singlet prediction for inelastic $\Upsilon(1S)$
 photoproduction at HERA as a function of the photon-proton
 centre-of-mass energy $\sqrt{s_{\gamma p}}$.  Parameters: MRST parton
 distribution functions~\cite{Martin:2000ww}; factorisation and
 renormalisation scale $\mu = \sqrt{2}m_b$~\cite{Kramer:1996nb}. The
 error band is obtained by varying the bottom quark mass and
 $\alpha_s$ in the range 4.5~GeV $\le m_c\le$ 5~GeV and 0.1225 $\le
 \alpha_s(M_Z)\le$ 0.1175, respectively.}  \label{figure18}
\end{figure}

\noindent
{\bf Bottomonium production$\;$} With the significant increase in
statistics at the upgraded HERA collider, it might be possible to
study inelastic photoproduction of bottomonium states for the first
time. The large value of the bottom quark mass makes the perturbative
QCD predictions more reliable than for charm production, and the
application of NRQCD should be on safer grounds for the bottomonium
system where $v^2 \approx 0.1$. However, the production rates for
$\Upsilon$ states are suppressed, compared with $J/\psi$, by a more
than two orders of magnitude at HERA, a consequence of the smaller
bottom electric charge and the phase space reduction by the large $b$
mass. The next-to-leading order colour-singlet cross section in the
inelastic region $p_t > 1$~GeV and $z < 0.8$ is shown in
Figure~\ref{figure18} as a function of the photon-proton
centre-of-mass energy.
Several years of data taking at the upgraded HERA may be required to
obtain accurate experimental information on the inelastic $\Upsilon$
photoproduction cross section.

\section{Conclusion and Outlook\label{sec_conc}}
The NRQCD factorisation approach is a systematic framework for
inclusive quarkonium production and provides a striking and elegant
explanation for the large direct $J/\psi$ and $\psi(2S)$ cross
sections measured in $p\bar{p}$ collisions at the Fermilab Tevatron.
Still, more theoretical work and experimental information is needed to
firmly establish the applicability of NRQCD factorisation to
charmonium production.  The absence of transversely polarised $J/\psi$
and $\psi(2S)$ at large transverse momentum, if confirmed with higher
statistics data, could imply that the conventional NRQCD power
counting rules have to be replaced by alternative schemes, or that
higher-order QCD effects are essential to describe the spin-dependence
of the charm cross section.

The substantial increase in luminosity at the Tevatron Run~II will
allow to assess the role of colour-octet contributions in bottomonium
production more reliably. The calculation of next-to-leading order QCD
corrections and a systematic treatment of soft-gluon radiation are,
however, required to describe the cross section in the $p_t$ region
probed by the experiments at the Tevatron. A comprehensive test of the
NRQCD factorisation approach in the bottomonium sector will have to
wait until the LHC starts operating. If the charmonium mass is indeed
not large enough for a non-relativistic approach to be reliable, the
onset of transverse $\Upsilon$ polarisation at $p_t\gg M_{\Upsilon}$
may become the most decisive test of the NRQCD factorisation approach.

\vspace*{1mm}

A global analysis of different production processes is needed to
verify the universality of long-distance matrix elements in NRQCD.
The colour-octet matrix elements obtained from Tevatron data are in
qualitative agreement with analyses of other production processes,
like charmonium hadroproduction at fixed-target experiments and
charmonium production in $B$ decays. However, the sizeable theoretical
uncertainties, mainly related to QCD effects in the short-distance
cross sections, prevent a more quantitative test at present.

\vspace*{1mm}

The experimental data on charmonium photoproduction are adequately
described by the colour-singlet channel, including next-to-leading
order corrections in $\alpha_s$. But there is a considerable amount of
uncertainty in the normalisation of the theoretical prediction, and no
constraints on the size of colour-octet processes can be deduced from
the photoproduction analysis at present. The upgraded HERA collider
has a remarkable potential to test the theoretical framework more
thoroughly. The analysis of resolved photon processes and charmonium
polarisation, in particular, could be essential in assessing the
importance of different quarkonium production mechanisms. 

\vspace*{1mm}

The theoretical issues to be addressed in the future include a
critical evaluation of the power counting rules and the calculation of
higher-order QCD corrections. The importance of non-vanishing initial
state transverse momentum has to be clarified by analysing various
high-energy QCD processes. Moreover, systematic effects inherent in
NQRCD, such as the inaccurate treatment of energy conservation in the
hadronisation of heavy quark pairs and the mismatch between partonic
and hadronic phase space need to be addressed quantitatively. Finally,
reliable estimates of non-factorisable corrections and production
mechanisms beyond NRQCD are required to improve the theoretical
analyses. Theoretical progress and the wealth of new results from the
future Tevatron and HERA experiments will lead to exciting and
comprehensive phenomenology and result in a more complete and rigorous
understanding of quarkonium physics.

\vspace*{5mm}
\noindent
{\bf Acknowledgements$\;$} It is a pleasure to thank Martin Beneke,
Matteo Cacciari, Mario Greco, Fabio Maltoni, Mikko V\"anttinen, Peter
Zerwas and J\"org Zunft for their collaboration on various topics
presented in this review, and Alessandro Bertolin, Katja Kr\"uger,
David Lamb and Beate Naroska for many discussions on quarkonium
production at HERA.


\begin{thebibliography}{999}

\bibitem{Berger:1981ni}
E.~L.~Berger and D.~Jones,
Phys.\ Rev.\  {\bf D23} (1981) 1521.

\bibitem{Baier:1981uk}
R.~Baier and R.~R\"uckl,
Phys.\ Lett.\  {\bf B102} (1981) 364.

\bibitem{Bodwin:1995jh}
G.~T.~Bodwin, E.~Braaten and G.~P.~Lepage,
Phys.\ Rev.\  {\bf D51} (1995) 1125
[hep-ph/9407339];
erratum {\em ibid.} 
{\bf D55} (1997) 5853.

\bibitem{Fritzsch:1977ay}
H.~Fritzsch,
Phys.\ Lett.\  {\bf 67B} (1977) 217.

\bibitem{Halzen:1977rs}
F.~Halzen,
Phys.\ Lett.\  {\bf 69B} (1977) 105.

\bibitem{Gluck:1978zm}
M.~Gl\"uck, J.~F.~Owens and E.~Reya,
Phys.\ Rev.\  {\bf D17} (1978) 2324.

\bibitem{Gavai:1995in}
R.~Gavai, D.~Kharzeev, H.~Satz, G.~A.~Schuler, K.~Sridhar and R.~Vogt,
Int.\ J.\ Mod.\ Phys.\  {\bf A10} (1995) 3043
[hep-ph/9502270].

\bibitem{Schuler:1996ku}
G.~A.~Schuler and R.~Vogt,
Phys.\ Lett.\  {\bf B387} (1996) 181
[hep-ph/9606410].

\bibitem{Amundson:1997qr}
J.~F.~Amundson, O.~J.~Eboli, E.~M.~Gregores and F.~Halzen,
Phys.\ Lett.\  {\bf B390} (1997) 323
[hep-ph/9605295].

\bibitem{Edin:1997zb}
A.~Edin, G.~Ingelman and J.~Rathsman,
Phys.\ Rev.\  {\bf D56} (1997) 7317
[hep-ph/9705311].

\bibitem{Brodsky:1997tv}
S.~J.~Brodsky,
Int.\ J.\ Mod.\ Phys.\  {\bf A12} (1997) 4087
[hep-ph/9609415].

\bibitem{Hoyer:1999ha}
P.~Hoyer and S.~Peigne,
Phys.\ Rev.\  {\bf D59} (1999) 034011
[hep-ph/9806424].

\bibitem{Hoyer:1999dr}
P.~Hoyer,
Nucl.\ Phys.\ Proc.\ Suppl.\  {\bf 75B} (1999) 153
[hep-ph/9809362].

\bibitem{Braaten:1996pv}
E.~Braaten, S.~Fleming and T.~C.~Yuan,
Ann.\ Rev.\ Nucl.\ Part.\ Sci.\  {\bf 46} (1996) 197
[hep-ph/9602374].

\bibitem{Braaten:1996ix}
E.~Braaten, Talk given at 3rd International Workshop on Particle
Physics Phenomenology, Taipei, Taiwan, 14-17 Nov 1996
[hep-ph/9702225].

\bibitem{Beneke:1997av}
M.~Beneke, Lectures given at the 24th Annual SLAC Summer Institute on
Particle Physics: The Strong Interaction, From Hadrons to Protons (SSI
96), Stanford, CA, 19-30 Aug 1996
[hep-ph/9703429].

\bibitem{Rothstein:1999vz}
I.~Z.~Rothstein, Talk given at the 8th International Symposium on
Heavy Flavor Physics (Heavy Flavors 8), Southampton, England, 25-29
Jul 1999
[hep-ph/9911276].

\bibitem{Maltoni:2000km}
F.~Maltoni, Talk given at the 5th Workshop on QCD (QCD 2000),
Villefranche-sur-Mer, France, 3-7 Jan 2000.
[hep-ph/0007003].

\bibitem{Caswell:1986ui}
W.~E.~Caswell and G.~P.~Lepage,
Phys.\ Lett.\ B {\bf 167} (1986) 437.

\bibitem{Collins:1989gx}
J.~C.~Collins, D.~E.~Soper and G.~Sterman,
in ``Perturbative QCD'' (A.H.\ Mueller, ed.) (World Scientific Publ., 1989),  
ITP-SB-89-31.

\bibitem{Braaten:1996jt}
E.~Braaten and Y.~Chen,
Phys.\ Rev.\ D {\bf 54} (1996) 3216
[hep-ph/9604237].

\bibitem{Lepage:1992tx}
G.~P.~Lepage, L.~Magnea, C.~Nakhleh, U.~Magnea and K.~Hornbostel,
Phys.\ Rev.\  {\bf D46} (1992) 4052
[hep-lat/9205007].

\bibitem{Schuler:1997is}
G.~A.~Schuler,
Int.\ J.\ Mod.\ Phys.\ A {\bf 12} (1997) 3951
[hep-ph/9702230].

\bibitem{Fleming:2000ib}
S.~Fleming, I.~Z.~Rothstein and A.~K.~Leibovich,
CMU-0005, Dec 2000 
[hep-ph/0012062].

\bibitem{Sanchis-Lozano:2001rr}
M.~A.~Sanchis-Lozano,
IFIC-01-09, Mar 2001
[hep-ph/0103140].

\bibitem{Bodwin:1992ye}
G.~T.~Bodwin, E.~Braaten and G.~P.~Lepage,
Phys.\ Rev.\  {\bf D46} (1992) 1914
[hep-lat/9205006].

\bibitem{Braaten:1994cq}
E.~Braaten, Talk given at the Tennessee International Symposium on
Radiative Corrections: Status and Outlook, Gatlinburg, TN, 27 Jun - 1
Jul 1994 
[hep-ph/9409286].

\bibitem{Grinstein:2000xb}
B.~Grinstein,
Int.\ J.\ Mod.\ Phys.\ A {\bf 15} (2000) 461
[hep-ph/9811264].

\bibitem{Bodwin:1997mf}
G.~T.~Bodwin, D.~K.~Sinclair and S.~Kim,
Int.\ J.\ Mod.\ Phys.\ A {\bf 12} (1997) 4019
[hep-ph/9609371].

\bibitem{Albajar:1991hf}
C.~Albajar {\it et al.}  [UA1 Collaboration],
Phys.\ Lett.\ B {\bf 256} (1991) 112.

\bibitem{Abe:1992ww}
F.~Abe {\it et al.}  [CDF Collaboration],
Phys.\ Rev.\ Lett.\  {\bf 69} (1992) 3704.

\bibitem{Glover:1988az}
E.~W.~Glover, A.~D.~Martin and W.~J.~Stirling,
Z.\ Phys.\ C {\bf 38} (1988) 473
[Erratum-ibid.\ C {\bf 49} (1988) 526].

\bibitem{Mangano:1993kh}
M.~L.~Mangano,
Z.\ Phys.\ C {\bf 58} (1993) 651
[hep-ph/9302279].

\bibitem{Sansoni:1996yg}
A.~Sansoni  [CDF Collaboration],
Nuovo Cim.\  {\bf 109A} (1996) 827.

\bibitem{Papadimitriou:1997ck}
V.~Papadimitriou  [CDF Collaboration],
Int.\ J.\ Mod.\ Phys.\ A {\bf 12} (1997) 3867.

\bibitem{Abe:1997jz}
F.~Abe {\it et al.}  [CDF Collaboration],
Phys.\ Rev.\ Lett.\  {\bf 79} (1997) 572.

\bibitem{Baier:1983va}
R.~Baier and R.~R\"uckl,
Z.\ Phys.\  {\bf C19} (1983) 251.

\bibitem{Gastmans:1987be}
R.~Gastmans, W.~Troost and T.~T.~Wu,
Nucl.\ Phys.\  {\bf B291} (1987) 731.

\bibitem{Braaten:1993rw}
E.~Braaten and T.~C.~Yuan,
Phys.\ Rev.\ Lett.\  {\bf 71} (1993) 1673
[hep-ph/9303205].

\bibitem{Cacciari:1994dr}
M.~Cacciari and M.~Greco,
Phys.\ Rev.\ Lett.\  {\bf 73} (1994) 1586
[hep-ph/9405241].

\bibitem{Braaten:1994xb}
E.~Braaten, M.~A.~Doncheski, S.~Fleming and M.~L.~Mangano,
Phys.\ Lett.\  {\bf B333} (1994) 548
[hep-ph/9405407].

\bibitem{Roy:1994ie}
D.~P.~Roy and K.~Sridhar,
Phys.\ Lett.\  {\bf B339} (1994) 141
[hep-ph/9406386].

\bibitem{Braaten:1995vv}
E.~Braaten and S.~Fleming,
Phys.\ Rev.\ Lett.\  {\bf 74} (1995) 3327
[hep-ph/9411365].

\bibitem{Cacciari:1995yt}
M.~Cacciari, M.~Greco, M.~L.~Mangano and A.~Petrelli,
Phys.\ Lett.\  {\bf B356} (1995) 553
[hep-ph/9505379].

\bibitem{Cho:1996vh}
P.~Cho and A.~K.~Leibovich,
Phys.\ Rev.\  {\bf D53} (1996) 150
[hep-ph/9505329].

\bibitem{Cho:1996ce}
P.~Cho and A.~K.~Leibovich,
Phys.\ Rev.\  {\bf D53} (1996) 6203
[hep-ph/9511315].

\bibitem{Beneke:1998re}
M.~Beneke, M.~Kr\"amer and M.~V\"anttinen,
Phys.\ Rev.\ D {\bf 57} (1998) 4258
[hep-ph/9709376].

\bibitem{Buchmuller:1981su}
W.~Buchm\"uller and S.~H.~Tye,
Phys.\ Rev.\  {\bf D24} (1981) 132.

\bibitem{Eichten:1995ch}
E.~J.~Eichten and C.~Quigg,
Phys.\ Rev.\  {\bf D52} (1995) 1726
[hep-ph/9503356].

\bibitem{Lai:1999wy}
H.~L.~Lai {\it et al.}  [CTEQ Collaboration],
hep-ph/9903282.

\bibitem{Beneke:1997yw}
M.~Beneke and M.~Kr\"amer,
Phys.\ Rev.\  {\bf D55} (1997) 5269
[hep-ph/9611218].

\bibitem{Abe:1997yz}
F.~Abe {\it et al.}  [CDF Collaboration],
Phys.\ Rev.\ Lett.\  {\bf 79} (1997) 578.

\bibitem{Papadimitriou:2001bb}
V.~Papadimitriou  [CDF Collaboration],
FERMILAB-CONF-00-308-E.

\bibitem{fabio}
F.~Maltoni, private communication.

\bibitem{Beneke:1997qw}
M.~Beneke, I.~Z.~Rothstein and M.~B.~Wise,
Phys.\ Lett.\  {\bf B408} (1997) 373
[hep-ph/9705286].

\bibitem{Kuhn:1993qw}
J.~H.~K\"uhn and E.~Mirkes,
Phys.\ Rev.\  {\bf D48} (1993) 179
[hep-ph/9301204].

\bibitem{Petrelli:1998ge}
A.~Petrelli, M.~Cacciari, M.~Greco, F.~Maltoni and M.~L.~Mangano,
Nucl.\ Phys.\  {\bf B514}, 245 (1998)
[hep-ph/9707223].

\bibitem{Petrelli:2000rh}
A.~Petrelli,
Nucl.\ Phys.\ Proc.\ Suppl.\  {\bf 86} (2000) 533
[hep-ph/9910274].

\bibitem{Kramer:1996nb}
M.~Kr\"amer,
Nucl.\ Phys.\  {\bf B459} (1996) 3
[hep-ph/9508409].

\bibitem{Collins:1985kg}
J.~C.~Collins, D.~E.~Soper and G.~Sterman,
Nucl.\ Phys.\  {\bf B250} (1985) 199.

\bibitem{Cacciari:2000sy}
M.~Cacciari,
Nucl.\ Phys.\ B {\bf 571} (2000) 185
[hep-ph/9910412].

\bibitem{Sridhar:1998rt}
K.~Sridhar, A.~D.~Martin and W.~J.~Stirling,
Phys.\ Lett.\  {\bf B438} (1998) 211
[hep-ph/9806253].

\bibitem{Cano-Coloma:1997rn}
B.~Cano-Coloma and M.~A.~Sanchis-Lozano,
Nucl.\ Phys.\  {\bf B508} (1997) 753
[hep-ph/9706270].

\bibitem{Sanchis-Lozano:2000um}
M.~A.~Sanchis-Lozano,
Nucl.\ Phys.\ Proc.\ Suppl.\  {\bf 86} (2000) 543
[hep-ph/9907497].

\bibitem{Kniehl:1999qy}
B.~A.~Kniehl and G.~Kramer,
Eur.\ Phys.\ J.\  {\bf C6} (1999) 493
[hep-ph/9803256].

\bibitem{Sjostrand:1994yb}
T.~Sj\"ostrand,
Comput.\ Phys.\ Commun.\  {\bf 82}, 74 (1994).

\bibitem{Hagler:2001eu}
P.~Hagler, R.~Kirschner, A.~Schafer, L.~Szymanowski and O.~V.~Teryaev,
Phys.\ Rev.\ D {\bf 63} (2001) 077501
[hep-ph/0008316].

\bibitem{Yuan:2001cp}
F.~Yuan and K.~Chao,
Phys.\ Rev.\ D {\bf 63} (2001) 034006
[hep-ph/0008302].

\bibitem{Braaten:2000qk}
E.~Braaten, B.~A.~Kniehl and J.~Lee,
Phys.\ Rev.\ D {\bf 62} (2000) 094005
[hep-ph/9911436].

\bibitem{Martin:1993zi}
A.~D.~Martin, W.~J.~Stirling and R.~G.~Roberts,
Phys.\ Lett.\ B {\bf 306} (1993) 145
[Erratum-ibid.\ B {\bf 309} (1993) 492].

\bibitem{Lai:1997mg}
H.~L.~Lai {\it et al.},
Phys.\ Rev.\ D {\bf 55} (1997) 1280
[hep-ph/9606399].

\bibitem{Gluck:1995uf}
M.~Gl\"uck, E.~Reya and A.~Vogt,
Z.\ Phys.\ C {\bf 67} (1995) 433.

\bibitem{Martin:1996as}
A.~D.~Martin, R.~G.~Roberts and W.~J.~Stirling,
Phys.\ Lett.\ B {\bf 387} (1996) 419
[hep-ph/9606345].

\bibitem{Martin:1998sq}
A.~D.~Martin, R.~G.~Roberts, W.~J.~Stirling and R.~S.~Thorne,
Eur.\ Phys.\ J.\ C {\bf 4} (1998) 463
[hep-ph/9803445].

\bibitem{Tung:1994ua}
W.~K.~Tung,
{\it Prepared for International Workshop on Deep Inelastic Scattering and Related Subjects, Eilat, Israel, 6-11 Feb 1994}.

\bibitem{Kwiecinski:1997ee}
J.~Kwiecinski, A.~D.~Martin and A.~M.~Stasto,
Phys.\ Rev.\ D {\bf 56} (1997) 3991
[hep-ph/9703445].

\bibitem{Hagler:2001dd}
P.~Hagler, R.~Kirschner, A.~Schafer, L.~Szymanowski and O.~V.~Teryaev,
Phys.\ Rev.\ Lett.\  {\bf 86} (2001) 1446
[hep-ph/0004263].

\bibitem{Beneke:1999ks}
M.~Beneke, F.~Maltoni and I.~Z.~Rothstein,
Phys.\ Rev.\ D {\bf 59} (1999) 054003
[hep-ph/9808360].

\bibitem{Ma:2000bz}
J.~P.~Ma,
Phys.\ Lett.\ B {\bf 488} (2000) 55
[hep-ph/0006060].

\bibitem{Cho:1995ih}
P.~Cho and M.~B.~Wise,
Phys.\ Lett.\  {\bf B346} (1995) 129
[hep-ph/9411303].

\bibitem{Beneke:1996yb}
M.~Beneke and I.~Z.~Rothstein,
Phys.\ Lett.\  {\bf B372} (1996) 157
[hep-ph/9509375];
erratum {\em ibid.} {\bf D54} (1997) 7082.

\bibitem{Leibovich:1997pa}
A.~K.~Leibovich,
Phys.\ Rev.\  {\bf D56} (1997) 4412
[hep-ph/9610381].

\bibitem{Affolder:2000nn}
T.~Affolder {\it et al.}  [CDF Collaboration],
Phys.\ Rev.\ Lett.\  {\bf 85} (2000) 2886
[hep-ex/0004027].

\bibitem{Kniehl:2000nn}
B.~A.~Kniehl and J.~Lee,
Phys.\ Rev.\ D {\bf 62} (2000) 114027
[hep-ph/0007292].

\bibitem{Yuan:2000qe}
F.~Yuan and K.~Chao, Sep 2000 
[hep-ph/0009224].

\bibitem{Yuan:2001xw}
F.~Yuan and K.~Chao,
Phys.\ Lett.\ B {\bf 500} (2001) 99
[hep-ph/0009252].

\bibitem{Marchal:2000wd}
N.~Marchal, S.~Peigne and P.~Hoyer,
Phys.\ Rev.\ D {\bf 62} (2000) 114001
[hep-ph/0004234].

\bibitem{Maul:2001fw}
M.~Maul,
Nucl.\ Phys.\ B {\bf 594} (2001) 89
[hep-ph/0009279].

\bibitem{Abe:1995an}
F.~Abe {\it et al.}  [CDF Collaboration],
Phys.\ Rev.\ Lett.\  {\bf 75} (1995) 4358.

\bibitem{Affolder:1999wm}
T.~Affolder {\it et al.}  [CDF Collaboration],
hep-ex/9910025.

\bibitem{CDF-ups}
CDF collaboration, CDF Note 5027, unpublished;

\bibitem{LHC-workshop}
M.~Kr\"amer and F.~Maltoni, 
in 'Bottom Production', 
P.~Nason, G.~Ridolfi, O.~Schneider, G.F.~Tartarelli, P.~Vikas et al.,
hep-ph/0003142, published in CERN-YR-2000/01, G.~Altarelli and 
M.L.~Mangano editors.

\bibitem{Braaten:2001cm}
E.~Braaten, S.~Fleming and A.~K.~Leibovich,
Phys.\ Rev.\ D {\bf 63} (2001) 094006
[hep-ph/0008091].

\bibitem{Mangano:1995yd}
M.~L.~Mangano, CERN-TH-95-190, Jul 1995, 
in Batavia Collider Workshop 1995:0120-136
[hep-ph/9507353].

\bibitem{Domenech:2000qg}
J.~L.~Domenech and M.~A.~Sanchis-Lozano,
Phys.\ Lett.\ B {\bf 476} (2000) 65
[hep-ph/9911332].

\bibitem{Braaten:2001gw}
E.~Braaten and J.~Lee,
Phys.\ Rev.\ D {\bf 63} (2001) 071501
[hep-ph/0012244].

\bibitem{Brown:2001bz}
C.~N.~Brown {\it et al.}  [FNAL E866 Collaboration],
Phys.\ Rev.\ Lett.\  {\bf 86} (2001) 2529
[hep-ex/0011030].

\bibitem{Kharchilava:1999wa}
A.~Kharchilava, T.~Lohse, A.~Somov and A.~Tkabladze,
Phys.\ Rev.\ D {\bf 59} (1999) 094023
[hep-ph/9811361].

\bibitem{Tkabladze:1999mb}
A.~Tkabladze,
Phys.\ Lett.\ B {\bf 462} (1999) 319
[hep-ph/9907210].

\bibitem{Mathews:1998nk}
P.~Mathews, P.~Poulose and K.~Sridhar,
Phys.\ Lett.\  {\bf B438} (1998) 336
[hep-ph/9803424].

\bibitem{Sridhar:1996vd}
K.~Sridhar,
Phys.\ Rev.\ Lett.\  {\bf 77} (1996) 4880
[hep-ph/9609285].

\bibitem{Qiao:1997wb}
C.~Qiao, F.~Yuan and K.~Chao,
Phys.\ Rev.\ D {\bf 55} (1997) 5437
[hep-ph/9701249].

\bibitem{Kim:1997bb}
C.~S.~Kim, J.~Lee and H.~S.~Song,
Phys.\ Rev.\  {\bf D55} (1997) 5429
[hep-ph/9610294].

\bibitem{Mathews:1999ye}
P.~Mathews, K.~Sridhar and R.~Basu,
Phys.\ Rev.\  {\bf D60} (1999) 014009
[hep-ph/9901276].

\bibitem{Barger:1996vx}
V.~Barger, S.~Fleming and R.~J.~Phillips,
Phys.\ Lett.\ B {\bf 371} (1996) 111
[hep-ph/9510457].

\bibitem{Braaten:1999th}
E.~Braaten, J.~Lee and S.~Fleming,
Phys.\ Rev.\  {\bf D60} (1999) 091501
[hep-ph/9812505].

\bibitem{Domenech:2001ri}
J.~L.~Domenech and M.~A.~Sanchis-Lozano,
Nucl.\ Phys.\ B {\bf 601} (2001) 395
[hep-ph/0012296].

\bibitem{Donnachie:1987pu}
A.~Donnachie and P.~V.~Landshoff,
Phys.\ Lett.\ B {\bf 185} (1987) 403.

\bibitem{Ryskin:1993ui}
M.~G.~Ryskin,
Z.\ Phys.\ C {\bf 57} (1993) 89.

\bibitem{Brodsky:1994kf}
S.~J.~Brodsky, L.~Frankfurt, J.~F.~Gunion, A.~H.~Mueller and M.~Strikman,
Phys.\ Rev.\ D {\bf 50} (1994) 3134
[hep-ph/9402283].

\bibitem{Keung:1983jb}
W.~Keung and I.~J.~Muzinich,
Phys.\ Rev.\ D {\bf 27} (1983) 1518.

\bibitem{Jung:1993cd}
H.~Jung, D.~Krucker, C.~Greub and D.~Wyler,
Z.\ Phys.\ C {\bf 60} (1993) 721.

\bibitem{Khan:1996pg}
H.~Khan and P.~Hoodbhoy,
Phys.\ Lett.\ B {\bf 382} (1996) 189
[hep-ph/9511360].

\bibitem{Paranavitane:2000if}
C.~B.~Paranavitane, B.~H.~McKellar and J.~P.~Ma,
Phys.\ Rev.\ D {\bf 61} (2000) 114502.

\bibitem{Kramer:1995zi}
M.~Kr\"amer, J.~Zunft, J.~Steegborn and P.~M.~Zerwas,
Phys.\ Lett.\ B {\bf 348} (1995) 657
[hep-ph/9411372].

\bibitem{Cacciari:1996dg}
M.~Cacciari and M.~Kr\"amer,
Phys.\ Rev.\ Lett.\  {\bf 76} (1996) 4128
[hep-ph/9601276].

\bibitem{Amundson:1997ik}
J.~Amundson, S.~Fleming and I.~Maksymyk,
Phys.\ Rev.\ D {\bf 56} (1997) 5844
[hep-ph/9601298].

\bibitem{Ko:1996xw}
P.~Ko, J.~Lee and H.~S.~Song,
Phys.\ Rev.\  {\bf D54} (1996) 4312 [hep-ph/9602223], 
Erratum-{\it ibid.} {\bf D60} (1999) 119902.

\bibitem{Maltoni:1998pt}
F.~Maltoni, M.~L.~Mangano and A.~Petrelli,
Nucl.\ Phys.\ B {\bf 519} (1998) 361
[hep-ph/9708349].

\bibitem{Godbole:1996ie}
R.~M.~Godbole, D.~P.~Roy and K.~Sridhar,
Phys.\ Lett.\ B {\bf 373} (1996) 328
[hep-ph/9511433].

\bibitem{Kniehl:1997fv}
B.~A.~Kniehl and G.~Kramer,
Phys.\ Lett.\ B {\bf 413} (1997) 416
[hep-ph/9703280].

\bibitem{Kniehl:1997gh}
B.~A.~Kniehl and G.~Kramer,
Phys.\ Rev.\ D {\bf 56} (1997) 5820
[hep-ph/9706369].

\bibitem{Gluck:1992jc}
M.~Gl\"uck, E.~Reya and A.~Vogt,
Phys.\ Rev.\ D {\bf 46} (1992) 1973.

\bibitem{H1inelastic}
H1 Collaboration, Contributed Paper 157aj, International
Europhysics Conference on High Energy Physics (EPS99), Tampere, 
Finland, 1999.

\bibitem{ZEUSinelastic}
ZEUS Collaboration, Contributed Paper 851, International
Conference on High Energy Physics (ICHEP2000), Osaka, Japan, 2000.

\bibitem{Beneke:2000gq}
M.~Beneke, G.~A.~Schuler and S.~Wolf,
Phys.\ Rev.\ D {\bf 62} (2000) 034004
[hep-ph/0001062].

\bibitem{Cacciari:1996dy}
M.~Cacciari and M.~Kr\"amer, Talk given at the Workshop on Future
Physics at HERA, Hamburg, Germany, 30-31 May 1996
[hep-ph/9609500].

\bibitem{H1lowz}
Katja Kr\"uger for the H1 Collaboration, Proceedings of the 
International Workshop on Deep Inelastic Scattering (DIS2000), 
Liverpool, England, 2000.

\bibitem{Martin:2000ww}
A.~D.~Martin, R.~G.~Roberts, W.~J.~Stirling and R.~S.~Thorne,
Eur.\ Phys.\ J.\ C {\bf 14} (2000) 133
[hep-ph/9907231].

\bibitem{Breitweg:1997we}
J.~Breitweg {\it et al.}  [ZEUS Collaboration],
Z.\ Phys.\ C {\bf 76} (1997) 599
[hep-ex/9708010].

\bibitem{Groom:2000in}
D.~E.~Groom {\it et al.}  [Particle Data Group Collaboration],
Eur.\ Phys.\ J.\ C {\bf 15} (2000) 1.

\bibitem{ZEUSpsii}
ZEUS Collaboration, Contributed Paper 504, International
Europhysics Conference on High Energy Physics (EPS99), Tampere, 
Finland, 1999.

\bibitem{Baier:1982zz}
R.~Baier and R.~R\"uckl,
Nucl.\ Phys.\ B {\bf 201} (1982) 1.

\bibitem{Korner:1982fk}
J.~G.~K\"orner, J.~Cleymans, M.~Kuroda and G.~J.~Gounaris,
Nucl.\ Phys.\ B {\bf 204} (1982) 6
[Erratum-ibid.\ B {\bf 213} (1982) 546].

\bibitem{Korner:1982fm}
J.~G.~K\"orner, J.~Cleymans, M.~Kuroda and G.~J.~Gounaris,
Phys.\ Lett.\ B {\bf 114} (1982) 195.

\bibitem{Merabet:1994sm}
H.~Merabet, J.~F.~Mathiot and R.~Mendez-Galain,
Z.\ Phys.\ C {\bf 62} (1994) 639.

\bibitem{Fleming:1998fq}
S.~Fleming and T.~Mehen,
Phys.\ Rev.\ D {\bf 57} (1998) 1846
[hep-ph/9707365].

\bibitem{Yuan:2001cn}
F.~Yuan and K.~Chao,
Phys.\ Rev.\ D {\bf 63} (2001) 034017
[hep-ph/0008301].

\bibitem{Adloff:1999zs}
C.~Adloff {\it et al.}  [H1 Collaboration],
Eur.\ Phys.\ J.\ C {\bf 10} (1999) 373
[hep-ex/9903008].

\bibitem{ZEUSpol}
R. Brugnera for the ZEUS Collaboration, Talk at the International
Workshop on Deep Inelastic Scattering (DIS2001), Bologna, Italy, 2001.

\bibitem{Baranov:1998af}
S.~P.~Baranov,
Phys.\ Lett.\ B {\bf 428} (1998) 377.

\bibitem{Ma:1996fd}
J.~P.~Ma,
Nucl.\ Phys.\ B {\bf 460} (1996) 109
[hep-ph/9510333].

\bibitem{Fleming:1998md}
S.~Fleming and T.~Mehen,
Phys.\ Rev.\ D {\bf 58} (1998) 037503
[hep-ph/9801328].

\bibitem{Hao:1999kq}
L.~Hao, F.~Yuan and K.~Chao,
Phys.\ Rev.\ Lett.\  {\bf 83} (1999) 4490
[hep-ph/9902338].

\bibitem{Hao:2000ci}
L.~Hao, F.~Yuan and K.~Chao,
Phys.\ Rev.\ D {\bf 62} (2000) 074023
[hep-ph/0004203].

\bibitem{Kim:1993at}
C.~S.~Kim and E.~Reya,
Phys.\ Lett.\ B {\bf 300} (1993) 298.

\bibitem{Mehen:1997vx}
T.~Mehen,
Phys.\ Rev.\ D {\bf 55} (1997) 4338
[hep-ph/9611321].

\bibitem{Cacciari:1997zu}
M.~Cacciari, M.~Greco and M.~Kr\"amer,
Phys.\ Rev.\ D {\bf 55} (1997) 7126
[hep-ph/9611324].

\end{thebibliography}
\end{document}